\documentclass[final,5p,twocolumn]{elsarticle}

\usepackage{amssymb}
\usepackage{amsmath}
\usepackage{amsthm}

\usepackage{amsfonts}
\usepackage{graphicx}
\usepackage{hyperref}   
\usepackage{bm}
\usepackage{mathtools}
\usepackage{color}
\usepackage{subcaption}   
\usepackage{booktabs}   
\usepackage{tabularx}   

\usepackage[sort&compress]{cleveref}

\usepackage{comment}

\biboptions{sort&compress}

\newtheorem{assumption}{Assumption}
\newtheorem{remark}{Remark}

\newtheorem{theorem}{Theorem}
\newtheorem{lemma}{Lemma}
\newtheorem{corollary}{Corollary}

\newcommand{\bx}{\bm{x}}
\newcommand{\bxb}{\bar{\bm{x}}}
\newcommand{\bof}{\bm{f}}

\newcommand{\bG}{\bm{G}}
\newcommand{\bGtil}{\bm{\widetilde{G}}}
\newcommand{\bu}{\bm{u}}
\newcommand{\bus}{\bm{u}_s}
\newcommand{\buh}{\bm{u}_h}
\newcommand{\bs}{\bm{s}}
\newcommand{\be}{\bm{e}}
\newcommand{\bz}{\bm{z}}
\newcommand{\dbz}{\dot{\bm{z}}}
\newcommand{\beh}{\bm{\hat{e}}}
\newcommand{\bvarTheta}{\bm{\varTheta}}
\newcommand{\bXi}{\bm{\Xi}}
\newcommand{\bI}{\bm{I}}
\newcommand{\bphi}{\bm{\phi}}
\newcommand{\bp}{\bm{p}}
\newcommand{\bS}{\bm{S}}
\newcommand{\bzeta}{\bm{\zeta}}
\newcommand{\bM}{\bm{M}}
\newcommand{\bD}{\bm{D}}
\newcommand{\bC}{\bm{C}}
\newcommand{\bd}{\bm{d}}
\newcommand{\ba}{\bm{a}}

\newcommand{\eh}{\hat{e}}
\newcommand{\epi}{\varepsilon}
\newcommand{\epib}{\bar{\varepsilon}}
\newcommand{\bepi}{\bm{\varepsilon}}
\newcommand{\dbepi}{\dot{\bm{\varepsilon}}}
\newcommand{\varphih}{\varphi_h}
\newcommand{\varphigam}{\varphi_{\gamma}}
\newcommand{\es}{e_s}
\newcommand{\des}{\dot{e}_s}
\newcommand{\epis}{\varepsilon_s}
\newcommand{\epibs}{\bar{\varepsilon}_s}
\newcommand{\depis}{\dot{\varepsilon}_s}
\newcommand{\epih}{\varepsilon_h}
\newcommand{\epibh}{\bar{\varepsilon}_h}
\newcommand{\depih}{\dot{\varepsilon}_h}

\newcommand{\krec}{k_{\rho}}
\newcommand{\bA}{\bm{A}}
\newcommand{\bR}{\bm{R}}

\newcommand{\cinv}{\bar{c}_1}

\newcommand{\mbR}{\mathbb{R}}
\newcommand{\mbRpos}{\mathbb{R}_{\geq 0}}
\newcommand{\mbRspos}{\mathbb{R}_{> 0}}
\newcommand{\mbRneg}{\mathbb{R}_{\leq 0}}
\newcommand{\mbN}{\mathbb{N}}
\newcommand{\mbI}{\mathbb{I}}

\newcommand{\mathdot}{\mathord{\cdot}}

\newcommand{\Ccal}{\mathcal{C}}
\newcommand{\Ical}{\mathcal{I}}
\newcommand{\Tcal}{\mathcal{T}}

\newcommand{\Linf}{\mathcal{L}_{\infty}}

\newcommand{\col}[1]{\mathrm{col}({#1})}
\newcommand{\cl}[1]{\mathrm{cl}({#1})}

\newcommand{\ngrad}[1]{\nabla_{\bx}{#1}}
\newcommand{\ngradd}[1]{\nabla_{\bx}^2{#1}}
\newcommand{\ngradt}[1]{\nabla_{t}{#1}}
\newcommand{\ntgrad}[1]{\nabla_{t \bx}{#1}}

\newcommand{\grad}[1]{\nabla_{\bx_1}{#1}}
\newcommand{\gradd}[1]{\nabla_{\bx_1}^2{#1}}
\newcommand{\gradt}[1]{\nabla_{t}{#1}}
\newcommand{\tgrad}[1]{\nabla_{t \bx_1}{#1}}

\newcommand{\diag}[1]{\mathrm{diag}({#1})}

\newcommand{\lambdamin}{\lambda_{\textit{min}}}
\newcommand{\psib}{\bar{\psi}}
\newcommand{\psih}[1]{\psi_{h_{#1}}}
\newcommand{\psis}[1]{\psi_{s_{#1}}}
\newcommand{\Ob}{\bar{\Omega}}
\newcommand{\Obs}{\bar{\Omega}_{s}}
\newcommand{\Obh}{\bar{\Omega}_{h}}
\newcommand{\Os}{\Omega_{s}}
\newcommand{\Osv}{\tilde{\Omega}_{s}}
\newcommand{\Oh}{\Omega_{h}}
\newcommand{\Oalph}{\Omega_{\alphh}}
\newcommand{\Ohi}{\Omega_{h}^{\mathrm{ir}}}
\newcommand{\Ohb}{\Omega_{h}^{\mathrm{br}}}
\newcommand{\Oz}{\Omega_{\bz}}
\newcommand{\alphb}{\bar{\alpha}}
\newcommand{\alphbh}{\bar{\alpha}_{h}}
\newcommand{\alphbs}{\bar{\alpha}_{s}}

\newcommand{\alphh}{\alpha_{h}}
\newcommand{\dalphh}{\dot{\alpha}_{h}}
\newcommand{\alphs}{\alpha_{s}}
\newcommand{\alphhstr}{\alpha_{h}^{\ast}}
\newcommand{\alphsstr}{\alpha_{s}^{\ast}}
\newcommand{\epsih}{\epsilon_{h}}
\newcommand{\epsis}{\epsilon_{s}}
\newcommand{\rhos}{\rho_{s}}
\newcommand{\drhos}{\dot{\rho}_{s}}
\newcommand{\rhosn}{\rho_{s}^{\text{n}}}
\newcommand{\drhosn}{\dot{\rho}_{s}^{\text{n}}}
\newcommand{\rhosr}{\rho_{s}^{\text{r}}}
\newcommand{\rhosrb}{\bar{\rho}_{s}^{\text{r}}}
\newcommand{\Orhosr}{\Omega_{\rhosr}}
\newcommand{\Ox}{\Omega_{\bx_1}}
\newcommand{\Oxsp}{\Omega_{\bx_1}^{s^\prime}}
\newcommand{\drhosr}{\dot{\rho}_{s}^{\text{r}}}
\newcommand{\lb}{\underline{b}}
\newcommand{\ub}{\overline{b}}
\newcommand{\taum}{\tau_\mathrm{max}}
\newcommand{\Ts}{T_{s}}

\newcommand{\rhol}{\underline{\rho}}
\newcommand{\rhou}{\overline{\rho}}

\journal{ArXiv}

\begin{document}

\begin{frontmatter}



\title{Robust Closed-Form Control for MIMO Nonlinear Systems under Generalized Conflicting Time-Varying Hard and Soft Constraints\tnoteref{label1} \\ (extended version)}

\tnotetext[label1]{This work is supported by ERC CoG LEAFHOUND, the KAW foundation, and the Swedish Research Council (VR).}

\author[1]{Farhad Mehdifar} \fnref{presentaddress} 
\ead{farhad.mehdifar@chalmers.se}
\author[2]{Charalampos P. Bechlioulis} \ead{chmpechl@upatras.gr}
\author[1]{Dimos V. Dimarogonas} \ead{dimos@kth.se}

\affiliation[1]{organization={Department of Decision and Control Systems},
            addressline={KTH Royal Institute of Technology}, 
            city={Stockholm},
            country={Sweden}}

\affiliation[2]{organization={Department of Electrical and Computer Engineering},
	addressline={University of Patras}, 
	city={Patras},
	country={Greece}}

\fntext[presentaddress]{Present address: 
Department of Mechanical Engineering, Chalmers University of Technology, Gothenburg, Sweden.}

\begin{abstract}
This paper introduces a novel robust closed-form control law to handle time-varying hard and soft constraints in uncertain high-relative-degree nonlinear MIMO systems. These constraints represent spatiotemporal specifications in mechanical systems' operational space, with hard constraints ensuring safety-critical requirements and soft constraints encoding performance or task objectives. Initially, all constraints are consolidated into two separate scalar time-varying hard and soft constraint functions, whose positive level sets define feasible regions. A closed-form control law is developed to enforce these constraints using appropriately designed reciprocal barriers and nonlinear transformation functions. When conflicts between hard and soft constraints arise, the control law prioritizes hard constraints by virtually relaxing soft constraints via a dynamic relaxation law. Notably, the proposed control law maintains low complexity, avoiding approximation schemes for system uncertainties. Simulation results confirm the effectiveness of the proposed method.
\end{abstract}




\end{frontmatter}



\section{Introduction}
\label{sec:intro}

Advances in science and technology have profoundly influenced control system design, shifting the focus beyond traditional objectives like stability and accuracy toward explicitly enforcing transient performance and safety requirements. These specifications are typically imposed by introducing (often time-varying) constraints in controller design for practical nonlinear systems, where violations may cause performance degradation, system damage, or safety hazards. A range of control strategies has been developed to guarantee performance and/or prevent unsafe behavior, including Model Predictive Control (MPC) \cite{mayne2014model}, Reference Governors (RGs) \cite{garone2017reference,nicotra2018explicit}, Control Barrier Functions (CBFs) \cite{ames2016control}, Barrier Lyapunov Functions (BLFs) \cite{tee2009barrier,tee2011control}, Funnel Control (FC) \cite{ilchmann2002tracking}, and Prescribed Performance Control (PPC) \cite{bechlioulis2008robust}. Each approach offers distinct strengths and limitations.

Among the aforementioned strategies, BLFs, FC, and PPC provide closed-form and robust designs for uncertain nonlinear systems, enforcing time-varying funnel-type constraints, $\rhol_i(t) < e_i = x_i - x_i^d(t) < \rhou_i(t)$, on independent tracking errors, where $x_i$ denotes independent state variables and $x_i^d(t)$ the desired trajectories. Properly designing the boundaries $\rhou_i(t) > \rhol_i(t)$ prescribes the desired transient and steady-state performance of tracking errors while satisfying certain classes of safety requirements. To achieve this, time-varying BLFs are constructed such that, as tracking errors approach the prescribed boundaries, the BLFs diverge to infinity, ensuring that system states respect the constraints \cite{tee2011control,jin2018adaptive}. PPC instead employs an error transformation mapping the constrained system into an unconstrained one, where uniform ultimate boundedness of the transformed system provides a necessary and sufficient solution \cite{bechlioulis2008robust,bechlioulis2014low}. The PPC transformation implicitly induces a constructive time-varying BLF. FC generalizes adaptive high-gain control by replacing the monotonically increasing gain with a time-varying, state-dependent function that enforces funnel constraints \cite{berger2018funnel,BERGER2025101024}.

Despite notable progress, existing BLF-, FC-, and PPC-based frameworks face two structural limitations under general time-varying constraints. First, they are tailored to collections of decoupled, funnel-type constraints, limiting their applicability to richer set representations and their deployment across diverse scenarios. Second, systematically handling potentially conflicting hard and soft constraints within them remains largely unresolved.

Throughout, we address \emph{explicitly defined}, and potentially conflicting, time-varying hard (safety) and soft (performance) constraints, i.e., both specifications are imposed on the same state variables. Hard and soft constraints may also arise implicitly, e.g., through hard actuator limits combined with soft performance objectives, or through conflicts between time-varying (soft) position bounds and (hard) velocity bounds in mechanical systems (see, e.g., \cite{ji2021saturation, fotiadis2023input, trakas2023robust, trakas2025input, lei2024adaptive}). These lie outside the scope of this article. Unlike optimization-based synthesis methods (e.g., MPC and CBF-QP-based controllers), which relax soft constraints via slack variables while preserving hard ones \cite{zeilinger2014soft, ames2019control}, deriving robust closed-form feedback laws that guarantee hard constraints while accommodating soft specifications is considerably harder.

Regarding the constraint classes considered by BLF, FC, and PPC-based methods, recent studies mainly integrate diverse time-varying boundary functions into funnel constraints: asymmetric \cite{jin2018adaptive}, fixed/finite-time \cite{yin2020robust,yang2025finite}, and tunnel \cite{ji2023tunnel,SHI2024111304} bounds. A well-known limitation of BLF- and PPC-based designs is the need for initial constraint satisfaction, requiring knowledge of the system’s initial condition to tune the bounds. To address this, many works employ large initial funnel bounds \cite{bu2022prescribed}, whereas deferred funnel constraints inherently remove this dependence \cite{sun2024global,song2018tracking}. This has been extended to irregular, intermittently active output funnel constraints in \cite{kong2022adaptive,shao2023unified,zhao2024decision}. In \cite{zhao2024decision}, the hard/soft terminology refers to the temporal regularization of a \emph{single} intermittently active funnel on a scalar output, whose smoothed (soft) form is in fact a tightening of the original; in this paper, as in \cite{zeilinger2014soft, ames2019control, mehdifar2022funnel}, these terms denote the priority between two simultaneously imposed and possibly conflicting specifications.

Collectively, these results address only funnel-type constraints, typically imposed on decoupled states or error signals, so satisfying one does not affect the others. Consequently, BLF/FC/PPC formulations yield time-varying boxes in the output/error space \cite{berger2021funnel,bechlioulis2014low,jin2018adaptive}. Our prior work \cite{mehdifar2023control, mehdifar2024low} overcame this by unifying multiple classes of time-varying constraints (beyond funnels) into a single scalar constraint function whose positive level set defines the feasible region, and by proposing a robust closed-form control law guaranteeing constraint satisfaction.

Concerning time-varying hard and soft constraints, most BLF-, FC-, and PPC-based results treat tracking-performance funnels as inviolable. This still permits satisfying certain hard (safety) constraints alongside tracking-error specifications when the former can be expressed in terms of system errors and are always compatible with the performance bounds. This is common, e.g., in multi-agent formation control, where performance bounds on formation errors are designed so that they also enforce hard constraints such as connectivity maintenance, inter-agent collision avoidance, and visibility constraints \cite{mehdifar2020prescribed, dai2020fixed, verginis2019robust, verginis2017robust, bechlioulis2019robust}.

For potentially conflicting time-varying hard and soft constraints, only a few results exist. In particular, \cite{zhu2025improved} studies trajectory tracking (as a soft constraint) under irregular, intermittent funnel-type hard constraints and proposes smooth modification of the desired trajectory to comply with the hard specifications. The work in \cite{mehdifar2022funnel} explicitly addresses both hard and soft funnel-type constraints and introduces a constraint-consistent online funnel-planning scheme that guarantees satisfaction of hard constraints at all times while enforcing soft constraints whenever feasible. This framework was later extended to accommodate reach-avoid-stay specifications \cite{das2024prescribed} and the control of flexible beam systems \cite{sun2025event}. The dynamic relaxation of soft funnel constraints proposed in \cite{mehdifar2022funnel} was further refined in \cite{zuo2026robust} through a static, event-triggered relaxation mechanism, improving transient performance during both the relaxation and recovery phases. A BLF-based cooperative tracking-in-formation design for marine vehicles is proposed in \cite{restrepo2024tracking}, treating inter-agent connectivity maintenance and collision avoidance as hard specifications and non-negativity of surge velocity as a soft requirement. The work in \cite{park2023collision} proposes flexible performance bounds for trajectory tracking of surface vehicles with moving-obstacle avoidance, which dynamically relaxes tracking performance when the collision-avoidance module activates; this is also extended to formation tracking in \cite{yoo2024adaptive}. Although these works provide closed-form robust control for hard and soft constraints, they either remain restricted to funnel-type specifications or are tailored to particular applications, limiting generalization.

\textbf{Contributions:} Motivated by these gaps, we propose a robust closed-form control law for uncertain high-relative-degree MIMO nonlinear systems under a generalized class of potentially conflicting time-varying hard and soft constraints. \textit{(i) Constraint class:} rather than one funnel per independent error signal, each specification is an arbitrary collection of time-varying inequalities on the same state variables, condensed into a single smooth scalar function whose positive level set is the feasible region. This admits coupled, arbitrarily shaped time-varying feasible sets, with decoupled funnels---yielding only time-varying boxes---recovered as a special case \cite[Remark 1]{mehdifar2024low}. \textit{(ii) Conflict handling:} when the two specifications cannot be met at once, the hard one is prioritized by enlarging the soft feasible region into a virtual soft set that contains it. A dynamic relaxation law drives this enlargement only while enforcing the soft specification would push the state towards the hard boundary, and contracts the virtual set back once the two are again compatible, without any online feasibility test or optimization. \textit{(iii) Controller design:} reciprocal barriers of the two scalar constraint functions yield closed-form velocity-level laws, lifted to the control input via tailored nonlinear transformations and a low-complexity backstepping-like recursion, without approximating or estimating the unknown dynamics and using no plant information beyond the system's known kinematics; shifting functions decouple the tuning from the initial condition. \textit{(iv) Guarantees:} Theorem \ref{th:main} establishes boundedness of all closed-loop signals and forward invariance of the intersection of the hard and virtual soft sets for all time; Corollary \ref{colo:main} renders this global, with the hard set invariant from the outset and the intersection after a settling time $\Ts \leq T$. Hard constraints are thus never violated, while soft ones are met within a user-prescribed finite time $T$ and recovered whenever compatible, barring the deadlocks discussed after Theorem \ref{th:main}. Relative to our earlier work, \cite{mehdifar2024low} treats generalized time-varying constraints but not the hard/soft interaction, while \cite{mehdifar2022funnel} is confined to box sets.

\textbf{Notations:} $\mbN$ and $\mbR$ denote the sets of natural and real numbers, respectively. $\mbRpos$, $\mbRspos$, and $\mbRneg$ denote non-negative, positive, and non-positive reals. $\mbR^n$ is the $n$-dimensional real vector space. Bold lowercase symbols denote vectors or vector-valued functions, bold uppercase matrices, and non-bold symbols scalars or scalar-valued functions. For instance, $\ba \in \mbR^n$ is an $n \times 1$ column vector, $\ba^\top$ its transpose, and $\|\ba\|$ its Euclidean norm. The column-concatenation operator is $\col{\ba_i}\coloneqq [\ba_1^\top,\ldots,\ba_m^\top]^\top \in \mbR^{mn}$, where $\ba_i \in \mbR^n$, $i\in\{1,\ldots,m\}$. The set of real $n \times m$ matrices is $\mbR^{n \times m}$. For $\bA \in \mbR^{n \times m}$, $\bA^\top$ is the transpose and $\|\bA\|$ the induced norm. The absolute value of a scalar is $|\cdot|$. $\bI_n \in \mbR^{n \times n}$ is the identity matrix and $\mathbf{0}_n \in \mbR^n$ the zero vector. $\diag{\cdot}$ forms a diagonal matrix. The closure of an open set $\Omega$ is $\cl{\Omega}$. The class of $n$-times continuously differentiable functions is $\Ccal^n$, and $\Linf$ denotes essentially bounded measurable functions. The index set $\mbI_i^j \coloneqq \{i,\ldots,j\}$ is defined for $i,j \in \mbN$ with $i\le j$. Let $f(t,\bx)$ be a scalar-valued function, $f:\mbR \times \mbR^n \rightarrow \mbR$. We also define: $\ngrad{f} \coloneqq \frac{\partial f(t,\bx)}{\partial \bx}^{\scriptscriptstyle \top} \in \mbR^n$,  $\ngradd{f} \coloneqq \frac{\partial}{\partial \bx} \ngrad{f}  \in \mbR^{n \times n}$, $\ntgrad{f} \coloneqq \frac{\partial}{\partial t} \ngrad{f}  \in \mbR^n$, $\ngradt{f} \coloneqq \frac{\partial f(t,\bx)}{\partial t}  \in \mbR$.

\section{Problem Formulation}
\label{sec:prob_formu}

Consider the following class of high-relative-degree MIMO nonlinear systems:
\begin{equation} \label{eq:sys_dynamics_mimo}
	\begin{aligned}
		\dot{\bx}_i &= \bof_i(t,\bxb_i) + \bG_i(t,\bxb_i) \bx_{i+1}, \; i \in \mbI_1^{r-1} \\
		\dot{\bx}_r &= \bof_r(t,\bxb_r) + \bG_r(t,\bxb_r) \bu,
	\end{aligned}
\end{equation}
where $\bx_i \coloneqq [x_{i,1}, x_{i,2}, \ldots, x_{i,n}]^\top \in \mbR^{n}$,  $\bxb_i \coloneqq [\bx_1^\top, \ldots, \bx_i^\top]^\top \in \mbR^{ni}, i \in \mbI_1^{r}$, and $r,n \in \mbN$. Moreover,  $\bx \coloneqq \bxb_r \in \mbR^{nr}$ is the state vector, available for measurement, and $\bu \in \mbR^n$ the control input. Furthermore, $\bof_i: \mbRpos \times  \mbR^{ni} \rightarrow \mbR^{n}$, and $\bG_i: \mbRpos \times \mbR^{ni} \rightarrow \mbR^{n \times n}, i \in \mbI_1^{r}$ are vector fields that are piece-wise continuous in $t$ and locally Lipschitz continuous in $\bxb_i$ for each $t \in \mbRpos$. Let $\bx(t;\bx(0), \bu)$ denote the solution of \eqref{eq:sys_dynamics_mimo} under initial condition $\bx(0)$ and control input $\bu$, and let $\bx_1(t;\bx(0)) \coloneqq \bx_1(t;\bx(0), \bu)$ be its $\bx_1$-component.

\begin{assumption} \label{assu:known_unknown}
	The matrix $\bG_1(t,\bx_1)$ is known, continuously differentiable, and uniformly nonsingular $\forall t \geq 0$ and $\forall \bx_1 \in \mbR^n$; i.e., there is a known continuous function $\cinv:\mbR^n \rightarrow \mbRspos$ such that $\|\bG_1^{-1}(t,\bx_1)\| \leq \cinv(\bx_1)$. Moreover, $\bof_i$, $i \in \mbI_1^{r}$, and $\bG_i$, $i \in \mbI_2^{r}$, are unknown.
\end{assumption} 

\begin{assumption} \label{assu:bounded_in_t_dyn}
	All elements of the maps $\bof_i$, and $\bG_i, i \in \mbI_1^{r}$, are bounded for each fixed $\bxb_i \in \mbR^{ni}$ and for all $t \geq 0$. In other words, there exist continuous functions $\bar{f}_i:\mbR^{ni} \rightarrow \mbR$ and $\bar{g}_i:\mbR^{ni} \rightarrow \mbR, i \in \mbI_1^{r}$, with unknown analytical expressions such that $\|\bof_i(t,\bxb_i)\| \leq \bar{f}_i(\bxb_i)$ and $\|\bG_i(t,\bxb_i)\| \leq \bar{g}_i(\bxb_i)$, for all $t \geq 0$ and all $\bxb_i \in \mbR^{ni}$. 
\end{assumption}

\begin{assumption}\label{assu:symmetric_com_pd}
	The matrices $\bGtil_i(t,\bxb_i) \coloneqq \frac{1}{2} \left(\bG_i^\top + \bG_i \right)$, $i \in \mbI_2^{r}$,
	are uniformly sign-definite with known signs $\forall t \geq 0$ and $\forall \bxb_i \in \mbR^{ni}$. Without loss of generality, we assume that all $\bGtil_i(t,\bxb_i)$ are positive definite, implying the existence of unknown strictly positive constants $\underline{\lambda}_i > 0, i \in \mbI_2^{r}$, such that $\lambdamin (\bGtil_i(t,\bxb_i)) \geq \underline{\lambda}_i > 0$,  $\forall t \geq 0$ and $ \forall \bxb_i \in \mbR^{ni}$.
\end{assumption}

Note that the class \eqref{eq:sys_dynamics_mimo} is square, and the nonsingularity of $\bG_1$ in Assumption~\ref{assu:known_unknown} together with the sign-definiteness in Assumption~\ref{assu:symmetric_com_pd} renders each $\bG_i$ nonsingular, establishing a global controllability property and identifying \eqref{eq:sys_dynamics_mimo} as a class of fully actuated systems with a well-defined vector relative degree $r$. Regarding Assumption~\ref{assu:known_unknown}, the case $\bG_1(t,\bx_1)=\bI_n$ holds whenever $\bx_2=\dot{\bx}_1$, e.g., for Euler--Lagrange models with $\bx_1=\bm{q}$, $\bx_2=\dot{\bm{q}}$ under \emph{any} generalized coordinates. More generally, when $\bx_2$ is a quasi-velocity in a different frame, $\bG_1$ is a configuration-dependent \emph{kinematic} map fixed by the chosen representation, known exactly and independent of the inertial, damping, friction, payload and disturbance terms; these instead act at the acceleration level and are absorbed into the unknown $\bof_i$ and $\bG_i$, $i\in\mbI_2^{r}$. Assumption~\ref{assu:known_unknown} thus amounts to knowledge of the system kinematics only, which is the sole plant information used by the control law; in the common case $\bG_1=\bI_n$, no plant information is required. A representative instance is a \emph{fully actuated} $3$-DOF surface vessel \cite{fossen2021handbook}, $\dot{\bm{\eta}}=\bm{J}(\psi)\bm{v}_b$, $\bM\dot{\bm{v}}_b+\bC(\bm{v}_b)\bm{v}_b+\bD(\bm{v}_b)\bm{v}_b=\bm{\tau}+\bm{d}(t)$, $\bm{\tau}\in\mathbb{R}^{3}$, with rotation matrix $\bm{J}(\psi)\in SO(3)$: here $\bG_1(\bx_1)=\bm{J}(\psi)$ is known and nonsingular for all $\psi$, with $\bm{J}^{-1}=\bm{J}^{\top}$, while $\bG_2=\bM^{-1}\succ0$ is unknown. Non-square or singular $\bG_1$, as in general underactuated, nonholonomic or flexible mechanical systems, falls outside \eqref{eq:sys_dynamics_mimo}, though some such systems admit a \emph{known} transformation into it, as in Section~\ref{sec:simu}.

Assumption \ref{assu:bounded_in_t_dyn} imposes a uniformity-in-time requirement rather
than a boundedness condition in the state: the unknown elements of $\bof_i(t,\bxb_i)$ and
$\bG_i(t,\bxb_i)$, $i \in \mbI_1^{r}$, may grow arbitrarily large with $\bxb_i$, but not with $t$; jumps or switching in $t$ of $\bof_i$, $i \in \mbI_1^{r}$, and of $\bG_i$, $i \in \mbI_2^{r}$, remain admissible.
The bound $\cinv$ in Assumption \ref{assu:known_unknown} is of the same nature: it does not
restrict how large $\|\bG_1^{-1}\|$ may become with $\bx_1$, but excludes $\bG_1$ from
becoming arbitrarily ill-conditioned as $t$ grows at a fixed $\bx_1$.

Let system \eqref{eq:sys_dynamics_mimo} be subject to the following generalized class of time-varying constraints on the $\bx_1$ states
\begin{subequations}\label{eq:hard_soft_constraints}
	\begin{align}
		\psih{j}(t,\bx_1) &> 0, \quad j \in \mbI_1^{m_h}, \label{eq:hard_constraints} \\
		\psis{j}(t,\bx_1) &> 0, \quad j \in \mbI_1^{m_s}, \label{eq:soft_constraints}
	\end{align}
\end{subequations}
where $m_h, m_s \in \mbN$, and $\psih{j}, \psis{j}: \mbRpos \times \mbR^n \rightarrow \mbR$ are $\Ccal^2$ in $\bx_1$ and $\Ccal^1$ in $t$, representing inequality-type hard and soft constraints, respectively. In practical mechanical systems modeled by \eqref{eq:sys_dynamics_mimo}, $\bx_1$ typically represents spatial coordinates (positions), so \eqref{eq:hard_soft_constraints} naturally encodes hard and soft spatiotemporal (space \& time) specifications. We make the following regularity assumptions for $\psi_{\star_j}$, $\star\in\{h,s\}$.

\begin{assumption} \label{assu:bounded_in_t_psi_fun}
	Functions $\psi_{\star_j}(\cdot,\cdot), j \in \mbI_1^{m_{\star}}$, $\star \in \{h,s\}$, are bounded for each fixed $\bx_1 \in \mbR^{n}$ and for all $t \geq 0$, i.e., there exist continuous functions $\psib_{\star_j}: \mbR^n \rightarrow \mbR$, such that $|\psi_{\star_j}(t,\bx_1)| \leq \psib_{\star_j}(\bx_1)$, for all $t \geq 0$ and all $\bx_1 \in \mbR^{n}$.
\end{assumption}

\begin{assumption} \label{assu:bounded_in_t_psi_derivatives}
	There exist continuous functions $\kappa_{\star_j}, \varkappa_{\star_j}, \mu_{\star_j} ,\iota_{\star_j}: \mbR^n \rightarrow \mbR$, $j \in \mbI_1^{m_{\star}}$, $\star \in \{h,s\}$, such that $\| \grad{\psi_{\star_j}} \| \leq \kappa_{\star_j}(\bx_1)$, $\| \gradd{\psi_{\star_j}} \| \leq \varkappa_{\star_j}(\bx_1)$, $\| \tgrad{\psi_{\star_j}} \| \leq \mu_{\star_j}(\bx_1)$, and $|\gradt{\psi_{\star_j}}| \leq \iota_{\star_j}(\bx_1)$, $\forall t \geq 0$ and $\forall \bx_1 \in \mbR^{n}$.
\end{assumption}

Assumptions \ref{assu:bounded_in_t_psi_fun} and \ref{assu:bounded_in_t_psi_derivatives} ensure that, for each fixed $\bx_1$, the functions $\psi_{\star_j}(\cdot,\cdot)$ and their partial derivatives remain bounded in $t$. Since the constraint functions in \eqref{eq:hard_soft_constraints} are known, the corresponding upper-bound functions in these assumptions are also computable and known. Nevertheless, these bounds are used solely for the stability analysis and do not enter the proposed control design.

\begin{remark}
	If $r=1$ in \eqref{eq:sys_dynamics_mimo}, it suffices that $\psih{j}, \psis{j}$, $j \in \mbI_1^{m_{\star}}$, $\star \in \{h,s\}$ be $\Ccal^1$ in both $\bx_1$ and $t$. The requirements on $\gradd{\psi_{\star_j}}$ and $\tgrad{\psi_{\star_j}}$ in Assumption \ref{assu:bounded_in_t_psi_derivatives} can be omitted.
\end{remark}

Define the hard and soft constraint sets as follows:
\begin{subequations}\label{eq:hard_soft_sets}
	\begin{align}
		\Obh(t) &\coloneqq \{ \bx_1 \in \mbR^n \mid \psih{j}(t,\bx_1) > 0 , \forall j \in \mbI_1^{m_h} \},  \label{eq:hard_set} \\
		\Obs(t) &\coloneqq \{ \bx_1 \in \mbR^n \mid \psis{j}(t,\bx_1) > 0 , \forall j \in \mbI_1^{m_s} \}. \label{eq:soft_set}
	\end{align}
\end{subequations}

\textbf{Objective:} 
We aim to design a closed-form, low-complexity robust continuous control law \( \bu(t, \bx) \) for \eqref{eq:sys_dynamics_mimo} that ensures \( \bx_1(t; \bx(0)) \) adheres to the hard constraints \eqref{eq:hard_constraints} at all times (i.e., \( \bx_1(t; \bx(0)) \in \Obh(t), \forall t \geq 0 \)). The soft constraints \eqref{eq:soft_constraints} must hold only when compatible with the hard ones (i.e., \( \Obh(t) \cap \Obs(t) \neq \emptyset \)). Specifically, in the absence of incompatibilities, if the soft constraints are not initially met, our goal is to achieve their satisfaction \( \forall t \geq T > 0 \), where \( T \) is a user-defined finite time.

\section{Main Results}
\label{sec:main_result}

We first introduce smooth consolidated hard and soft constraint functions, whose positive level sets provide smooth inner-approximations of the constrained sets in \eqref{eq:hard_soft_sets}.  Next, we propose a low-complexity control design strategy to achieve the stated objectives. Finally, the main results are summarized in Theorem \ref{th:main} and Corollary \ref{colo:main}.

\subsection{Smooth Consolidated Hard/Soft Constraint Functions}
\label{subsec:hard_soft_consolidated}

Note that \eqref{eq:hard_constraints} and \eqref{eq:soft_constraints} are satisfied whenever
\begin{subequations}\label{eq:consolidated_hard_soft}
	\begin{align}
		\alphbh(t,\bx_1) &\coloneqq \min \{ \psih{1}(t,\bx_1), \ldots , \psih{m_h}(t,\bx_1) \} > 0, \label{eq:consolidated_hard_constraints} \\
		\alphbs(t,\bx_1) &\coloneqq \min \{ \psis{1}(t,\bx_1), \ldots , \psis{m_s}(t,\bx_1) \} > 0, \label{eq:consolidated_soft_constraints}
	\end{align}
\end{subequations}
hold, respectively. Therefore, the sets in \eqref{eq:hard_soft_sets} can be equivalently expressed as $\Ob_{\star}(t) = \{ \bx_1 \in \mbR^n \mid \alphb_{\star}(t,\bx_1) > 0 \}$, where $\star \in \{h,s\}$. In general, $\alphb_\star(\cdot,\cdot)$ is continuous yet nonsmooth in $\bx_1$. Using the Log-Sum-Exp function, we introduce \textit{smooth consolidated constraint functions} as
\begin{subequations}\label{smooth_alph}
	\begin{align}
		\alphh(t,\bx_1) &\coloneqq -\frac{1}{\nu} \ln \Big( \sum_{j=1}^{m_h} e^{- \nu  \, \psih{j}(t,\bx_1)} \Big), \label{smooth_alpha_hard_def} \\
		\alphs(t,\bx_1) &\coloneqq -\frac{1}{\nu} \ln \Big( \sum_{j=1}^{m_s} e^{- \nu  \, \psis{j}(t,\bx_1)} \Big), \label{smooth_alpha_soft_def}
	\end{align}
\end{subequations}
where $\nu>0$ is a tuning coefficient whose larger values give a closer (under) approximation (i.e., $\alpha_{\star}(t,\bx_1) \rightarrow \alphb_{\star}(t,\bx_1)$ as $\nu \rightarrow \infty$, $\star \in \{h,s\}$). It is known that $\alpha_{\star}(t,\bx_1) \leq \alphb_{\star}(t,\bx_1) \leq \alpha_{\star}(t,\bx_1) + \frac{1}{\nu} \ln(m_{\star})$, $\star \in \{h,s\}$ \cite{gilpin2020smooth}. Ensuring $\alpha_{\star}(t,\bx_1) > 0$ guarantees $\alphb_{\star}(t,\bx_1) > 0, \star \in \{h,s\}$, and thus the satisfaction of \eqref{eq:hard_soft_constraints}. Define
\begin{subequations}\label{eq:smooth_hard_soft_sets}
	\begin{align}
		\Oh(t) &\coloneqq \{\bx_1 \in \mbR^n \mid \alphh(t,\bx_1) > 0 \} \subseteq \Obh(t), \label{eq:smoth_hard_set} \\
		\Os(t) &\coloneqq \{\bx_1 \in \mbR^n \mid \alphs(t,\bx_1) > 0 \} \subseteq \Obs(t), \label{eq:smoth_soft_set}
	\end{align}
\end{subequations}
which are \textit{smooth inner-approximations} of the sets in \eqref{eq:hard_soft_sets}.

\begin{assumption} \label{assu:coercive_alpha_funs}
	Functions $-\alphb_{\star}(t,\bx_1), \star \in \{h,s\}$ are coercive (radially unbounded) in $\bx_1$ and uniformly in $t$, i.e., $-\alphb_{\star}(t,\bx_1) \rightarrow +\infty$ as $\|\bx_1\| \rightarrow +\infty, \forall t \geq 0$. 
\end{assumption}

This assumption is necessary and sufficient for the level curves of $\alphb_{\star}(t,\bx_1), \star \in \{h,s\}$ (and consequently $\alpha_{\star}(t,\bx_1)$) to be compact for all $t \geq 0$. Thus, Assumption \ref{assu:coercive_alpha_funs} guarantees boundedness of the sets in \eqref{eq:hard_soft_sets} (and also \eqref{eq:smooth_hard_soft_sets}), establishing a well-posedness condition for time-varying hard and soft constrained sets; see \cite[Lemma 1]{mehdifar2024low}. 

From \eqref{eq:consolidated_hard_soft} (resp. \eqref{smooth_alph}), Assumption \ref{assu:coercive_alpha_funs} can be verified by checking whether at least one of the functions $\psi_{\star_j}(t,\bx_1), \star \in \{h,s\}$ in \eqref{eq:hard_soft_constraints} approaches $-\infty$ as $\|\bx_1\| \rightarrow +\infty$ (along any path in $\mbR^n$). As outlined in \cite[Section III.A]{mehdifar2024low}, it can also be ensured by introducing auxiliary (hard/soft) constraints of the form $\psi_{\star_\mathrm{aux}}(\bx_{1}) \coloneqq c_{\mathrm{aux}} - \|\bx_1\|^2$, with  sufficiently large constant $c_{\mathrm{aux}}>0$. 

Assumption \ref{assu:coercive_alpha_funs} also guarantees the existence of at least one global maximizer for $\alphb_\star(t,\bx_1)$ (resp. for $\alpha_\star(t,\bx_1)$), $\star \in \{h,s\}$, for all $t \geq 0$, see \cite[Proposition 2.9]{Grippo2023intro}. Accordingly, for $\Oh(t)$ and $\Os(t)$, we define $\alphhstr(t) \coloneqq \max_{\bx_1 \in \mbR^n} \alphh(t,\bx_1)$ and $\alphsstr(t) \coloneqq \max_{\bx_1 \in \mbR^n} \alphs(t,\bx_1)$ for each instant $t$. Clearly, $\alphhstr(t^\prime) > 0$ and $\alphsstr(t^\prime) > 0$ signify the feasibility of $\Oh(t)$ and $\Os(t)$ in \eqref{eq:smooth_hard_soft_sets} at time $t = t^\prime$, respectively.

\begin{assumption} \label{assu:feasible_constrained_sets}
	The inner-approximated sets $\Oh(t)$ and $\Os(t)$, are nonempty (feasible) for all $t \geq 0$.
\end{assumption}

The above assumption ensures the existence of sufficiently small positive constants $\epsih > 0$ and $\epsis > 0$ such that $\alphhstr(t) \geq \epsih > 0$ and $\alphsstr(t) \geq \epsis > 0$ for all $t \geq 0$.

\begin{assumption} \label{assu:safe_ini}
	The initial condition $\bx(0)$ of \eqref{eq:sys_dynamics_mimo} satisfies $\alphh(0, \bx_1(0)) > 0$, i.e., the hard constraints hold at  $t =0$.
\end{assumption}

\begin{remark}
	Note that the feasibility of the inner-approximated hard and soft constrained sets $\Oh(t)$ and $\Os(t)$ ensures the feasibility of $\Obh(t)$ and $\Obs(t)$ in \eqref{eq:hard_soft_sets}. Furthermore, if the original time-varying hard and soft constrained sets $\Obh(t)$ and $\Obs(t)$ are tightly feasible (i.e., relatively small), one can always select a sufficiently large $\nu$ in \eqref{smooth_alph} to ensure that Assumption \ref{assu:feasible_constrained_sets} remains valid.
\end{remark}

\begin{remark}\label{rem:soft_infeasible}
	Assumption \ref{assu:feasible_constrained_sets} requires both constrained sets to stay feasible at all times. This is indispensable for the hard constrained set, but the soft constrained set may become infeasible in some scenarios. If this happens, one can design $\bus$ in Section \ref{subsec:control_design} following \cite[Section IV]{mehdifar2024low}, while the rest of the proposed scheme stays the same.
\end{remark}

\subsection{Hard and Soft Constraints Satisfaction Strategy}
\label{subsec:control_strategy}

We begin with designing two scalar time-varying constraints for system \eqref{eq:sys_dynamics_mimo}, which, when they are satisfied, guarantee the hard constraints \eqref{eq:hard_constraints} are met at all times and the soft constraints \eqref{eq:soft_constraints} are met only when compatible with the hard constraints. In the next subsection, we propose a control law to enforce these time-varying constraints along the closed-loop system trajectories. Hereafter, for brevity, we refer to $\Oh(t)$ and $\Os(t)$ in \eqref{eq:smooth_hard_soft_sets} as the time-varying hard and soft constrained sets (excluding the inner-approximation term) when there is no ambiguity.

Based on the formulation in \eqref{eq:smoth_hard_set}, enforcing 
\begin{equation} \label{eq:hard_const_satisfaction_cond}
	0 < \alphh(t,\bx_1), \quad \forall t \geq 0,
\end{equation}
along trajectories of \eqref{eq:sys_dynamics_mimo} guarantees that the hard constraints \eqref{eq:hard_constraints} hold for all \(t \geq 0\). Notice that, \eqref{eq:hard_const_satisfaction_cond} is a feasible constraint at all times under Assumptions \ref{assu:feasible_constrained_sets} and \ref{assu:safe_ini}.

Next, we introduce the time-varying constraint  
\begin{equation} \label{eq:soft_const_satisfaction_cond}  
	\rhos(t) < \alphs(t,\bx_1), \quad \forall t \geq 0,  
\end{equation}  
where \(\rhos: \mbRpos \rightarrow \mbRneg\) is a continuously differentiable non-positive function. Note that \eqref{eq:soft_const_satisfaction_cond} is feasible at all times by Assumption \ref{assu:feasible_constrained_sets} and since \(\rhos(t) \leq 0\). Define
\begin{equation} \label{eq:smoth_virtual_soft_set}
	\Osv(t) \coloneqq \{\bx_1 \in \mbR^n \mid \alphs(t,\bx_1) - \rhos(t) > 0 \},
\end{equation}
which is the nonempty set where \eqref{eq:soft_const_satisfaction_cond} holds. Since \(\alphs(t,\bx_1)\) has compact level curves (i.e., $-\alphs(t,\bx_{1})$ is radially unbounded in $\bx_{1}$) and \(\rhos(t) \leq 0\), we have \(\Os(t) \subseteq \Osv(t), \forall t \geq 0\). We refer to \(\Osv(t)\) as the \textit{virtual soft constrained set}.

We propose a low‐complexity control design ensuring \eqref{eq:hard_const_satisfaction_cond} and \eqref{eq:soft_const_satisfaction_cond} simultaneously along the closed‐loop trajectories of \eqref{eq:sys_dynamics_mimo}. In particular, when \(\rhos(t)=0\) in \eqref{eq:soft_const_satisfaction_cond} we have \(\Os(t)=\Osv(t)\), so enforcing \eqref{eq:soft_const_satisfaction_cond} ensures the soft constraints. However, since soft and hard constraints may become incompatible over some time intervals, enforcing both \(0 < \alphh(t,\bx_1)\) and \(0 < \alphs(t,\bx_1)\) might not be feasible for all \(t \geq 0\). Thus, \(\rhos(t)\) must be designed so that \eqref{eq:hard_const_satisfaction_cond} and \eqref{eq:soft_const_satisfaction_cond} can be simultaneously satisfied. Therefore, besides the controller design, a key aspect is to design \(\rhos(t)\) such that \(\Osv(t)\) remains compatible with the hard constrained set \(\Oh(t)\) (i.e., \(\Oh(t) \cap \Osv(t) \neq \emptyset\)) at all times. For this, \(\rhos(t) \leq 0\) should decrease whenever the soft and hard constraints are incompatible and return to zero whenever \(\Os(t)\) is compatible with \(\Oh(t)\). Hence, we define \(\rhos(t)\) as  
\begin{equation} \label{eq:soft_const_set_lower_bound}  
	\rhos(t) \coloneqq \rhosn(t) - \rhosr(t),  
\end{equation}  
where \(\rhosn: \mbRpos \rightarrow \mbRneg\) is a non-positive, continuously differentiable nominal lower bound in \eqref{eq:soft_const_satisfaction_cond} for enforcing soft constraint satisfaction. The term \(\rhosr: \mbRpos \rightarrow \mbRpos\) is a continuously differentiable non-negative relaxation signal with \(\rhosr(0) = 0\), increasing as necessary to ensure compatibility between \eqref{eq:hard_const_satisfaction_cond} and \eqref{eq:soft_const_satisfaction_cond}. In particular, \(\rhosr\) grows only when enforcing the soft constraints (i.e., \(\alphs(t,\bx_1) > 0\)) conflicts with enforcing the hard constraints (i.e., \(\alphh(t,\bx_1) > 0\)). As detailed in Section \ref{subsec:control_design}, this conflict is assessed locally along the trajectory $\bx_1(t; \bx(0))$, without checking the compatibility condition $\Oh(t)\cap\Os(t)\neq\emptyset$, whose verification amounts to a global, generally non-convex optimization over $\mbR^n$ at each instant and would not indicate by how much $\Os(t)$ should be relaxed.

We design \(\rhosn(t)\) assuming \(\rhosr(t)\equiv 0\) (i.e., hard and soft constraints are always compatible), so that \(\rhosn(t)=\rhos(t)\). Specifically, \(\rhosn(t)\) is chosen to ensure soft constraints are met (at least) within the user-defined finite time \(T\) by enforcing \eqref{eq:soft_const_satisfaction_cond} on \eqref{eq:sys_dynamics_mimo}. To this end, \(\rhosn(t)\) should satisfy:
\begin{itemize}
	\item[(i)] If \(\alphs(0,\bx_1(0)) > 0\) (i.e., soft constraints are initially met), then \(\rhosn(t)=0\) for all \(t \geq 0\).
	\item[(ii)] If \(\alphs(0,\bx_1(0)) \leq 0\) (i.e., soft constraints are initially violated), then \(\rhosn(0) < \alphs(0,\bx_1(0))\) and \(\rhosn(t)\) should increase and converge to zero by \(T\) seconds (i.e., \(\rhosn(t\geq T)=0\)).
\end{itemize}
One possible choice for \(\rhosn(t)\) is
\begin{equation}\label{eq:alpha_lower_bound}
	\rhosn(t) = \begin{cases}
		\left(\frac{T-t}{T}\right)^{\frac{1}{1-\beta}} \rho_0, & 0 \leq t \leq T,\\
		0, & t > T,
	\end{cases}
\end{equation}
where \(\beta\in (0,1)\) and \(\rho_0 \coloneqq \rhosn(0)=\rhos(0) \leq 0\) are tunable constants. The design of \(\rhosr(t)\) is given next.

\subsection{Low-Complexity Controller Design}
\label{subsec:control_design}

We design a low‐complexity robust state feedback controller for \eqref{eq:sys_dynamics_mimo} guaranteeing \eqref{eq:hard_const_satisfaction_cond} and \eqref{eq:soft_const_satisfaction_cond} at all times. Exploiting the lower triangular structure of \eqref{eq:sys_dynamics_mimo}, we adopt a backstepping‐like approach inspired by \cite{bechlioulis2014low, mehdifar2024low}. For brevity, some function arguments are dropped when unambiguous. The design steps are:

\textbf{Step 1-a.} Compute \(\alphs(0,\bx_1(0))\) and select \(\rho_0\) in \eqref{eq:alpha_lower_bound} with \(\rhosn(0) = \rhos(0) < \alphs(0,\bx_1(0))\), as in Section \ref{subsec:control_strategy}.

\textbf{Step 1-b.} Define
\begin{equation} \label{eq:es}
	\es(t,\bx_1) \coloneqq \alphs(t,\bx_1) - \rhos(t),
\end{equation}
so that \(\es(t,\bx_1)>0\) for all \(t\ge0\) is equivalent to \eqref{eq:soft_const_satisfaction_cond}. Next, consider the reciprocal barrier functions:
\begin{align} 
	\epih(t,\bx_1) &\coloneqq \frac{1}{\alphh(t,\bx_1)}, \label{eq:mapped_alphh} \\ 
	\epis(t,\bx_1) &\coloneqq \frac{1}{\es(t,\bx_1)}. \label{eq:mapped_alphs} 
\end{align}
By \textit{Step 1-a} and Assumption \ref{assu:safe_ini}, we have $\es(0,\bx_1(0))>0$ and $\alphh(0,\bx_1(0))>0$. Therefore, boundedness of \(\epih\) and \(\epis\) for all time ensures that \(\alphh(t,\bx_1)\) and \(\es(t,\bx_1)\) remain positive, thereby guaranteeing the forward invariance\footnote{A time-varying
		set $\Omega(t)\subseteq\mbR^n$ is \textit{forward invariant} from
		$t_0\ge 0$ if $\bx_1(t_0;\bx(0))\in\Omega(t_0)$ implies $\bx_1(t;\bx(0))\in\Omega(t)$ for
		all $t\ge t_0$.} of sets $\Oh(t)$ in \eqref{eq:smoth_hard_set} and $\Osv(t)$ in \eqref{eq:smoth_virtual_soft_set}, respectively \cite{ames2016control}.

\textbf{Step 1-c.} Define
\begin{equation} \label{eq:uh}
	\buh \coloneqq k_h\, \epih^2 \,\bG_1^{-1} \grad{\alphh},
\end{equation}
where \(k_h > 0\). Let $\gamma(t,\bx_1) \coloneqq \epih \, \grad{\alphs}^\top \grad{\alphh}$. We design the dynamics governing \(\rhosr(t)\) in \eqref{eq:soft_const_set_lower_bound} as
\begin{equation} \label{eq:rho_relax_dyn}
	\begin{cases}
		\drhosr = \phi_{\rho}(t, \bx_1, \rhosr) \coloneqq - \varphigam \varphih k_h  \epih^2  \grad{\alphs}^\top \grad{\alphh} -\krec \rhosr,  \\
		\rhosr(0) = 0, 
	\end{cases}
\end{equation}
where \( \krec > 0\), and \(\varphih \coloneqq \varphi(\alphh, \delta_h, 0) \) and $\varphigam \coloneqq \varphi(\gamma, 0, -\delta_{\gamma})$ are $\Ccal^1$ switch functions, refer to \ref{appen:cont_dif_switch_fun} for the definition of $\varphi(\mathdot,\mathdot,\mathdot)$, and $\delta_h, \delta_{\gamma} > 0$ are some sufficiently small tunable constants. Next, define
\begin{equation} \label{eq:us}
	\bus \coloneqq k_s\, \epis^2 \,\bG_1^{-1} \grad{\alphs},
\end{equation}
where \(k_s > 0\). We note that $\buh$ and $\bus$ are $\bG_1^{-1}$-preconditioned scaled negative gradients of the reciprocal barrier functions $\epih$ and $\epis$ with respect to $\bx_1$, respectively. Explicit gradients of \(\alphh\) and \(\alphs\) are given in \ref{appen:gradients}. Finally, the first intermediate (virtual) control law is designed as
\begin{equation} \label{eq:1st_intermed_ctrl}
	\bs_1(t,\bx_1) \coloneqq \bus +  \varphih \buh.
\end{equation}

\textbf{Step $\mathbf{i}$-a ($\mathbf{2 \leq i \leq r}$).} Define the $i$-th intermediate error vector as
\begin{equation}\label{eq:intermediate_err}
	\be_i = \col{e_{i,j}} \coloneqq \bx_i - \bs_{i-1}(t, \bxb_{i-1}),
\end{equation}  
where \( \be_i \in \mbR^n \). The goal is to design the $i$-th intermediate (virtual) control law \( \bs_i(t, \be_i) \) for \eqref{eq:sys_dynamics_mimo} to compensate for \( e_{i,j}(t, \bxb_i) \), \( j \in \mbI_1^n \), by enforcing narrowing intermediate constraints
\begin{equation} \label{eq:i-th_inter_funnels}
	-\vartheta_{i,j}(t) < e_{i,j}(t, \bxb_i) < \vartheta_{i,j}(t), \quad j \in \mbI_1^n,
\end{equation}  
for all \( t \geq 0 \), where \( \vartheta_{i,j}: \mbRpos \rightarrow \mbRspos \) are continuously differentiable \textit{strictly positive performance functions} decaying to a neighborhood of zero. One option for \( \vartheta_{i,j}(t) \) is:  
\begin{equation}\label{exponential_performance_fun}
	\vartheta_{i,j}(t) \coloneqq (\vartheta_{i,j}^0 - \vartheta_{i,j}^{\infty}) \exp(-l_{i,j} t) + \vartheta_{i,j}^{\infty},
\end{equation}  
where \( l_{i,j}, \vartheta_{i,j}^{\infty}, \vartheta_{i,j}^0 \) are user-defined positive constants. To ensure \eqref{eq:i-th_inter_funnels} holds at $t=0$, select \( \vartheta_{i,j}^0 > |e_{i,j}(0, \bxb_i(0))| \).

\textbf{Step $\mathbf{i}$-b ($\mathbf{2 \leq i \leq r}$).} Define the diagonal matrix \( \bvarTheta_i(t) \coloneqq \diag{\vartheta_{i,j}(t)} \in \mbR^{n \times n} \) and consider:
\begin{equation}\label{eq:normalized_err_vec}
	\beh_i(t, \be_i) = \col{\eh_{i,j}} \coloneqq \bvarTheta_i^{-1}(t) \, \be_i,
\end{equation}  
as the vector of normalized errors, with elements:
\begin{equation} \label{eq:normal_e_i_j}
	\eh_{i,j}(t, e_{i,j}) = \frac{e_{i,j}}{\vartheta_{i,j}(t)}, \quad j \in \mbI_1^n.
\end{equation}  
Note that \( \eh_{i,j} \in (-1, 1) \) iff \( e_{i,j} \in (-\vartheta_{i,j}(t), \vartheta_{i,j}(t)) \). Next, introduce the following nonlinear transformations 
\begin{equation} \label{eq:mapping_fun}
	\epi_{i,j}(t, e_{i,j}) = \Tcal(\eh_{i,j}) \coloneqq \ln \left( \frac{1 + \eh_{i,j}}{1 - \eh_{i,j}} \right), \quad j \in \mbI_1^n,
\end{equation}  
where \( \epi_{i,j} \) is the unconstrained transformed signal of \( e_{i,j}(t, \bxb_i) \), and \( \Tcal: (-1, 1) \rightarrow (-\infty, +\infty) \) is a smooth, strictly increasing bijection with \( \Tcal(0) = 0 \). Boundedness of \( \varepsilon_{i,j} \) ensures \( \eh_{i,j} \) remains in \( (-1, 1) \), thus satisfying \eqref{eq:i-th_inter_funnels}.

\textbf{Step $\mathbf{i}$-c ($\mathbf{2 \leq i \leq r}$).}  Design the $i$-th intermediate control as
\begin{equation} \label{eq:i-th_intermed_ctrl}
	\bs_i(t, \be_i) \coloneqq  -k_i \, \bXi_i \, \bepi_i,
\end{equation}  
where \( k_i > 0 \) is a control gain, $\bepi_i \coloneqq \col{\epi_{i,j}} \in \mbR^n$, and \( \bXi_i \coloneqq \diag{\xi_{i,j}}  \in \mbR^{n \times n} \) is a diagonal matrix with entries
\begin{equation} \label{eq:xi_i,j}
	\xi_{i,j}(t, e_{i,j}) \coloneqq \frac{2}{\vartheta_{i,j}(t) \, (1 - \eh_{i,j}^2)}, \quad j \in \mbI_1^n.
\end{equation}  
Note that \( \bs_i(t, \be_i) \) depends on \( t \) and \( \bxb_i \) because \( \be_i \) itself depends on \( \bxb_i \), as seen in \eqref{eq:intermediate_err}. Thus, we can write \( \bs_i(t, \bxb_i) \).

\textbf{Step $\mathbf{r+1}$.} The control input \( \bu(t, \bx) \) is designed as:  
\begin{equation} \label{eq:control_law}
	\bu(t, \bx) \coloneqq \bs_r(t, \bx).
\end{equation}

\textbf{\underline{\textit{Design Philosophy for Step 1:}}} The design begins with an intermediate control \(\bs_1(t,\bx_1)\) for the \(\bx_1\) dynamics in \eqref{eq:sys_dynamics_mimo}, in which \( \buh \) and \( \bus \) are constructed to meet \eqref{eq:hard_const_satisfaction_cond} and \eqref{eq:soft_const_satisfaction_cond}, respectively. Since \(\varphih\) becomes active (i.e., \(\varphih > 0\)) only when \( \alphh(t, \bx_1(t; \bx(0))) < \delta_h \), the enforcement of hard constraints through \(\buh\) is applied only when the trajectory approaches the boundary of \(\cl{\Oh(t)}\). 

\begin{figure}[!tbp]
	\centering
	\begin{subfigure}[t]{0.49\linewidth}
		\centering
		\includegraphics[width=\linewidth]{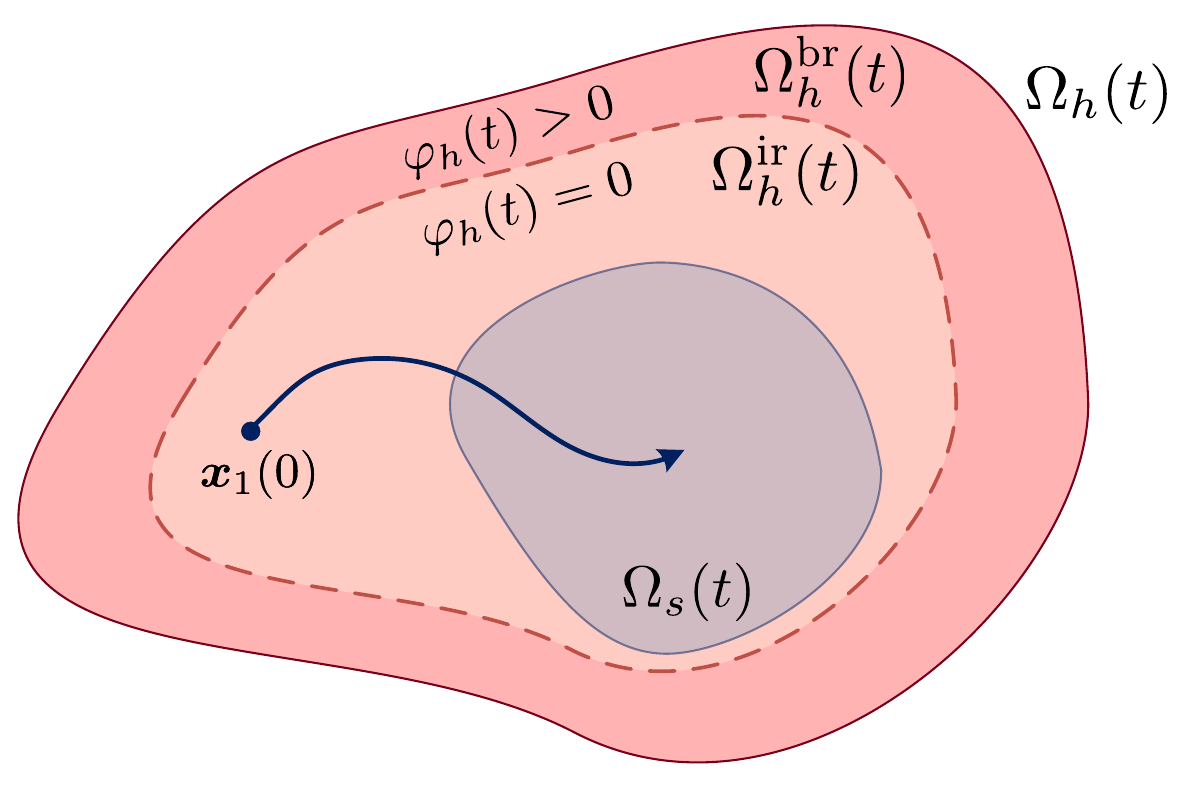}
		\caption{}
		\label{ctrl_philo_us_only}
	\end{subfigure}%
	\hfill
	\begin{subfigure}[t]{0.49\linewidth}
		\centering
		\includegraphics[width=\linewidth]{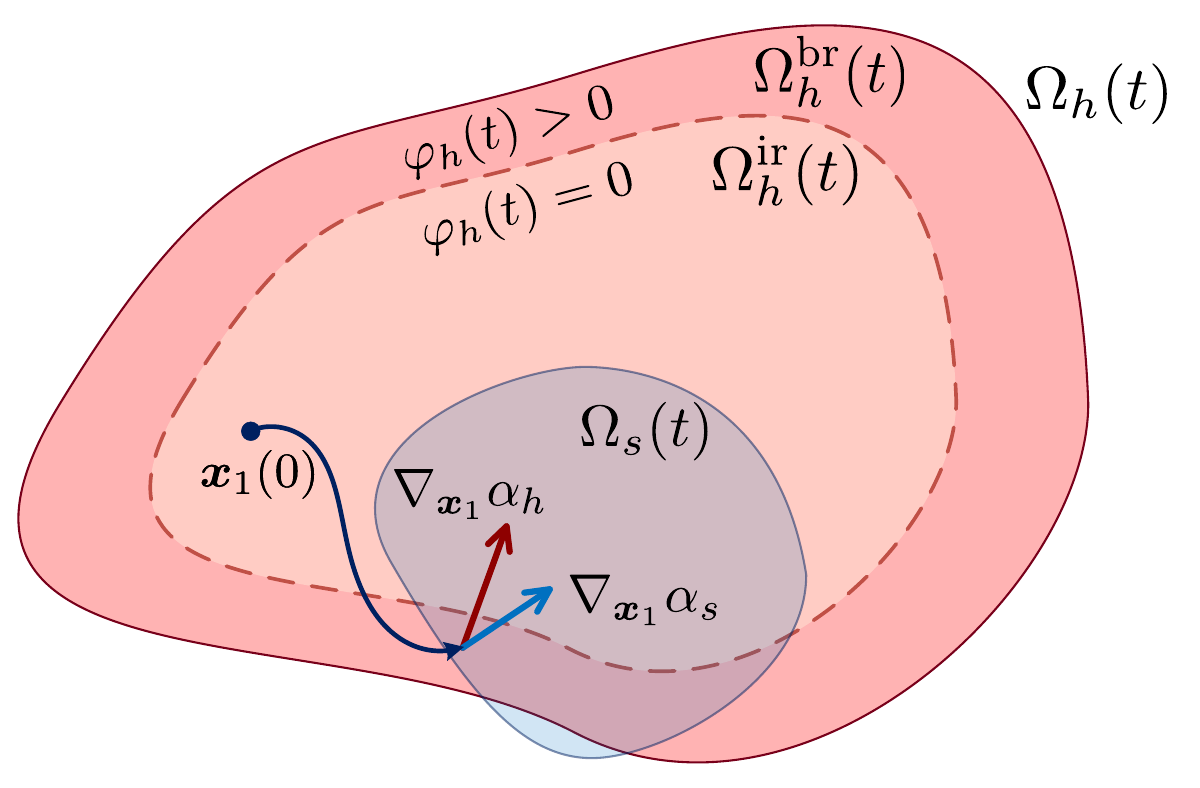}
		\caption{}
		\label{ctrl_philo_uh_active}
	\end{subfigure}
	\begin{subfigure}[t]{0.61\linewidth}
		\centering
		\includegraphics[width=\linewidth]{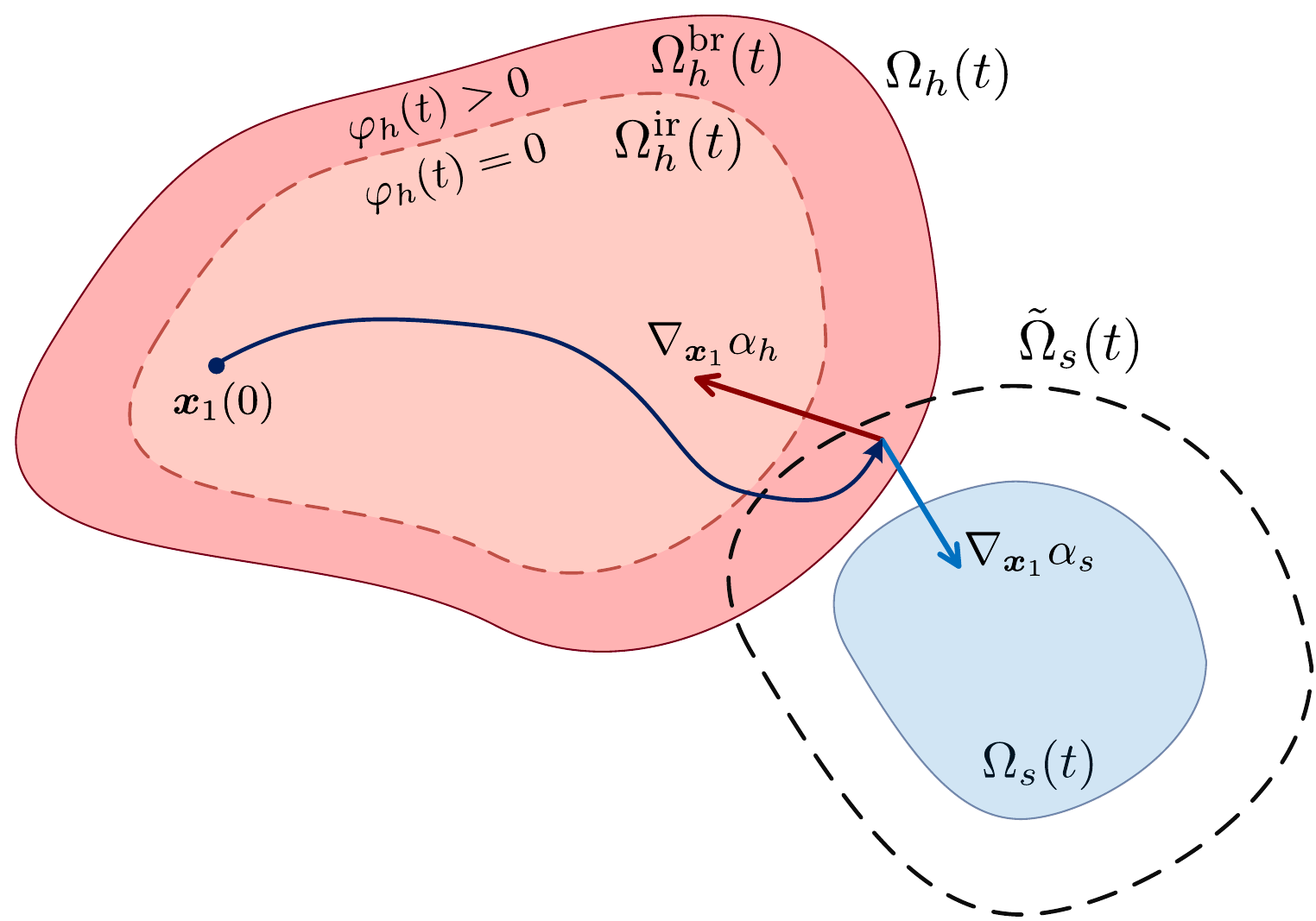}
		\caption{}
		\label{ctrl_philo_uh_active_relaxation}
	\end{subfigure}
	\caption[Illustration of the control design philosophy for Step 1]{Illustration of the control design philosophy for Step~1, showing the three regimes of the conflict indicator in \eqref{eq:rho_relax_dyn}. (a) $\bx_1(t;\bx(0))\in\Ohi(t)$, so $\varphih=0$, $\bs_1(t,\bx_1)=\bus$, and $\buh$ does not influence $\bs_1(t,\bx_1)$. (b) and (c) $\bx_1(t;\bx(0))\in\Ohb(t)$, so $\varphih>0$ and $\varphih\buh$ becomes active to keep $\bx_1(t;\bx(0))$ inside $\Oh(t)$; the arrows are the gradients $\grad{\alphh}$ and $\grad{\alphs}$ at the current state, which are also the directions of the velocities imparted to the $\bx_1$ dynamics by $\varphih\buh$ and $\bus$, since $\varphih\bG_1\buh=\varphih k_h\epih^2\grad{\alphh}$ and $\bG_1\bus=k_s\epis^2\grad{\alphs}$. In (b) the two gradients form an acute angle, so $\gamma\geq0$ and $\varphigam=0$: although part of $\Os(t)$ lies outside $\Oh(t)$, the two terms do not oppose one another at the current state and $\Osv(t)=\Os(t)$. In (c) they form an obtuse angle, so $\gamma<0$ and $\varphigam>0$: $\rhosr$ increases and $\Osv(t)$ expands away from $\Os(t)$ to ensure $\Oh(t)\cap\Osv(t)\neq\emptyset$, thereby balancing the opposing efforts $\bus$ and $\varphih\buh$.}
	\vspace{-0.5cm}
	\label{fig:ctrl_philo_step1}
\end{figure}

For future reference, let us define the sets  
\begin{subequations}
	\begin{align}
		\Ohi(t) &\coloneqq \{\bx_1 \in \mbR^n \mid \alphh(t, \bx_1) \geq \delta_h \}, \label{eq:interior_region_hard} \\
		\Ohb(t) &\coloneqq \{\bx_1 \in \mbR^n \mid 0 < \alphh(t, \bx_1) < \delta_h \}, \label{eq:boundary_region_hard}
	\end{align}
\end{subequations}
which we call the \textit{interior} and \textit{boundary regions} of the (inner-approximated) time-varying hard-constrained set $\Oh(t)$, respectively. Note that \(\Oh(t) = \Ohi(t) \cup \Ohb(t)\) and that \(\Ohi(t)\) is nonempty for sufficiently small \(\delta_h > 0\) (in particular, when \(\alphhstr(t) > \delta_h\); see Assumption \ref{assu:feasible_constrained_sets}). From \eqref{eq:1st_intermed_ctrl}, when \( \bx_1(t;\bx(0)) \) lies in \(\Ohi(t)\), we have \( \bs_1(t, \bx_1) = \bus \); thus, \(\bus\) remains unaltered to satisfy soft constraints. This is illustrated in Fig.~\ref{ctrl_philo_us_only}. However, once \( \bx_1(t;\bx(0)) \) enters \(\Ohb(t)\), the term \(\varphih \buh\) becomes active and influences $\bs_1(t,\bx_1)$ in \eqref{eq:1st_intermed_ctrl} so as to keep the trajectory $\bx_1(t; \bx(0))$ within $\Oh(t)$; see Fig.~\ref{ctrl_philo_uh_active}. Simultaneously, a virtual relaxation of the soft constraints is enacted by expanding \(\Osv(t)\) (defined in \eqref{eq:smoth_virtual_soft_set}) away from \(\Os(t)\) through an increase in \(\rhosr\) as dictated by \eqref{eq:rho_relax_dyn}. This virtual relaxation of soft constraints serves two purposes:  
(i) ensuring that \(\Osv(t)\) remains compatible with \(\Oh(t)\) (i.e., \(\Oh(t) \cap \Osv(t) \neq \emptyset\)), and  
(ii) balancing \(\varphih \buh\) with \(\bus\) in \eqref{eq:1st_intermed_ctrl}, thereby confining \( \bx_1(t;\bx(0)) \) within the set \(\Oh(t) \cap \Osv(t)\); see Fig.~\ref{ctrl_philo_uh_active_relaxation}.

The first term on the right-hand side (RHS) of \eqref{eq:rho_relax_dyn} increases \(\rhos(t)\) when necessary. This term is active and positive when both \(\varphigam > 0\) (i.e., when \(\gamma = \epih \, \grad{\alphs}^\top \grad{\alphh} < 0\)) and \(\varphih > 0\), and it grows unbounded as \(\epih \rightarrow +\infty\) (i.e., as \(\alphh \rightarrow 0\)). In particular, \(\rhosr(t)\) increases only when the barrier gradients \(\grad{\alphs}\) and \(\grad{\alphh}\) partially oppose (i.e., \(\gamma(t,\bx_1) < 0\)) and the trajectory is close to the boundary of \(\cl{\Oh(t)}\), i.e., \( \bx_1(t; \bx(0)) \in \Ohb(t)\). The condition \(\gamma < 0\) has a direct geometric meaning: since \(\bs_1\) enters the \(\bx_1\) dynamics of \eqref{eq:sys_dynamics_mimo} in place of \(\bx_2\), it imparts to them the velocity \(\bG_1\bs_1 = k_s\epis^{2}\grad{\alphs} + \varphih k_h\epih^{2}\grad{\alphh}\), so that the soft and hard terms act along \(\grad{\alphs}\) and \(\grad{\alphh}\), respectively, and \(\gamma < 0\) states that these two contributions form an obtuse angle. The pair \((\varphih,\varphigam)\) therefore acts as a local online conflict indicator, built from \(\alphh\), \(\grad{\alphh}\) and \(\grad{\alphs}\), which \eqref{eq:uh} and \eqref{eq:us} already evaluate: no feasibility test and no online optimization are performed. An illustration is given in Fig.~\ref{fig:ctrl_philo_step1}. Selecting a smaller $\delta_h$ and a larger $\delta_\gamma$ in \eqref{eq:rho_relax_dyn} lets $\bx_1(t;\bx(0))$ approach the boundary of \(\cl{\Oh(t)}\) more closely before \(\Osv(t)\) begins to expand.

The second term on the RHS of \eqref{eq:rho_relax_dyn} is a stabilizing term driving \(\rhos(t)\) back to zero exponentially fast when the first term is inactive (i.e., when the hard and soft sets are compatible). Moreover, a larger \( \krec \) in \eqref{eq:rho_relax_dyn} accelerates recovery of the soft constrained set, i.e., \(\Osv(t) \rightarrow \Os(t)\), once the soft and hard constrained sets become compatible again.

\textbf{\underline{\textit{Design Philosophy for Step $\mathbf{2 \leq i \leq r}$:}}} 
As in backstepping, the design continues by crafting a second intermediate control \(\bs_2(t,\bxb_2)\) for the \(\bx_2\) dynamics in \eqref{eq:sys_dynamics_mimo} so that \(\bx_2\) closely tracks \(\bs_1(t,\bx_1)\). In particular, \(\bs_2(t,\be_2)\) ensures that all components \(e_{2,j}\), \(j \in \mbI_1^n\), of the error \(\be_2 = \bx_2 - \bs_1(t,\bx_1)\) become sufficiently small by satisfying \eqref{eq:i-th_inter_funnels}. The process continues until we obtain \(\bu(t,\bx)\) at \textit{Step $r+1$}. Notably, unlike classical backstepping, we use no derivatives of \(\be_i, i \in \mbI_2^r\) and no filtering when designing the intermediate control laws \(\bs_i(t,\be_i), i \in \mbI_2^r\) \cite{bechlioulis2014low}. Moreover, the only plant-side quantity in \eqref{eq:control_law} is the known kinematic map $\bG_1$, via $\bG_1^{-1}$ in \eqref{eq:uh} and \eqref{eq:us}; for $\bG_1=\bI_n$ the control law uses no plant information at all. Neither the bounds $\cinv, \bar{f}_i,\bar{g}_i$ nor any derivative of the plant nonlinearities is required, making the approach low-complexity and tractable.

\begin{remark}
	Note that the proposed constraints in \eqref{eq:hard_const_satisfaction_cond} and \eqref{eq:soft_const_satisfaction_cond}, as well as \eqref{eq:i-th_inter_funnels} for the intermediate error signals \(e_{i,j}, i \in \mbI_2^r, j \in \mbI_1^n\), are satisfied merely by keeping \(\epih\), \(\epis\), and \(\|\bepi_i\|\) bounded, respectively. This is achieved by applying the control input \eqref{eq:control_law} in \eqref{eq:sys_dynamics_mimo}. This key observation will aid the stability analysis of the closed-loop system.
\end{remark}

\begin{remark}
	Note that the choice of the nonlinear transformation \(\epi_{i,j} = \Tcal(\eh_{i,j})\) in \eqref{eq:mapping_fun} is not limited to the log-type function shown in \eqref{eq:mapping_fun}. Indeed, any smooth, strictly increasing bijective mapping \(\Tcal: (-1, 1) \rightarrow (-\infty, +\infty)\) may be employed. For example, one may opt for the tan-type transformation \(\epi_{i,j} = \tan(\frac{\pi}{2}\eh_{i,j})\). Similarly, other reciprocal barrier functions may replace those in \eqref{eq:mapped_alphh} and \eqref{eq:mapped_alphs}. For instance, the log-type reciprocal barrier function \cite{ames2016control}: \(-\log(\frac{\star}{1+\star})\), with \(\star \in \{\es, \alphh\}\), shares similar properties with the inverse-type reciprocal barrier function used in \eqref{eq:mapped_alphh} and \eqref{eq:mapped_alphs}.
\end{remark}

\subsection{Stability Analysis}
\label{subsec:stability_analysis}

Recall that $\bG_1^{-1}\grad{\alphh(t,\bx_1)}$ in \eqref{eq:uh} and $\bG_1^{-1}\grad{\alphs(t,\bx_1)}$ in \eqref{eq:us} denote the control directions in the first intermediate control law $\bs_1(t,\bx_1)$ for satisfying time-varying hard and soft constraints, respectively. Since $\bG_1$ is nonsingular, these directions vanish precisely when $\grad{\alphh(t,\bx_1)}$ and $\grad{\alphs(t,\bx_1)}$ vanish, respectively. Note that $\bus = \mathbf{0}_n$ may occur at undesirable (time-varying) critical points of $\alphs(t,\bx_1)$ outside the soft-constrained set $\Os(t)$, even when $\Os(t)$ is compatible with $\Oh(t)$, potentially rendering $\bs_1(t,\bx_1)$ incapable of ensuring soft constraint satisfaction. This issue is avoided if $\grad{\alphs(t,\bx_1)}= \mathbf{0}_n$ only holds at points where the soft constraints are already met. To this end, we adopt the following technical assumption from \cite{mehdifar2024low}.  
\begin{assumption}\label{assu:alpha_globalmax}
	For all $t \geq 0$, the function $-\alphs(t,\bx_1)$ is invex, i.e., each critical point of $\alphs(t,\bx_1)$ is a (time-varying) global maximizer (see \cite[Theorem 2.2]{mishra2008invexity}). 
\end{assumption}

\cite[Lemma 3]{mehdifar2024low} provides sufficient conditions on the structure and composition of time-varying soft constraints for Assumption \ref{assu:alpha_globalmax}. Moreover, from \cite[Remark 8]{mehdifar2024low}, one can infer that this assumption is only sufficient to prevent $\grad{\alphs(t,\bx_1)}$ from vanishing outside $\Os(t)$; in practice, scenarios often arise where $\bus$ in \eqref{eq:1st_intermed_ctrl} effectively satisfies soft constraints even without Assumption \ref{assu:alpha_globalmax}. A similar requirement is unnecessary for $\grad{\alphh(t,\bx_1)}$, since $\bx_1(t;\bx(0))$ must always remain inside $\Oh(t)$. Note that Assumption \ref{assu:alpha_globalmax} is used only via $\grad{\alphs}=\mathbf{0}_n \Rightarrow \alphs>0$, and may thus be weakened to requiring every non-positive real number to be a regular value of $\alphs(t,\mathdot)$, i.e., $\grad{\alphs(t,\bx_1)} \neq \mathbf{0}_n$ whenever $\alphs(t,\bx_1) \leq 0$.

The time-varying hard constrained set \(\Oh(t)\) can, in general, decompose into disconnected components as time evolves. Hence, mere overall feasibility (Assumption \ref{assu:feasible_constrained_sets}) does not guarantee continuous satisfaction of the hard constraints whenever any component disappears. For example, suppose \(\Oh(t)=\Oh^1(t)\cup\Oh^2(t)\) for $t \geq t^\prime$, with \(\Oh^1\) and \(\Oh^2\) connected but mutually disjoint. Assume further that, for \(t \geq t^{\prime \prime} > t^\prime\), the component \(\Oh^1(t)\) vanishes while \(\Oh^2(t)\) persists. Although \(\Oh(t)\) remains feasible for \(t \geq t^{\prime \prime}\), a trajectory confined to \(\Oh^1\) during \(t^\prime \le t<t^{\prime \prime}\) cannot remain in \(\Oh(t)\) for \(t>t^{\prime \prime}\) without violating the hard constraints. To avoid this scenario, we adopt the following assumption.

\begin{assumption} \label{assu:hard_set_connected_compo_feasibility}
	The set \(\Oh(t)\) stays connected \(\forall t \geq 0\).
\end{assumption}

Finally, a regularity assumption on the boundary of the hard-constrained set streamlines the analysis.

\begin{assumption} \label{assu:hard_set_boundary_non_zero_grad}
	For all $\bx_1 \in \{\bx_{1} \in \mbR^n  \mid \alphh(t,\bx_1) = 0\}$ and all $t \geq 0$, it holds that $\grad{\alphh(t,\bx_1)} \neq 0$.
\end{assumption}

This precludes critical points on the boundary of the hard‑constrained set at all times. Such a non‑degenerate boundary is generic and typically satisfied by standard combinations of hard constraints. Moreover, Assumption \ref{assu:hard_set_boundary_non_zero_grad} implicitly supports Assumption \ref{assu:hard_set_connected_compo_feasibility}. To preserve connectivity of the feasible hard‑constrained set, the boundary $\alphh(t,\bx_1) = 0$  must stay free of (i) saddle points separating regions with $\alphh(t,\bx_1)>0$ and $\alphh(t,\bx_1)<0$ when $\bx_1\in\mathbb{R}^n$, $n \geq 2$, and (ii) both saddle points and local minima when $n=1$.

The following theorem summarizes our main result.
\begin{theorem} \label{th:main}
	Consider the MIMO nonlinear system \eqref{eq:sys_dynamics_mimo} subject to time-varying hard and soft constraints \eqref{eq:hard_soft_constraints}. Let $\rhos(t)$ in \eqref{eq:soft_const_set_lower_bound} be designed according to Subsection \ref{subsec:control_strategy} (i.e., \textit{Step 1-a}) and \eqref{eq:rho_relax_dyn}. Moreover, suppose that constants $\vartheta_{i,j}^0, i \in \mbI_2^r, j \in \mbI_1^n$ in \eqref{exponential_performance_fun} are selected such that $\vartheta_{i,j}^0 > |e_{i,j}(0,\bxb_i(0))|$ (as explained in \textit{Step i-a} in Subsection \ref{subsec:control_design}).  Under Assumptions 1-11, the feedback control law \eqref{eq:control_law} ensures the boundedness of all closed-loop signals as well as the forward invariance of the  set \(\Oh(t) \cap \Osv(t)\) for all time.
\end{theorem}
\begin{proof}
	See \ref{appen:main_theorem_proof}. 
\end{proof}

Theorem \ref{th:main} guarantees that the set \(\Oh(t)\cap\Osv(t)\), with \(\Oh(t)\) and \(\Osv(t)\) defined in \eqref{eq:smoth_hard_set} and \eqref{eq:smoth_virtual_soft_set}, respectively, is forward invariant for all \(t\ge 0\). Consequently, the partial state trajectory \(\bx_1(t;\bx(0))\) always stays inside the hard-constraint set. This, however, does not by itself ensure that the soft constraints are respected whenever \(\Oh(t) \cap \Os(t) \neq \emptyset \) (i.e., compatibility of hard and soft constraints). The relaxation signal \(\rhos(t)\) in \eqref{eq:soft_const_set_lower_bound} is designed so that, when \(\Oh(t)\) and \(\Os(t)\) are compatible, the virtual soft constrained set \(\Osv(t)\) converges back to the true one, \(\Os(t)\). Indeed, Lemma \ref{lem:nonnegative_rho_s_relax} in \ref{appen:lemma}
guarantees that $\rhosr(t)$ remains non-negative and bounded, irrespective of how long
the constraints stay in conflict, while $\varphih\varphigam = 0$ on some interval
$[t_1,\infty)$ reduces \eqref{eq:rho_relax_dyn} to $\drhosr = -\krec\rhosr$, giving
$\Osv(t)\to\Os(t)$ at the rate $\krec$. Yet this convergence can fail in certain deadlock situations. Consider, for example, \eqref{eq:sys_dynamics_mimo} reduced to a single integrator, i.e., $r=1$, $\bof_1(t,\bx_{1})=\mathbf{0}_n$, and $\bG_1(t,\bx_{1})=\bI_{n}$, so that $\dot{\bx}_1=\bu=\bs_{1}(t,\bx_{1})$. Suppose the trajectory enters and evolves in the boundary region of the hard-constraint set, i.e., $\bx_1(t;\bx_1(0))\in\Ohb(t)$ with $\Ohb(t)$ defined in \eqref{eq:boundary_region_hard}. If the terms $\bus$ and $\varphih\,\buh$ in \eqref{eq:1st_intermed_ctrl} cancel, we obtain $\bs_1(t,\bx_1)=\mathbf{0}_n$, and the state can stall at an undesirable equilibrium.
Then $\rhosr(t)$, governed by \eqref{eq:rho_relax_dyn}, may converge to a positive
constant, which keeps $\Osv(t)$ apart from $\Os(t)$ even when $\Oh(t)$ and $\Os(t)$ are
compatible. Such a stall is precisely the failure of the condition above: a nonzero
steady state of \eqref{eq:rho_relax_dyn} requires its first term to remain active, so
$\varphih\varphigam$ does not vanish on any interval $[t_1,\infty)$.
The time-varying nature of the hard or soft constrained sets, combined with
the internal dynamics in \eqref{eq:sys_dynamics_mimo}, can perturb such a boundary
equilibrium with $\bs_1(t,\bx_1)=\mathbf{0}_n$, thereby re-activating the evolution of
$\rhosr(t)$ and restoring $\Osv(t)\to\Os(t)$, as Example 3 in Section~\ref{sec:simu}
illustrates. Such deadlocks are, however, not excluded in general.
	
\begin{remark} \label{rem:finite_time_recovery}
	Recovery of the soft constraints is achieved in finite time whenever the conflict
	permanently ceases, i.e., $\varphih\varphigam = 0$ for all $t \ge t_1$ and some
	$t_1 \ge 0$, thus excluding the deadlocks described above. Indeed,
	\eqref{eq:rho_relax_dyn} then gives $\rhosr(t) = \rhosr(t_1)e^{-\krec(t-t_1)}$, while
	$\taum = \infty$ renders \eqref{eq:bounded_epih_epis_step1} valid for all $t \ge 0$, so
	that $\es \ge 1/\epibs$. Since $\rhosn(t) = 0$ for all $t \ge T$ (see
	\eqref{eq:alpha_lower_bound}), \eqref{eq:es} and \eqref{eq:soft_const_set_lower_bound}
	yield $\alphs(t,\bx_1(t)) \ge 1/\epibs - \rhosr(t_1)e^{-\krec(t-t_1)}$,
	$\forall t \ge \max\{t_1,T\}$, whose right-hand side increases in $t$ and is positive for
	all $t > t^\star \coloneqq \max\{T,\, t_1 + \krec^{-1}\max\{\ln(\epibs\rhosr(t_1)),\,0\}\}$.
	Hence $\bx_1(t;\bx(0)) \in \Os(t)$, $\forall t > t^\star$: the partial state trajectory
	re-enters the true soft constrained set in finite time and remains therein, by
	\eqref{eq:rhosr_strict_bounds} within $\krec^{-1}\max\{\ln(\epibs\rhosrb),\,0\}$ of $t_1$
	when $t_1 \ge T$, irrespective of how long the constraints stayed in conflict. Note,
	however, that $\epibs$ and $\rhosrb$ follow from contradiction and comparison arguments
	rather than explicit constructions, and depend on the unknown dynamics as well as on the
	control gains, the initial condition, and the constraint geometry in
	\eqref{eq:hard_soft_constraints}. Accordingly, $t^\star$ is finite but not computable
	here; explicit estimates in special cases, e.g., under full model knowledge, are left for
	future study. The rate $\krec$, in contrast, is a design parameter, and a larger $\krec$
	shortens the recovery interval.
\end{remark}

\begin{remark}\label{rem:sufficient_assum_modified}
	Assumptions \ref{assu:bounded_in_t_psi_fun}--\ref{assu:hard_set_boundary_non_zero_grad} constrain only the \emph{known} constraint functions \eqref{eq:hard_soft_constraints}, not the unknown dynamics, and impose no implementation requirement beyond the evaluation of $\alpha_{\star}$ and $\grad{\alpha_{\star}}$ already used by \eqref{eq:control_law}; they are hence verifiable offline. Assumptions \ref{assu:feasible_constrained_sets}, \ref{assu:safe_ini}, \ref{assu:hard_set_connected_compo_feasibility} are necessary for the hard constraints to be enforceable by any controller, while Assumptions \ref{assu:alpha_globalmax} and \ref{assu:hard_set_boundary_non_zero_grad} are sufficient non-degeneracy conditions safeguarding, respectively, soft-constraint satisfaction and forward invariance of $\Oh(t)$. Assumption \ref{assu:hard_set_boundary_non_zero_grad} is conservative but sufficient for Theorem \ref{th:main}; Example 4 in the simulation study  (Section \ref{sec:simu}) illustrates that the theorem may still hold when it is relaxed.
\end{remark}

\begin{remark}\label{rem:safety_filter}
	The architecture of Section \ref{subsec:control_design} also enables trajectory tracking under time-varying hard constraints. In \eqref{eq:1st_intermed_ctrl}, $\bus$ may be replaced by any nominal tracking controller that drives $\bx_1$ along the reference when $\varphih \buh = 0$. Since this nominal control law remains bounded, Theorem \ref{th:main} still applies, making the architecture akin to a safety filter. For example, the minimum-norm CBF-QP method of \cite{ames2016control} serves
	as such a filter but requires a system model and online optimization; a closed-form variant for \emph{parametrically} uncertain strict-feedback systems \cite{an2026cbf} still uses the known regressors, input matrix and parameter bounds under a single time-invariant constraint. In contrast, our approach requires neither the system dynamics nor online optimization: the only plant-side quantity entering the control law is the known kinematic map $\bG_1$, and none at all when $\bG_1=\bI_n$. Furthermore, it not only ensures safety but also enforces prescribed performance in stabilization or tracking by encoding time-varying soft constraints, capabilities beyond standard CBF-CLF QP designs \cite{ames2016control,ames2019control}.
\end{remark}

\begin{remark}\label{rem:general_composition}
	For brevity, this article focuses on the constraint sets in \eqref{eq:smooth_hard_soft_sets}. Each set arises solely from intersecting the corresponding hard or soft constraint functions in \eqref{eq:hard_soft_constraints}, implemented via the $\min$ operator in \eqref{eq:consolidated_hard_soft}. In practice, specifications are often more intricate, requiring unions or nested combinations as well, i.e., both $\min$ (for intersections) and $\max$ (for unions) across different functions \cite{wiltz2022handling}. For time-invariant constraints, \cite{molnar2023composing} proposes a method for constructing a single smooth constraint function whose positive level sets capture such general combinations. Building on this idea, the robust feedback-control strategy developed in this paper may be extendable to handle a broader class of constraints. A detailed investigation of this possibility is left to future work.
\end{remark}

\begin{remark} \label{rem:disjoint_hard_set}
	Assumption \ref{assu:hard_set_connected_compo_feasibility} can be weakened to allow $\Oh(t)$ to become disjoint, provided every connected component of $\Oh(t)$ remains nonempty for all $t\ge 0$. This removes the concern, noted before Assumption~\ref{assu:hard_set_connected_compo_feasibility}, about continuous satisfaction of the hard constraints. Because $\Oh(t)$ has compact level curves (Assumption \ref{assu:coercive_alpha_funs}) and is nonempty for all time (Assumption \ref{assu:feasible_constrained_sets}), each component must contain at least one positive local maximizer of $\alphh(t,\bx_1)$. A complete stability analysis under this relaxed assumption is left for future study.
\end{remark}

\subsection{From Semi-Global to Global Stability Results}
\label{subsec:global_stab}

The stability results in Theorem~\ref{th:main} are only semi‑global. This is because selecting $\rho_0 = \rhos(0) < \alphs(0,\bx_1(0))$ in \eqref{eq:alpha_lower_bound} and $\vartheta_{i,j}^0 > |e_{i,j}(0,\bar{\bx}_{i}(0))|$ in \eqref{exponential_performance_fun} requires knowledge of the system's initial condition $\bx(0)$. Besides tuning these parameters, $\bx(0)$ is also needed to implement the state‑feedback law \eqref{eq:control_law}. In practice, $\bx(0)$ is available at the initial time, so one can automatically tune the above parameters before applying the control signal. Nevertheless, certain applications benefit from eliminating parameter tuning at the initial time.

Inspired by \cite{song2018tracking,sun2024global}, we make minor modifications to the design of Section~\ref{subsec:control_design}. This eliminates reliance on the initial condition, extending Theorem~\ref{th:main} from semi-global to global. To this end, define the \textit{shifting function}
\begin{equation}\label{eq:shifting_fun}
	\eta(t) = \begin{cases}
		\varpi(t), & 0 \leq t \leq \Ts,\\
		1, & t > \Ts,
	\end{cases}
\end{equation}
where $\varpi(t)$ is any strictly increasing, continuously differentiable function satisfying $\varpi(0)=0$, $\varpi(\Ts)=1$, and $\dot{\varpi}(\Ts)=0$, and $\Ts>0$ is a prescribed settling time. Numerous designs for $\varpi(t)$ are possible; a convenient example is $\varpi(t)=\sin(\frac{\pi}{2}\frac{t}{T_s})$. One can verify that the shifting function $\eta(t)$ is continuously differentiable and that both $\eta(t)$ and its derivative $\dot{\eta}(t)$ remain bounded for all time.

\textbf{\textit{Modified Controller Design Scheme:}} First, set $\rho_0$ in \eqref{eq:alpha_lower_bound} to an \textit{arbitrary} negative constant, ignoring whether $\alphs(0,\bx_1(0))>0$ (condition~(i) above \eqref{eq:alpha_lower_bound}). Using $\eta(t)$, we redefine $\es$ in \eqref{eq:es} as
\begin{equation}\label{eq:es_modif}
	\es(t,\bx_1) \coloneqq \eta(t) \, \alphs(t,\bx_1) - \rhos(t).
\end{equation}
Because $\eta(0)=0$ and $\rho_0=\rhos(0)<0$, $\es(0,\bx_1(0))>0$, so $\epis$ in \eqref{eq:mapped_alphs} is well defined and positive at $t=0$. Note that, for $t\geq\Ts$, \eqref{eq:es_modif} reduces to $\es(t,\bx_1)=\alphs(t,\bx_1)-\rhos(t)$, which matches the original definition in \eqref{eq:es}. Unlike Section~\ref{subsec:control_design}, keeping $\epis$ in \eqref{eq:mapped_alphs} bounded, and hence $\es(t,\bx_1)$ in \eqref{eq:es_modif} strictly positive, guarantees \eqref{eq:soft_const_satisfaction_cond} only for $t\geq\Ts$. Moreover, the settling time $\Ts$ must satisfy $\Ts\leq T$, ensuring that enforcing $\es(t,\bx_1)>0$ in \eqref{eq:es_modif} can ideally satisfy the soft constraints by the user-defined nominal time $T$.

The modification in \eqref{eq:es_modif} removes the need to tune $\rho_0$ using $\alphs(0,\bx_1(0))$. Adopting \eqref{eq:es_modif} in place of \eqref{eq:es}, one may omit \textit{Step 1‑a} in Section~\ref{subsec:control_design} and proceed directly with \textit{Steps 1‑b} and \textit{1‑c}, giving $\bs_1(t,\bx_1)$ in \eqref{eq:1st_intermed_ctrl}.

Next, to avoid tuning $\vartheta_{i,j}^0$ in \eqref{exponential_performance_fun} on the basis of $|e_{i,j}(0,\bar{\bx}_{i}(0))|$ we modify \eqref{eq:i-th_inter_funnels} to
\begin{equation} \label{eq:i-th_inter_funnels_modif}
	-\vartheta_{i,j}(t) < \eta(t) \, e_{i,j}(t,\bxb_i) < \vartheta_{i,j}(t), \quad j \in \mbI_1^n,
\end{equation}
and redefine $\beh_i$ in \eqref{eq:normalized_err_vec} and its components in \eqref{eq:normal_e_i_j} as
$\beh_i(t,\be_i)=\col{\eh_{i,j}}\coloneqq\eta(t) \bvarTheta_i^{-1}(t) \be_i$ and
\begin{equation} \label{eq:normal_e_i_j_modif}
	\eh_{i,j}(t,e_{i,j})=\eta(t)\frac{e_{i,j}}{\vartheta_{i,j}(t)}, \quad j \in \mbI_1^n,
\end{equation}
respectively. Because $\eta(0)=0$, the modified constraints \eqref{eq:i-th_inter_funnels_modif} hold automatically at $t=0$, even if \eqref{eq:i-th_inter_funnels} is violated. Specifically, $\eh_{i,j}(0,e_{i,j}(0,\bxb_i(0)))=0$ for any initial $e_{i,j}(0,\bxb_i(0))$, so the transformed signals $\epi_{i,j}$ in \eqref{eq:mapping_fun} vanish at $t=0$. For $t\ge T_s$, \eqref{eq:i-th_inter_funnels_modif} coincides with the original intermediate constraints in \eqref{eq:i-th_inter_funnels}.

With any positive $\vartheta_{i,j}^0$ and the replacements \eqref{eq:i-th_inter_funnels_modif}–\eqref{eq:normal_e_i_j_modif}, one proceeds directly with \textit{Steps i‑a} and \textit{i‑c} ($2\le i\le r$) in Section~\ref{subsec:control_design}, yielding $\bs_i(t,\be_i)$ in \eqref{eq:i-th_intermed_ctrl}, whose diagonal matrix $\bXi_i$ has entries
\begin{equation}\label{eq:xi_i,j_modif}
	\xi_{i,j}(t,e_{i,j})\coloneqq\frac{2\eta(t)}{\vartheta_{i,j}(t) (1-\eh_{i,j}^2)}, \quad j \in \mbI_1^n.
\end{equation}
Finally, the control input follows from \textit{Step r+1} as in \eqref{eq:control_law}.

The shifting function $\eta$ guarantees that \eqref{eq:soft_const_satisfaction_cond} and \eqref{eq:i-th_inter_funnels} hold at $t=0$ for any initial state, without parameter tuning. Consequently, $\epi_s$ in \eqref{eq:mapped_alphs} and $\epi_{i,j}$ in \eqref{eq:mapping_fun} are initially bounded and well defined. If they remain bounded for all $t\ge0$, then \eqref{eq:soft_const_satisfaction_cond} and \eqref{eq:i-th_inter_funnels} hold for all $t\ge\Ts$. Thus, $\eta$ removes any dependence of the controller parameters on the initial condition, at the cost of postponing the original requirements \eqref{eq:soft_const_satisfaction_cond} and \eqref{eq:i-th_inter_funnels} until the settling time $\Ts$. The above modifications yield the following corollary, a global stability result complementing Theorem \ref{th:main}.

\begin{corollary} \label{colo:main}
	Consider the MIMO nonlinear system \eqref{eq:sys_dynamics_mimo} subject to the time‑varying hard and soft constraints \eqref{eq:hard_soft_constraints}.
	Design $\rhos(t)$ in \eqref{eq:soft_const_set_lower_bound} as described in Subsection \ref{subsec:control_strategy} and \eqref{eq:rho_relax_dyn}, and choose $\rho_0$ in \eqref{eq:alpha_lower_bound} to be any negative constant.
	Furthermore, select arbitrary positive constants $\vartheta_{i,j}^0$, $i \in \mbI_2^r$, $j \in \mbI_1^n$, in \eqref{exponential_performance_fun} and for the shifting function in \eqref{eq:shifting_fun}, pick $\Ts \leq T$.
	Under Assumptions 1‑11, the feedback control law obtained via the modified design scheme of Section \ref{subsec:global_stab} guarantees that all closed‑loop signals remain bounded for all $t \geq 0$ and renders $\Oh(t)$ forward invariant for all $t \geq 0$. Moreover, it ensures the forward invariance of the time‑varying set $\Oh(t) \cap \Osv(t)$ for all $t \geq \Ts$.
\end{corollary}
\begin{proof}
	The proof follows the three phases of the proof of Theorem~\ref{th:main} in
	\ref{appen:main_theorem_proof}. Since $\eta(t)$ vanishes at $t=0$, three modifications are
	required, which we detail below.
	
	\textit{Phase I (state, domain, and globality).} As $\beh_i=\eta\,\bvarTheta_i^{-1}\be_i$
	degenerates at $t=0$, the map $\be_i\mapsto\beh_i$ is not invertible there and
	$\bx_{i+1}=\eta^{-1}\bvarTheta_{i+1}\beh_{i+1}+\bs_i$ is undefined at $t=0$. Accordingly, in
	\eqref{eq:z_dynamics} the states $\beh_i$ are replaced by $\be_i$, and the time-invariant set
	$\Oz$ by the open set
	\begin{equation}\label{eq:D_domain}
		\begin{aligned}
			\mathcal{D}\coloneqq\big\{(t,\bz)&\in\mbRpos\times\mbR^{nr+3}\,:\,\alphh>0,\ \es>0, \\ 
			&|\eta(t)\,e_{i,j}|<\vartheta_{i,j}(t),\ i\in\mbI_2^r,\ j\in\mbI_1^n\big\},
		\end{aligned}
	\end{equation}
	so that $\bphi$ contains no factor $\eta^{-1}$. Since $\xi_{i,j}$ in \eqref{eq:xi_i,j_modif}
	is well defined (and vanishes) at $t=0$, and $\bG_1^{-1}$ is $\Ccal^1$ by
	Assumption~\ref{assu:known_unknown}, $\bphi$ is locally Lipschitz in $\bz$ and piece-wise
	continuous in $t$ on $\mathcal{D}$; applying \cite[Theorem~54]{sontag1998mathematical} to the
	augmented system $(\dot t,\dbz)=(1,\bphi(t,\bz))$ on $\mathcal{D}$ yields a unique maximal
	solution on $[0,\taum)$. Because $\eta(0)=0$, the last condition in \eqref{eq:D_domain} holds
	for \emph{every} $\be_i\in\mbR^n$ at $t=0$, and $\es(0,\bx_1(0))=-\rho_0>0$ for any negative
	$\rho_0$; hence $(0,\bz(0))\in\mathcal{D}$ for every $\bx(0)$ satisfying
	Assumption~\ref{assu:safe_ini} and arbitrary $\bx_i(0)$, $i\in\mbI_2^r$, which is the source
	of the global nature of the result. The bounds $\alphh\in(0,b_h]$, $\bx_1\in\Linf$ and
	$\alphs\in\Linf$ follow as in Theorem~\ref{th:main}, being unaffected by $\eta$.
	
	\textit{Phase II (absence of escape on $[0,\Ts]$).} Membership in $\mathcal{D}$ ensures
	$\alphh>0$, $\es>0$ and $|\eh_{i,j}|<1$; consequently, and since $\bx_1$ evolves in a compact
	set, the trajectory can leave every compact subset of $\mathcal{D}$ only through divergence
	of $\epih$, $\epis$ or $\bepi_i$. Such divergence at a time $t'$ is precluded by the
	following two facts.
	
	\emph{(F1) No escape at $t'=0$.} By Assumption~\ref{assu:safe_ini},
	$\epih(0)=1/\alphh(0,\bx_1(0))<\infty$; by $\eta(0)=0$, $\es(0,\bx_1(0))=-\rho_0$ so that
	$\epis(0)=-1/\rho_0<\infty$; and $\eh_{i,j}(0)=0$ so that $\epi_{i,j}(0)=0$. All three are
	finite for \emph{every} $\bx(0)$ and for arbitrary $\rho_0<0$ and $\vartheta_{i,j}^0>0$, and
	are continuous; hence no divergence occurs at the initial time.
	
	\emph{(F2) No escape at any $t'>0$.} Unlike in Theorem~\ref{th:main}, \eqref{eq:D_domain}
	yields only $|e_{i,j}|<\vartheta_{i,j}(t)/\eta(t)$, which is not uniformly bounded as
	$t\to0^{+}$. However, as $\eta$ is non-decreasing, for every $\delta\in(0,\taum)$, we have 	$\eta(t)\ge\eta(\delta)\eqqcolon\eta_\delta>0$, and thus
	\begin{equation}\label{eq:localisation}
		 |e_{i,j}(t)|\le\vartheta_{i,j}^{0}/\eta_\delta\eqqcolon M_\delta<\infty,
		\quad \forall t\in[\delta,\taum).
	\end{equation}
	The constant $M_\delta$ grows as $\delta$ decreases, but is finite, which is all the
	arguments of Theorem~\ref{th:main} require. Indeed, on $[\delta,\taum)$: $\sigma_h$ in
	\eqref{eq:epih_dot} is bounded by \eqref{eq:localisation}, which is the only term in which
	$\eta^{-1}$ exists; in \eqref{eq:epis_dot} the factor $\eta$ arising from
	\eqref{eq:es_modif} cancels the $\eta^{-1}$ in $\be_2$, giving
	\begin{equation}\label{eq:epis_dot_modif}
	\begin{aligned}
		\depis=-\epis^2\Big[&\eta\,k_s\epis^2\|\grad{\alphs}\|^2 \\
		&+(\eta-\varphigam)\,\varphih k_h\epih^2\,\grad{\alphs}^\top\grad{\alphh}+\tilde\sigma_s\Big],
	\end{aligned}
	\end{equation}
	with $\tilde\sigma_s\coloneqq\dot\eta\,\alphs+\eta\grad{\alphs}^\top\bof_1
	+\grad{\alphs}^\top\bG_1\bvarTheta_2\beh_2+\eta\gradt{\alphs}-\drhosn-\krec\rhosr$ bounded on
	all of $[0,\taum)$, exactly as $\sigma_s$ in \eqref{eq:epis_dot}; and
	$\xi_{i,j}\ge2\eta_\delta/\vartheta_{i,j}^{0}$ implies
	$\|\bXi_i\bepi_i\|\ge c_i\|\bepi_i\|$ for some $c_i>0$, preserving the Lyapunov argument of
	steps $2\le i\le r$. Since the divergent terms in \eqref{eq:epih_dot} and
	\eqref{eq:epis_dot_modif} are quartic in $\epih$ and $\epis$ whereas $\sigma_h$ and
	$\tilde\sigma_s$ are bounded, the contradiction arguments of \textit{Modes A, B1} and
	\textit{B2} carry over verbatim for any $\eta_\delta>0$, however small. Three points are
	specific to the present setting: in \textit{Mode A}, if $\grad{\alphs}=\mathbf{0}_n$ then
	Assumption~\ref{assu:alpha_globalmax} gives $\alphs>0$ and hence
	$\es\ge\eta_\delta\,\alphs>0$; in \textit{Mode B2} one has $(\eta-\varphigam)\le0$ while
	$\grad{\alphs}^\top\grad{\alphh}<0$, so that the associated term in
	\eqref{eq:epis_dot_modif} is non-negative and reinforces the contradiction, while
	$\rhosr=\rhosn-\eta\alphs+\es$ remains bounded as $\epis\to+\infty$ because $\eta\le1$; and
	for $2\le i\le r$, \eqref{eq:epi_i_dot} acquires the additional term
	$\dot\eta\,\bD_i\bvarTheta_i^{-1}\be_i$, with $\bD_i\coloneqq\diag{2/(1-\eh_{i,j}^2)}$, which
	is free of singularities in the $\be_i$ coordinates, whereas
	$\bXi_i\bG_i\bvarTheta_{i+1}\beh_{i+1}/\eta$ is independent of $\eta$. Lemma
	\ref{lem:nonnegative_rho_s_relax} applies unchanged, \eqref{eq:rho_relax_dyn} being
	unmodified.
	
	\textit{Phase III (forward completeness).} Note that the bounds furnished by (F2) depend on
	$\delta$ through $M_\delta$; the compact set is therefore constructed \emph{after} assuming
	finiteness of $\taum$, rather than before as in Theorem~\ref{th:main}. Suppose
	$\taum<\infty$ and set $\delta\coloneqq \tfrac{\taum}{2}$. By (F1)--(F2) the barrier variables are
	bounded on $[\delta,\taum)$, while on the compact interval $[0,\delta]$ the solution exists
	and is continuous, so that $\alphh$, $\es$ and $\vartheta_{i,j}-|\eta e_{i,j}|$ attain
	positive minima there. Hence the trajectory remains within a compact subset of
	$\mathcal{D}$ on $[0,\taum)$, contradicting
	\cite[Proposition C.3.6]{sontag1998mathematical}; therefore $\taum=\infty$. Repeating (F2)
	with the fixed choice $\delta=\Ts/2$ then yields time-uniform bounds on $[\Ts/2,\infty)$,
	which, combined with continuity on $[0,\Ts/2]$, gives boundedness of all closed-loop signals
	for all $t\ge0$. In particular $\alphh(t,\bx_1(t))>0$ for all $t\ge0$, so that $\Oh(t)$ is
	forward invariant from the initial time, the shifting function affecting neither
	\eqref{eq:mapped_alphh} nor \eqref{eq:uh}. Finally, since $\eta(\Ts)=1$, the modified scheme
	coincides with that of Section~\ref{subsec:control_design} for $t\ge\Ts$, and
	$\alphh(\Ts,\bx_1(\Ts))>0$, $\rhos(\Ts)<\alphs(\Ts,\bx_1(\Ts))$ and
	$|e_{i,j}(\Ts,\bxb_i(\Ts))|<\vartheta_{i,j}(\Ts)$ hold; Theorem~\ref{th:main} therefore
	applies with initial time $\Ts$, which establishes the forward invariance of
	$\Oh(t)\cap\Osv(t)$ for all $t\ge\Ts$.
\end{proof}

\begin{remark} \label{rem:corollary_explain}
Corollary~\ref{colo:main} strengthens Theorem~\ref{th:main} to global,
yet guarantees forward invariance of $\Omega_h(t)\cap\tilde\Omega_s(t)$ only for
$t\ge T_s$. This is because $\eta(t)$ in \eqref{eq:shifting_fun} acts solely on $e_s$
in \eqref{eq:es_modif} and on \eqref{eq:i-th_inter_funnels_modif}, leaving $\varepsilon_h$ unaffected:
the boundedness of $\varepsilon_h$ renders $\Omega_h(t)$ forward invariant for all
$t\ge 0$, so that safety is never deferred, whereas that of $\varepsilon_s$ renders
$\tilde\Omega_s(t)$ invariant only for $t\ge T_s$. Nevertheless, since
$\eta(t)>0$ for $t>0$, on $(0,T_s)$ the closed loop still satisfies the inflated
conditions $\alpha_s(t,x_1)>\rho_s(t)/\eta(t)$ and
$|e_{i,j}(t,\bar x_i)|<\vartheta_{i,j}(t)/\eta(t)$, which tighten continuously to \eqref{eq:soft_const_satisfaction_cond}  
and \eqref{eq:i-th_inter_funnels} at $t=T_s$. Finally, if the soft constraints hold at $t=0$ and are compatible
with the hard ones, the modified scheme may not preserve them on $0<t<T_s$; a smaller
$T_s$ shortens this interval at the cost of a more aggressive transient.	
\end{remark}

\section{Simulation Results}
\label{sec:simu}

This section presents several simulation examples to validate the proposed approach. We consider the position control problem of a unicycle mobile robot on a 2D plane subject to soft and hard spatiotemporal specifications.\footnote{Simulation videos: \url{https://youtu.be/EAbVfd9_BlE}}

The uncertain mobile robot dynamics are given by
\begin{equation} \label{eq:robot_kin_dyn}
	\begin{cases}
		\dot{\bp}_c = \bS(\theta)\bzeta \\ 
		\bar{\bM} \dot{\bzeta} + \bar{\bD} \bzeta = \bar{\bu} + \bar{\bd}(t) 
	\end{cases}\!\!\!\!\!\!, ~
	\begin{aligned}
		\bS(\theta)\! =\!\! \left[ \!\!
		\begin{array}{ccc}
			\cos \theta &\!\! \sin \theta &\!\! 0 \\
			0 &\!\! 0 &\!\! 1
		\end{array}\!\! \right]^{\top}\!\!,
	\end{aligned}
\end{equation}
where $\bp_c=[x_c,y_c,\theta]^{\top}$ is the robot position and orientation, and $\bzeta=[v_T,\dot{\theta}]^{\top}$ collects the translational speed $v_T$ along $\theta$ and the angular speed $\dot{\theta}$ about the vertical axis through the robot’s center of mass. Here $\bar{\bD}=\diag{\bar{D}_1,\bar{D}_2}$ is an unknown constant damping matrix, and $\bar{\bM}=\diag{m_R,I_R}$ with $m_R>0$, $I_R>0$ the unknown mass and inertia. The input $\bar{\bu}$ is the force/torque vector, and $\bar{\bd}(t)$ collects bounded external disturbances.

Selecting a virtual control point (VCP) transforms the underactuated dynamics \eqref{eq:robot_kin_dyn} into a fully actuated system \cite{cai2014adaptive}; applying the VCP transformation gives
\begin{flalign}\label{eq:robot_transformed_dyn}
	&\dot{\bx}_1 = \bx_2,& \\ 
	&\dot{\bx}_2 = \bM(\bx_1)^{-1} \left(-\bC(\bx_1, \bx_2) \bx_2 - \bD(\bx_1) \bx_2 + \bu + \bd(t) \right),& \nonumber
\end{flalign} 
where $\bx_1=[x_{1,1},x_{1,2}]^{\top}\coloneqq[x_c,y_c]^{\top}+L[\cos\theta,\sin\theta]^{\top}$ is the VCP located $L$ units ahead of the center of mass along the heading, and $\bx_2$ is its velocity. Details of the transformation matrix and the correspondence between the models in \eqref{eq:robot_transformed_dyn} and \eqref{eq:robot_kin_dyn} can be found in \cite{mehdifar2024low,cai2014adaptive}. The mappings $\bM(\bx_1)$, $\bC(\bx_1,\bx_2)$, and $\bD(\bx_1)$ are locally Lipschitz, with $\bM$ positive definite. Note that \eqref{eq:robot_transformed_dyn} is a particular form of \eqref{eq:sys_dynamics_mimo} with $n=2$ and $r=2$, satisfying Assumptions 1–3, with $\bG_1 = \bI_2$, so that $\buh$ and $\bus$ in \eqref{eq:uh} and \eqref{eq:us} reduce to scaled negative gradients and the implemented control law uses no information about the robot dynamics. In the simulations we set $m_R=3.6$, $I_R=0.0405$, $\bar{D}_1=0.3$, $\bar{D}_2=0.04$, $L=0.2$, and $\bar{\bd}(t)=[0.75\sin(3t+\tfrac{\pi}{3})+1.5\cos(t+\tfrac{3\pi}{7}),-0.8\exp(\cos(t+\tfrac{\pi}{3})+1)\sin(t)]^{\top}$.

\begin{table}[t]            
	\footnotesize                      
	\setlength{\tabcolsep}{3pt} 
	\renewcommand{\arraystretch}{1} 
	\centering
	\begin{tabularx}{\linewidth}{@{} l X @{}}
		\toprule
		\textbf{Eq.\ No.} & \textbf{Parameter(s)} \\
		\midrule
		\eqref{smooth_alph}            & $\nu = 10$ \\[2pt]
		\eqref{eq:alpha_lower_bound}   & $T = 4$, $\beta = 0.3$, $\rho_0 < \alphs(0,\bx_1(0))$ \\[2pt]
		\eqref{eq:uh}                  & $k_h = 1$ \\[2pt]
		\eqref{eq:rho_relax_dyn}       & $\delta_h = 0.5$, $\delta_\gamma = 10$, $\krec = 1.5$ \\[2pt]
		\eqref{eq:us}                  & $k_s = 1$ \\[2pt]
		\eqref{exponential_performance_fun}
		& $\vartheta_{2,j}^{\infty}=0.1$, $l_{2,j}=1$,  
		$\vartheta_{2,j}^0 > |e_{2,j}(0,\bar{\bx}_{2}(0))|$, $j\in\mbI_1^2$ \\[2pt]
		\eqref{eq:i-th_intermed_ctrl}  & $k_2 = 1$ \\
		\bottomrule
	\end{tabularx}
	\caption{Control-law parameters. \vspace{-0.2cm}}
	\label{tab:contr_pram_simu1}
\end{table}

\textbf{Example 1}:
The ground robot should stay within a circular zone directly beneath an aerial survey robot in a confined workspace. We model the workspace with the (time-invariant) $\mathcal{C}^{2}$ hard constraints $\psih{1}(\bx_1)=x_{1,1}+4.5>0$, $\psih{2}(\bx_1)=x_{1,2}-0.3x_{1,1}+4.5>0$, and $\psih{3}(\bx_1)=\tanh\bigl(0.1(36-\|\bx_1\|^{2})\bigr)>0$.\footnote{The monotone map $\tanh(\mathdot)$ is used only for clearer plots; its zero super-level set matches that of $36-\|\bx_1\|^{2}$. Its scaling can also help numerical conditioning \cite{molnar2023composing}.}
The task is the time-varying soft constraint $\psi_s(t,\bx_1)=1-\|\bx_1-\bx_{\mathrm{a}}^{\mathrm{p}}(t)\|^{2}>0$, i.e., a unit-radius disk centered at the aerial robot's ground projection $\bx_{\mathrm{a}}^{\mathrm{p}}(t) \coloneqq [x_{a,1}(t), x_{a,2}(t)]^\top$.

Table~\ref{tab:contr_pram_simu1} lists the values used in the control law~\eqref{eq:control_law}. The parameters $\rho_{0}$ and $\vartheta_{2,j}^{0}$, $j\in\mbI_{1}^{2}$, are tuned from the initial state $\bx(0)$ of \eqref{eq:robot_transformed_dyn}. Henceforth, for brevity we write $\bx_{1}(t)$ instead of $\bx_{1}(t;\bx(0))$.

\begin{figure}[!tbp]
	\centering
	\begin{subfigure}[t]{0.55\linewidth}
		\vspace{0pt}
		\includegraphics[width=\linewidth]{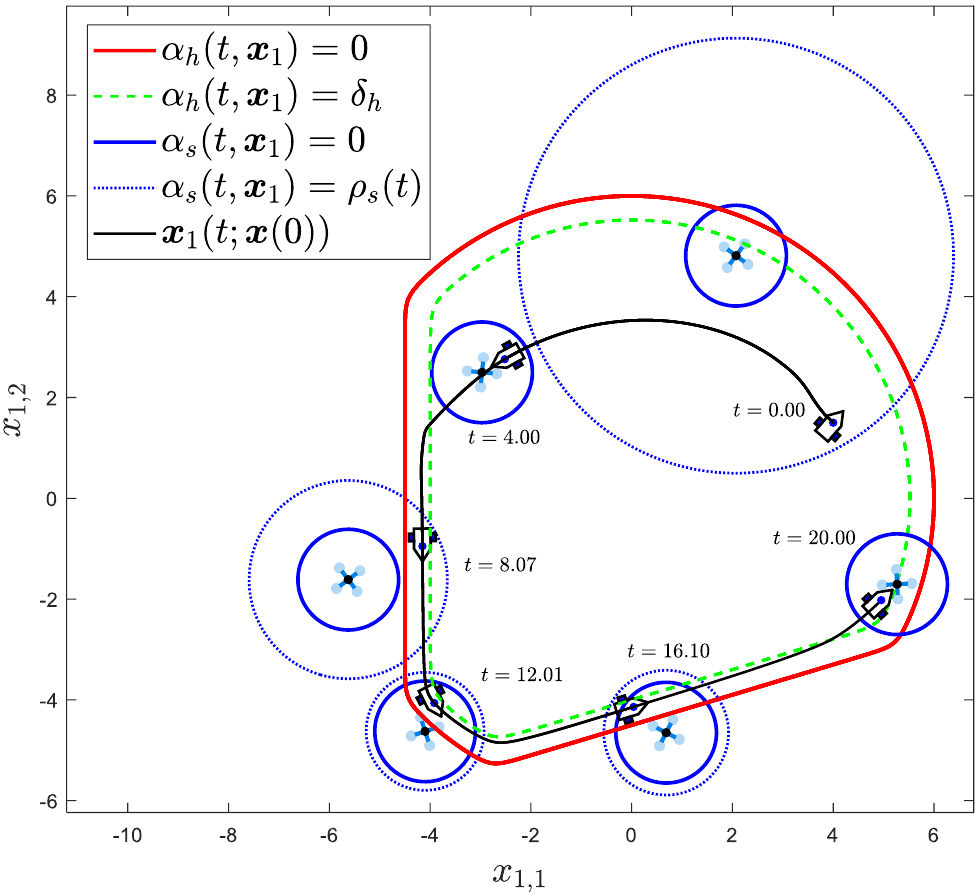}
		\caption{\vspace{-0.2cm}}
		\label{fig:confined_robot}
	\end{subfigure} 
	\begin{subfigure}[t]{0.43\linewidth}
		\vspace{0pt}
		\includegraphics[width=\linewidth]{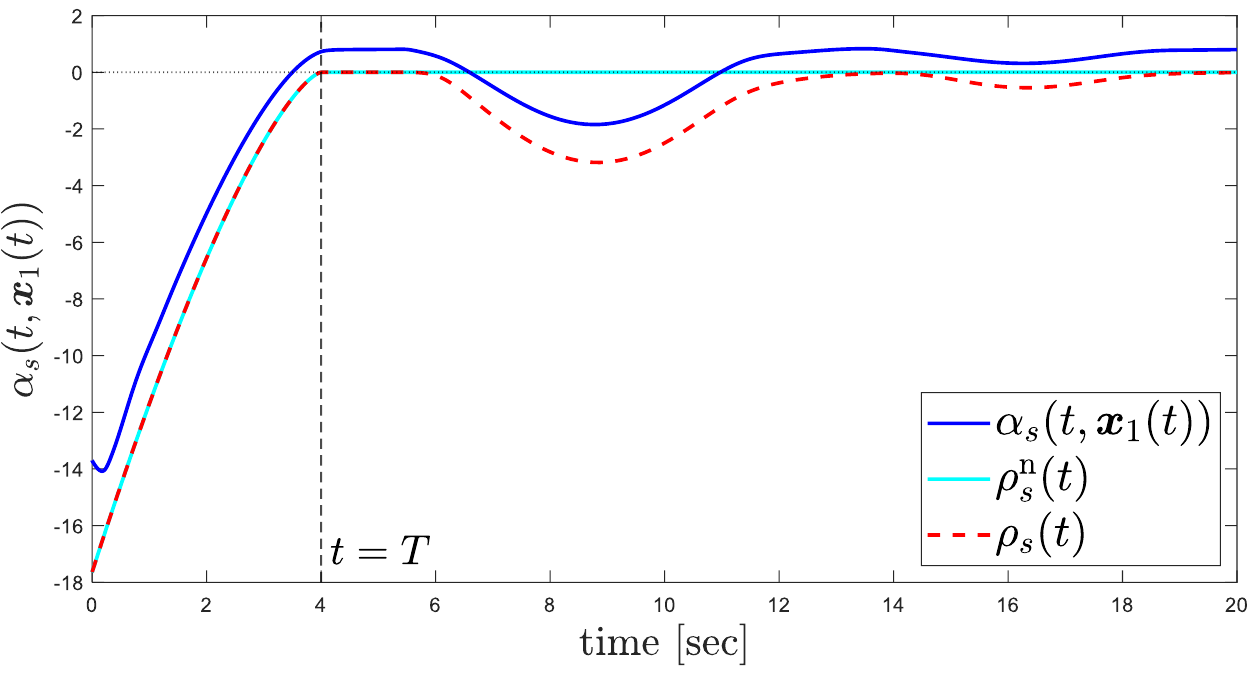}\\[9pt]
		\includegraphics[width=\linewidth]{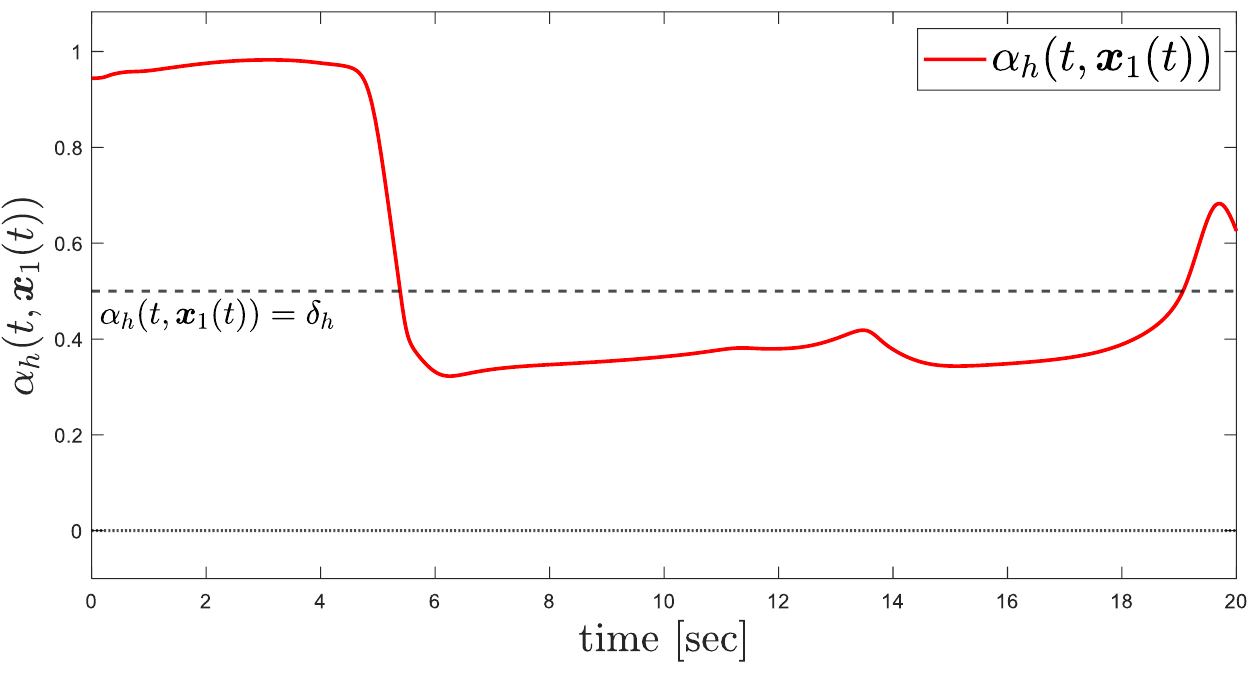}
		\caption{\vspace{-0.2cm}}
		\label{fig:confined_alpha_s_h}
	\end{subfigure}	
	\caption{Example 1: initial-condition-dependent tuning. \vspace{-0.2cm}}
	\label{fig:ex1}
\end{figure}

\begin{figure}[!tbp]
	\centering
	\begin{subfigure}[t]{0.55\linewidth}
		\vspace{0pt}
		\includegraphics[width=\linewidth]{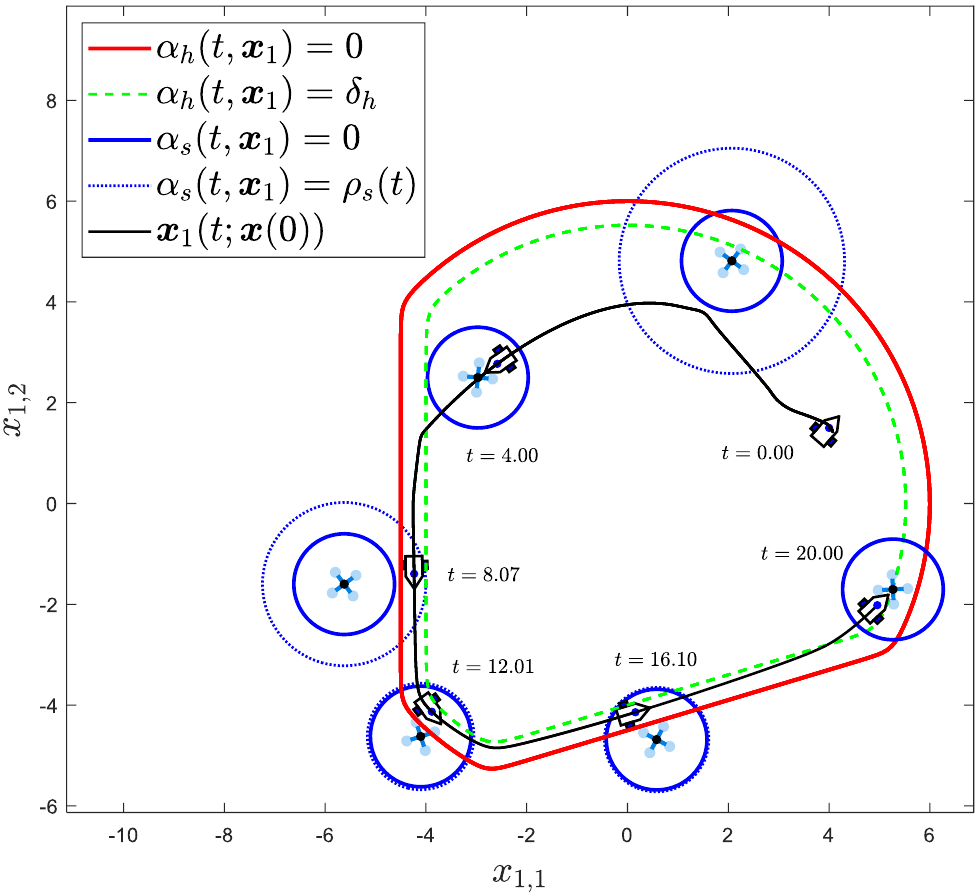}
		\caption{\vspace{-0.2cm}}
		\label{fig:confined_robot_ini_free}
	\end{subfigure} 
	\begin{subfigure}[t]{0.43\linewidth}
		\vspace{0pt}
		\includegraphics[width=\linewidth]{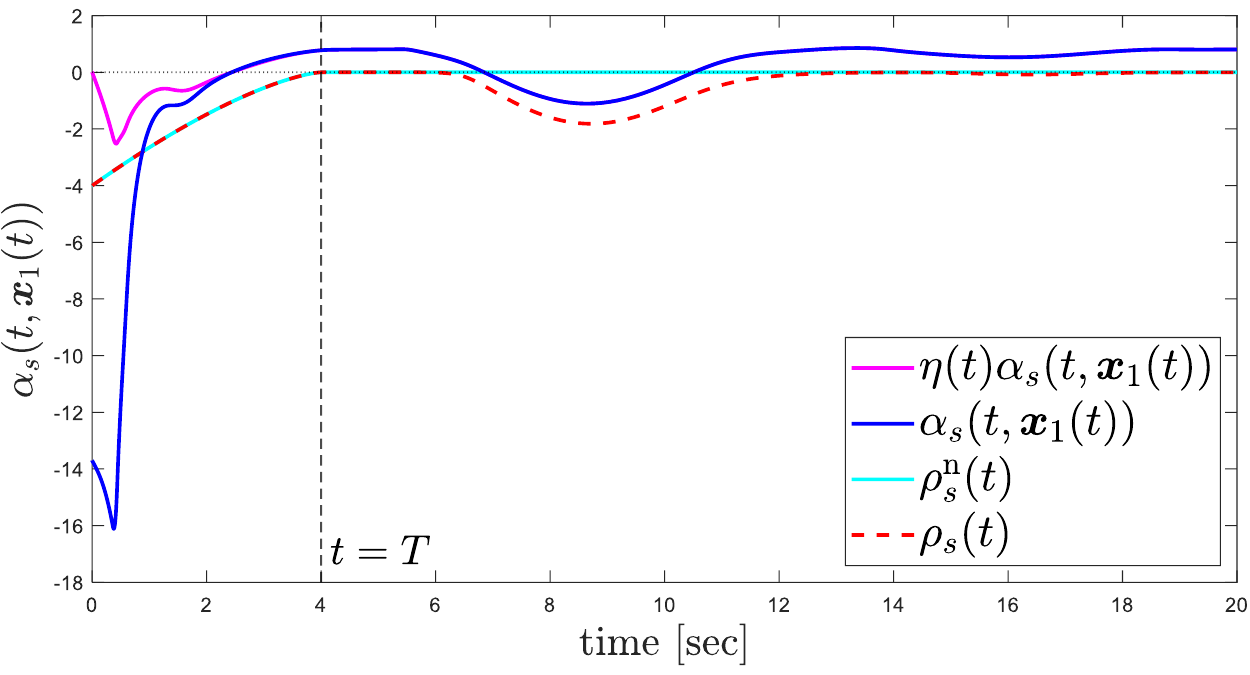}\\[9pt]
		\includegraphics[width=\linewidth]{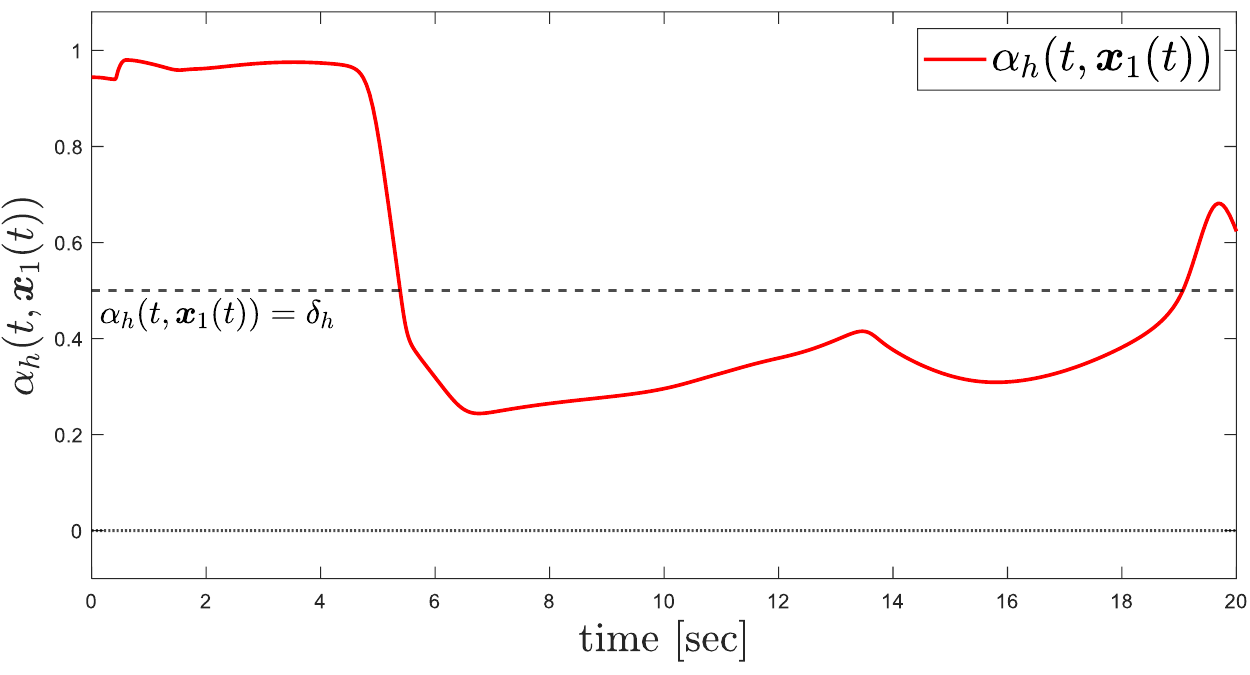}
		\caption{\vspace{-0.2cm}}
		\label{fig:confined_alpha_s_h_ini_free}
	\end{subfigure}	
	\caption{Example 1: initial-condition-free tuning with larger $\delta_\gamma$. \vspace{-0.5cm}}
	\label{fig:ex1_ini_free}
\end{figure}

Fig.~\ref{fig:confined_robot} shows snapshots of the robot's VCP trajectory under the proposed controller. The VCP remains inside the workspace, whose boundary $\alphh=0$ (solid red) encloses the hard-constraint set $\Oh$, while the robot strives to enter and stay within the $1$\,m horizontal radius of the aerial robot, whose boundary $\alphs=0$ (solid blue) encloses the (time-varying) soft-constraint set $\Os(t)$. Fig.~\ref{fig:confined_alpha_s_h} depicts the evolution of the consolidated soft constraint $\alphs(t,\bx_1(t))$ (top) and hard constraint $\alphh(t,\bx_1(t))$ (bottom) functions along the robot's VCP trajectory.

As shown in Fig.~\ref{fig:confined_robot}, when the robot enters the boundary region $\Ohb$ with $0<\alphh<\delta_h$ (green dashed curve), the term $\buh$ in \eqref{eq:1st_intermed_ctrl} switches on and keeps the robot inside the safe region. If the soft set tries to leave the hard set, active deviation of $\rhos(t)$ from its nominal value $\rhosn(t)$ (Fig.~\ref{fig:confined_alpha_s_h}, top) enlarges the virtual soft set $\Osv(t)$ in \eqref{eq:smoth_virtual_soft_set} (dotted blue curve in Fig.~\ref{fig:confined_robot}) just enough to keep it compatible with the hard constraints. Figs.~\ref{fig:confined_robot} and \ref{fig:confined_alpha_s_h} (top) confirm that the soft constraint is met within the user-defined time $T=4$. Fig.~\ref{fig:confined_alpha_s_h} (bottom) shows that the hard constraint is never violated, as $\alphh(t,\bx_1(t))$ stays strictly positive under the proposed controller.

To illustrate the effect of $\delta_\gamma$ in \eqref{eq:rho_relax_dyn} and of the initial-condition-free control law of Section \ref{subsec:global_stab}, we repeated the simulation; see Fig.~\ref{fig:ex1_ini_free}. Parameters match Table \ref{tab:contr_pram_simu1} except for the arbitrary choice of $\rho_0=-4$, $\vartheta_{2,1}^{0}=\vartheta_{2,2}^{0}=1$, and setting the shifting function's settling time to $T_s=T=4$. Moreover, we set $\delta_\gamma = 40$.

Fig.~\ref{fig:confined_robot_ini_free} shows the robot initially outside $\Osv(0)$ (dashed blue circle); the shifting function $\eta(t)$ keeps the modified controller well-posed and restores the desired behavior for $t>T_s=4$. Fig.~\ref{fig:confined_alpha_s_h_ini_free} (top) depicts the evolution of $\eta(t) \,\alphs(t,\bx_1(t))$ which was used in \eqref{eq:es_modif}; it is a scaled version of $\alphs(t,\bx_1(t))$ for $0\le t\le T_s$ and equals it once $\eta(t)=1$. Clearly, the proposed modified control scheme keeps $\eta(t)\alphs(t,\bx_1(t))-\rhos(t)>0$, $\forall t>0$. Here $\rho_0=-4$ is chosen irrespective of $\alphs(0,\bx_1(0)) \approx -14$, and $\Ts=T$ is the largest value Corollary \ref{colo:main} admits; a smaller $\Ts$ only shortens the interval where $\eta(t)\alphs$ differs from $\alphs$.

As noted in Section \ref{subsec:control_design}, a larger $\delta_\gamma$ yields a less conservative expansion of $\Osv(t)$ as the VCP approaches the hard set's boundary; compare the snapshots at $t=8.07$, $12.01$, and $16.10$ in Figs.~\ref{fig:confined_robot} and \ref{fig:confined_robot_ini_free}. Therefore, the soft constraint is violated over a shorter interval, roughly $t\in[7,10.5]$ in Fig.~\ref{fig:confined_alpha_s_h_ini_free} (top) versus $t\in[6.5,11.5]$ in Fig.~\ref{fig:confined_alpha_s_h} (top).

Finally, the funnel-based hard/soft control approaches
in \cite{mehdifar2022funnel,zuo2026robust} cannot handle the
constraints of this example, as their formulations are
restricted to box-shaped time-varying hard and soft sets.

\textbf{Example 2}: We study a scenario similar to Example 1, where a time-varying hard constrained set bounds the workspace and accounts for moving obstacles. We also impose soft constraints requiring the ground robot to stay within a rectangular region beneath the aerial robot. In particular, the soft constraints are defined as $\psis{1}(t,\bx_1)=1 - \bigl( x_{1,1} - x_{a,1}(t) \bigr) >0$, $\psis{2}(t,\bx_1)= \bigl( x_{1,1} - x_{a,1}(t) \bigr) + 1 >0$, $\psis{3}(t,\bx_1)= 1 - \bigl( x_{1,2} - x_{a,2}(t) \bigr) >0$, $\psis{4}(t,\bx_1)= \bigl( x_{1,2} - x_{a,2}(t) \bigr) + 1 >0$. The hard constraints are $\psih{1}(\bx_1) = 7 - x_{1,1} > 0$, $\psih{2}(\bx_1) = x_{1,1} + 7 > 0$, $\psih{3}(\bx_1) = 4 - x_{1,2} > 0$, $\psih{4}(\bx_1) = x_{1,2} + 4 > 0$, $\psih{5}(t,\bx_1) = \tanh\bigl(0.7(\|\bx_1 - \bx_{c_1}(t)\|^{2} - r_1^2 )\bigr) > 0$, $\psih{6}(t,\bx_1) = \tanh\bigl(0.7(\|\bx_1 - \bx_{c_2}(t)\|^{2} - r_2(t)^2)\bigr) > 0$, $\psih{7}(t,\bx_1) = \bigl(\bx_{1} - \bx_{e}(t)\bigr)^\top \bA(t) \bigl(\bx_{1} - \bx_{e}(t)\bigr) - 1 > 0$, where $\psih{1},\ldots,\psih{4}$ define the workspace boundary, $\psih{5}$ and $\psih{6}$ moving circular obstacles, and $\psih{7}$ a moving, rotating elliptical obstacle. In simulation we choose $r_1 = 1.2$ and $r_2(t)=1.2+0.3\sin(0.5t)$ for the obstacle radii, with centers $\bx_{c_1}(t) = [0,-1.7\cos(0.3t)]^\top$, $\bx_{c_2}(t) = [-4,-1.3\sin(0.15t)]^\top$, and $\bx_{e}(t) = [4,1.5\sin(0.3t)]^\top$. The matrix $\bA(t) = \bR(\theta_e(t)) \, \diag{a_e,b_e} \, \bR(\theta_e(t))^\top$, where $\bR(\theta_e(t))\in SO(2)$ is a rotation matrix with a time-varying rotation of $\theta_e(t)=2\pi\cos(0.2t)$, and $a_e=1.6$, $b_e=1$ are the ellipse semi-axes. All controller parameters match Table~\ref{tab:contr_pram_simu1}. Fig.~\ref{fig:ex2} illustrates snapshots of the mobile robot's trajectory along with the evolution of $\alphs(t,\bx_1(t))$ (bottom left) and $\alphh(t,\bx_1(t))$ (bottom right). 

\textbf{Example 3}:
As noted below Theorem \ref{th:main}, the closed-loop system can settle at undesired equilibria on the boundary region of the hard constrained set. Although Theorem \ref{th:main} still guarantees the hard constraints for all time, it does not ensure that $\bx_{1}(t)$ respects the soft constraint, even when the two are compatible. To illustrate this point, we present a simulation in which the ground robot's VCP is subject to the time-invariant hard constraints $\psih{1}(\bx_1) = 7 - x_{1,1} > 0$, $\psih{2}(\bx_1) = x_{1,1} + 7 > 0$, $\psih{3}(\bx_1) = 4 - x_{1,2} > 0$, $\psih{4}(\bx_1) = x_{1,2} + 4 > 0$, and $\psih{5}(\bx_1) = \tanh\bigl(0.7(\|\bx_1 - \bx_{c}\|^{2} - r_1^2)\bigr) > 0$ with $r_1 = 1.2$ and $\bx_{c} = [0,0]^\top$, which model the bounded workspace together with a stationary circular obstacle at the origin. As in Example~1, the soft constraint is defined as $\psi_s(t,\bx_1)=1-\|\bx_1-\bx_{\mathrm{a}}^{\mathrm{p}}(t)\|^{2}>0$.

\begin{figure}[tbp!]
	\centering
	\begin{subfigure}[t]{0.52\linewidth}
		\vspace{0pt}
		\includegraphics[width=\linewidth]{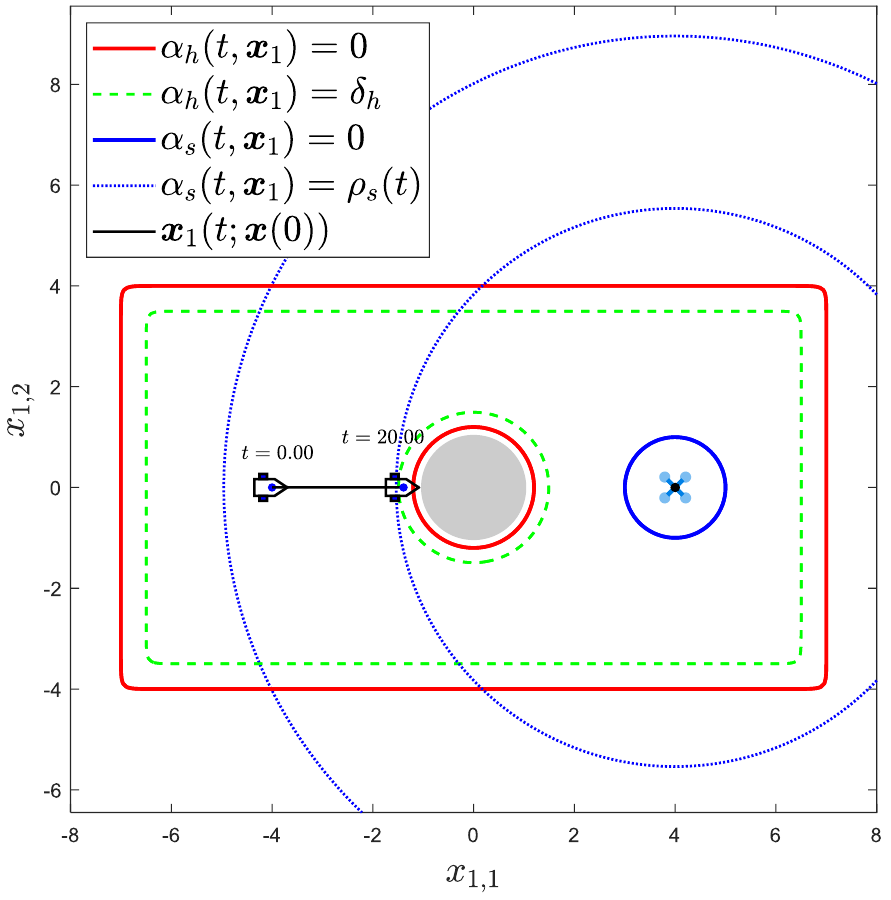}
	\end{subfigure} 
	\begin{subfigure}[t]{0.45\linewidth}
		\vspace{0pt}
		\includegraphics[width=\linewidth]{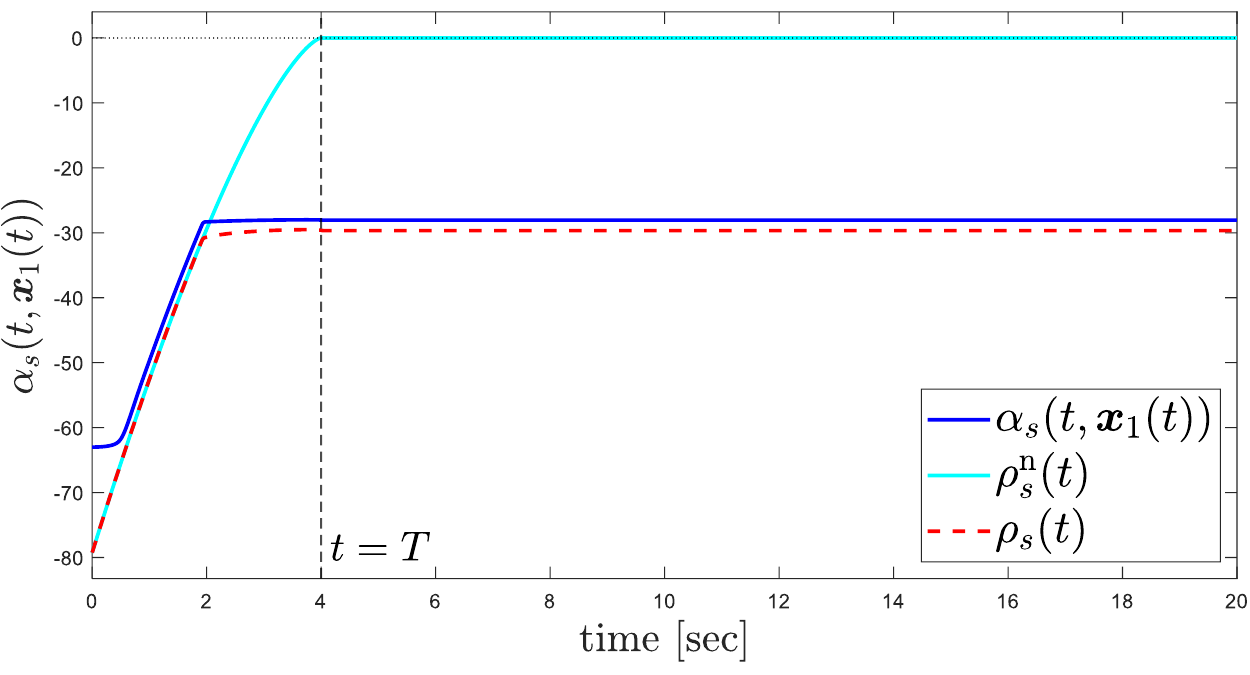}\\[7pt]
		\includegraphics[width=\linewidth]{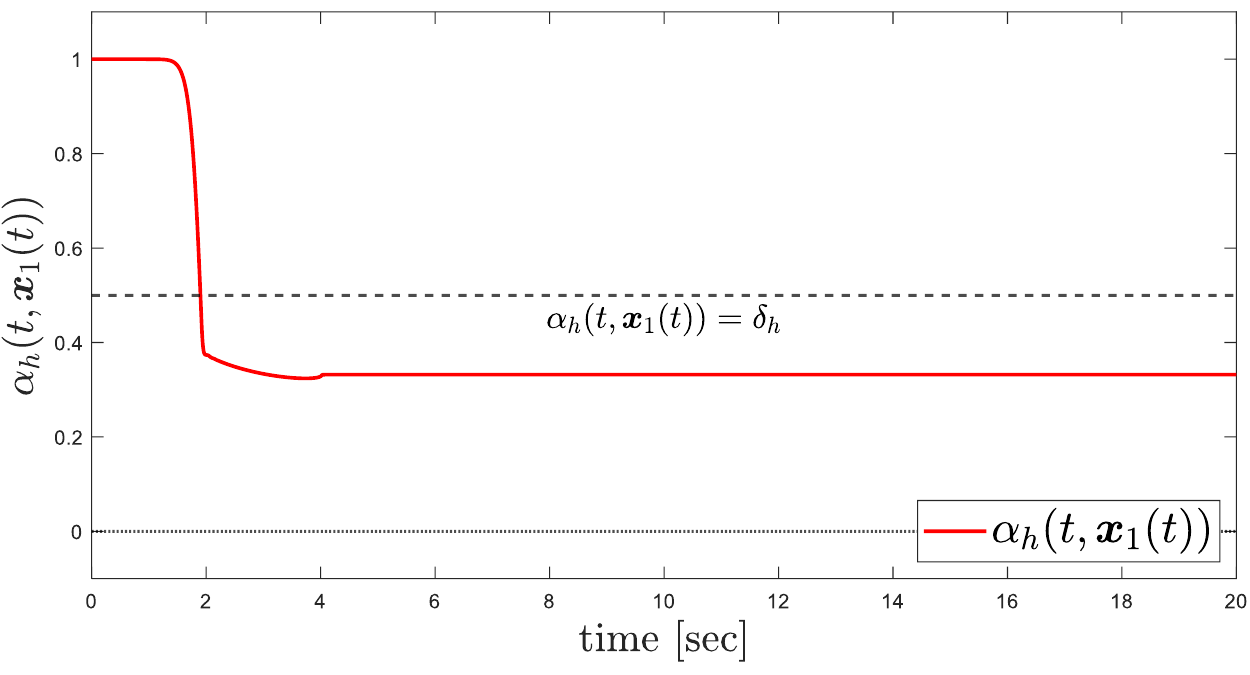}
	\end{subfigure}	
	\caption{Simulation results for Example 3: existence of an undesired equilibrium in the closed-loop system under stationary constraints. \vspace{-0.1cm}}
	\label{fig:ex3}
\end{figure}

Consider now the special case in which the external disturbance vector in~\eqref{eq:robot_kin_dyn} is zero, the aerial robot remains stationary at $\bx_{\mathrm{a}}^{\mathrm{p}} = [4,0]^\top$, and the ground robot's initial position lies on the line through the obstacle center and $\bx_{\mathrm{a}}^{\mathrm{p}}$. The controller parameters are identical to those in Example~2. Fig.~\ref{fig:ex3} shows the resulting trajectories. Despite the compatibility of hard and soft constraints, the ground robot fails to reach its goal and converges to an equilibrium on the boundary of the hard constrained set (evident from the evolution of $\alphh(t,\bx_{1}(t))$). At this equilibrium, the terms $\bus$ and $\varphih \buh$ cancel in~\eqref{eq:1st_intermed_ctrl}. The domain of att\begin{figure}[!tbp]
	\centering
	\begin{subfigure}[t]{0.492\linewidth}
		\includegraphics[width=\linewidth]{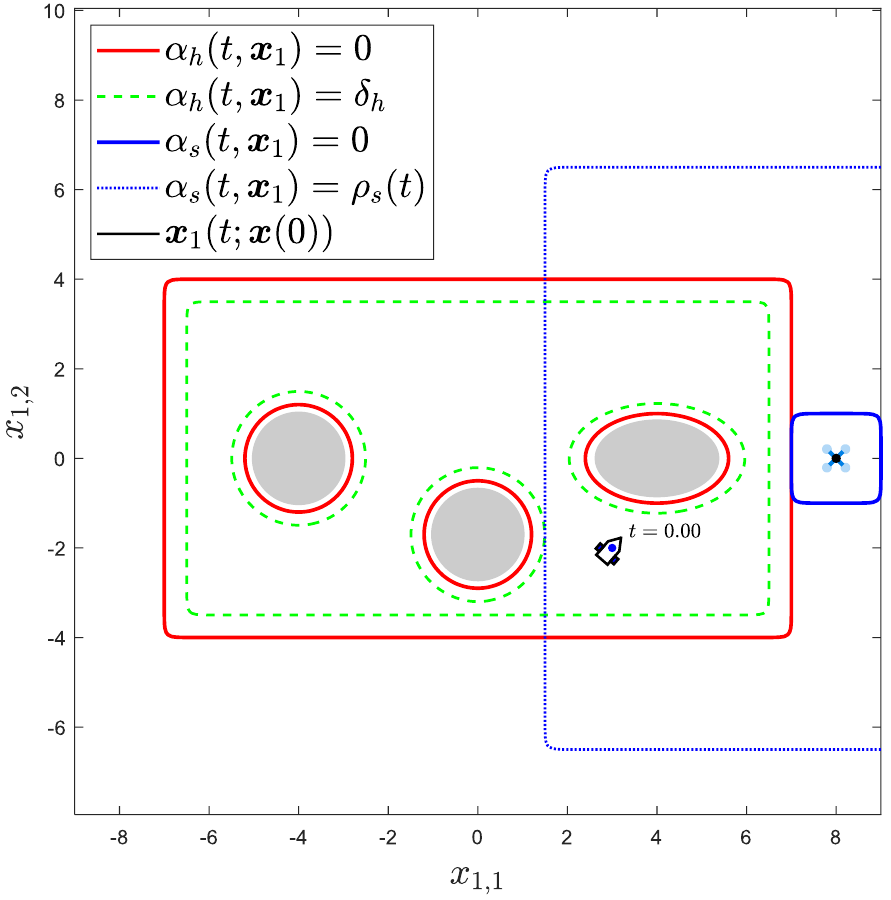}\\[2pt]
		\includegraphics[width=\linewidth]{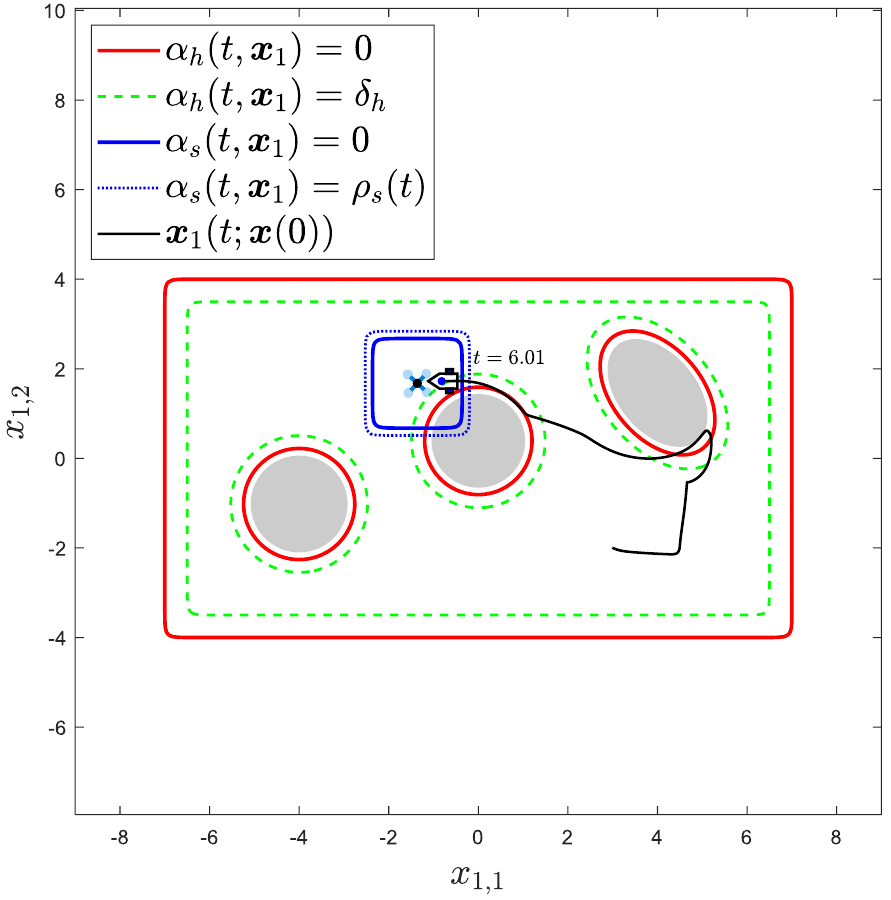}\\[2pt]
		\includegraphics[width=\linewidth]{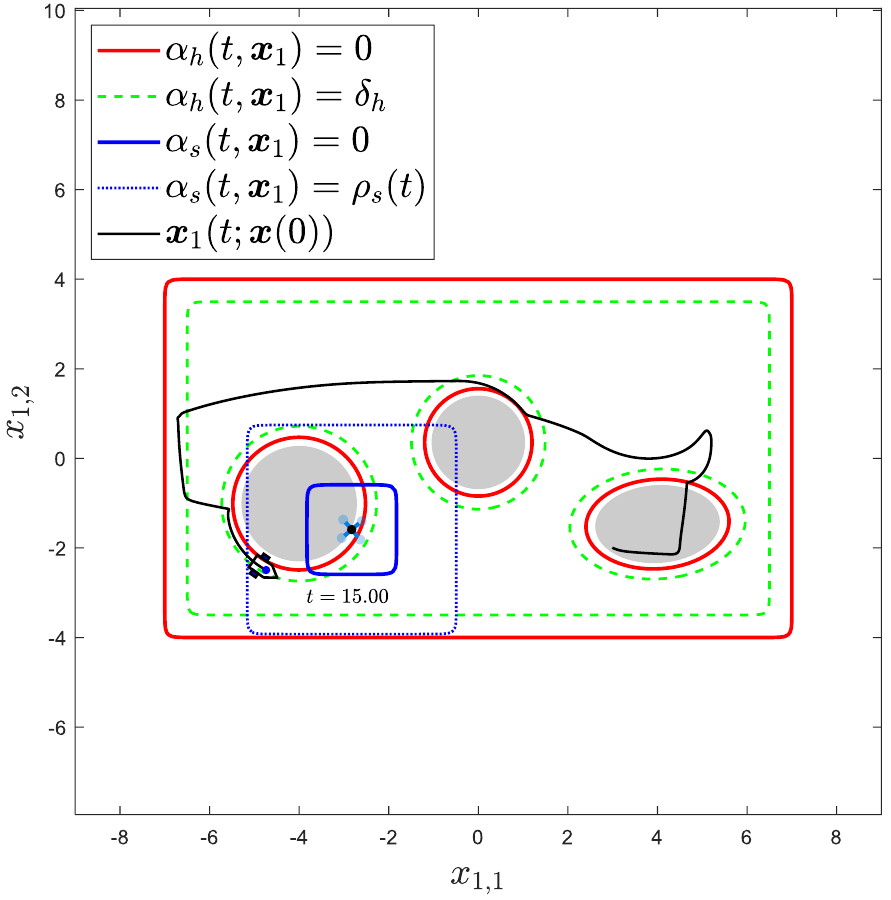}\\[2pt]
		\includegraphics[width=\linewidth]{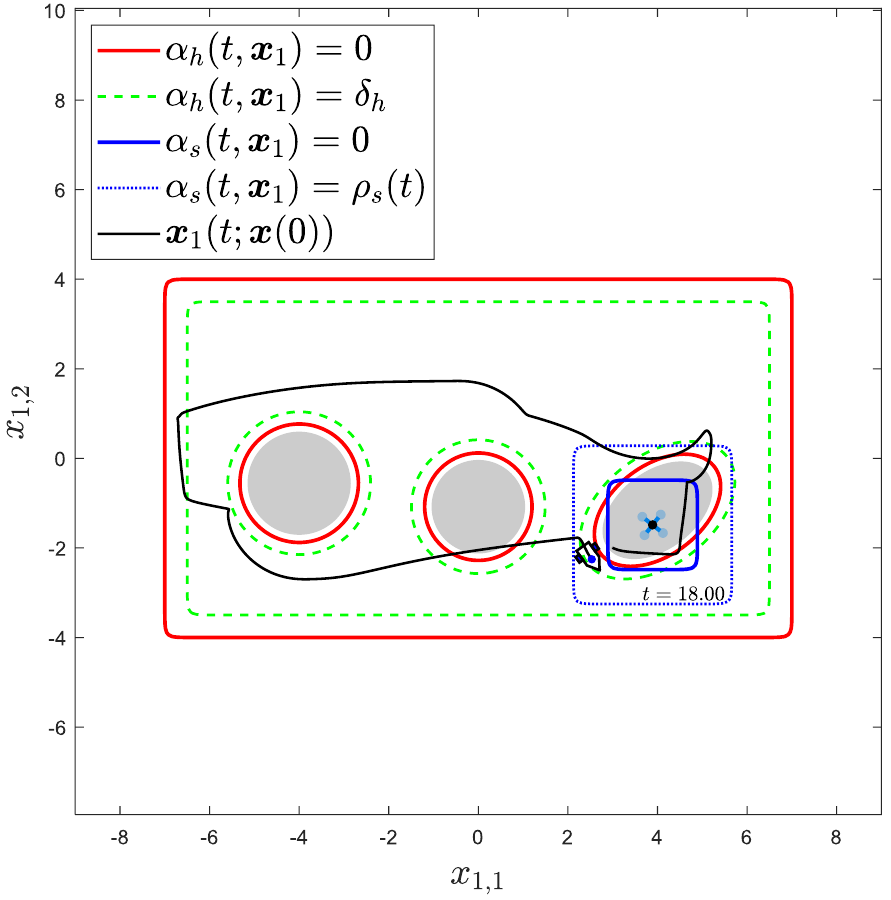}\\[2pt]
		\includegraphics[width=\linewidth]{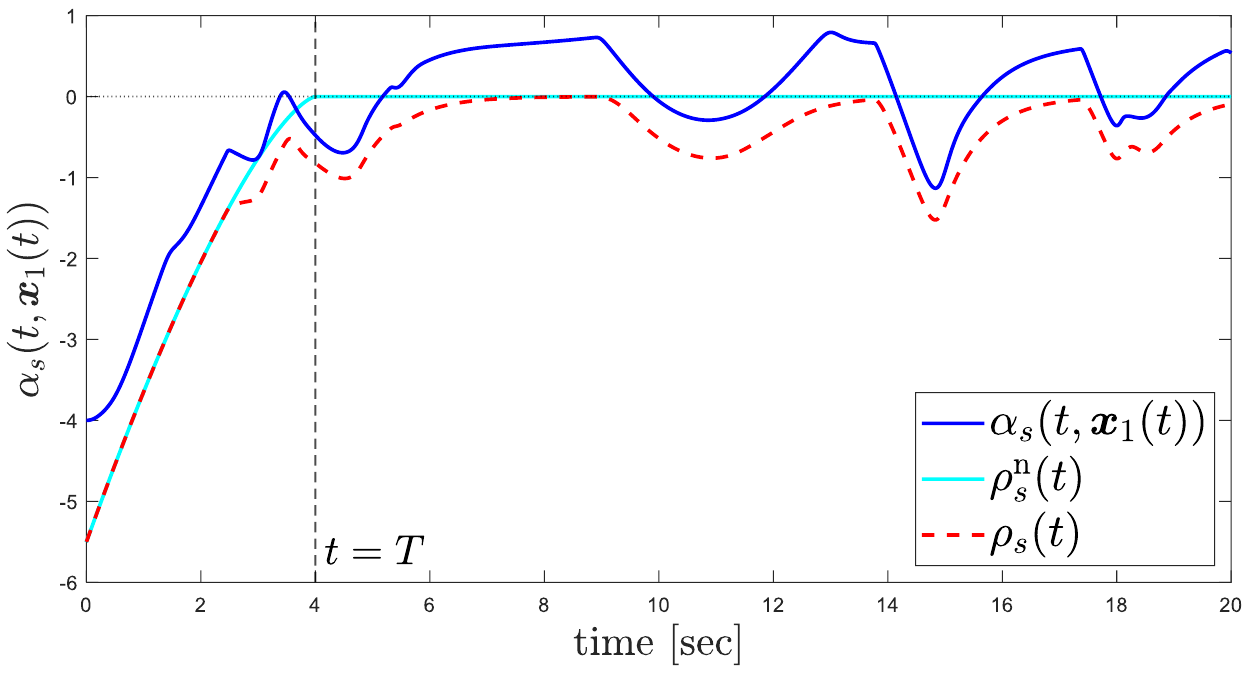}
	\end{subfigure} 
	\begin{subfigure}[t]{0.492\linewidth}
		\includegraphics[width=\linewidth]{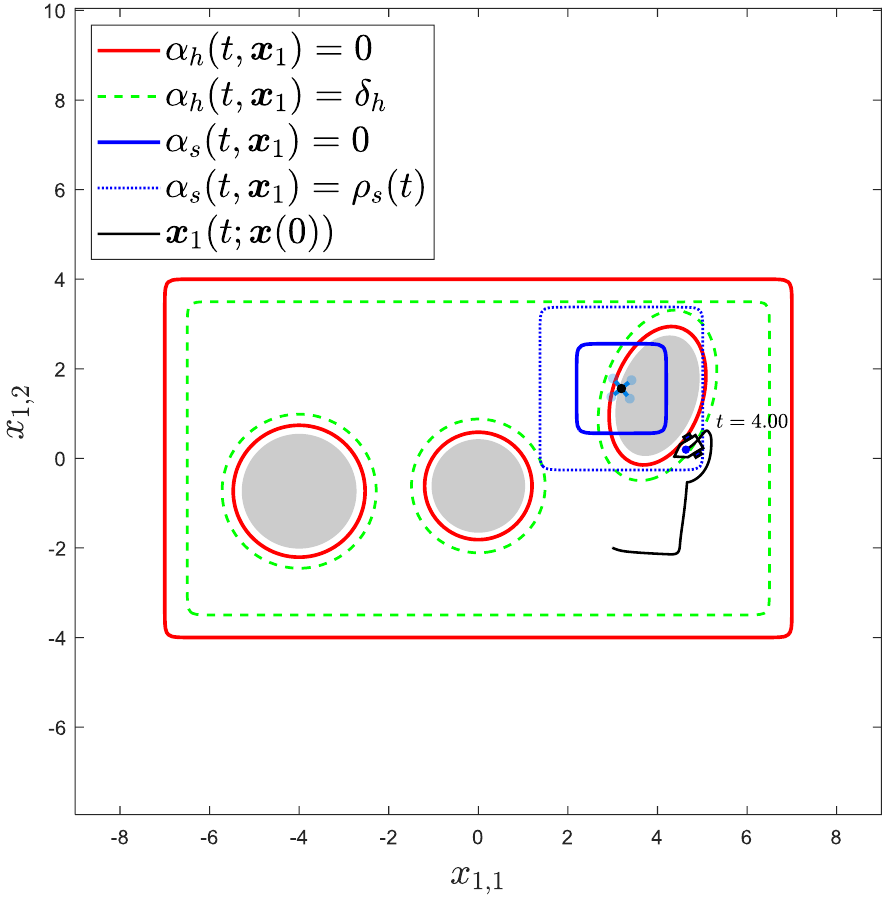}\\[2pt]
		\includegraphics[width=\linewidth]{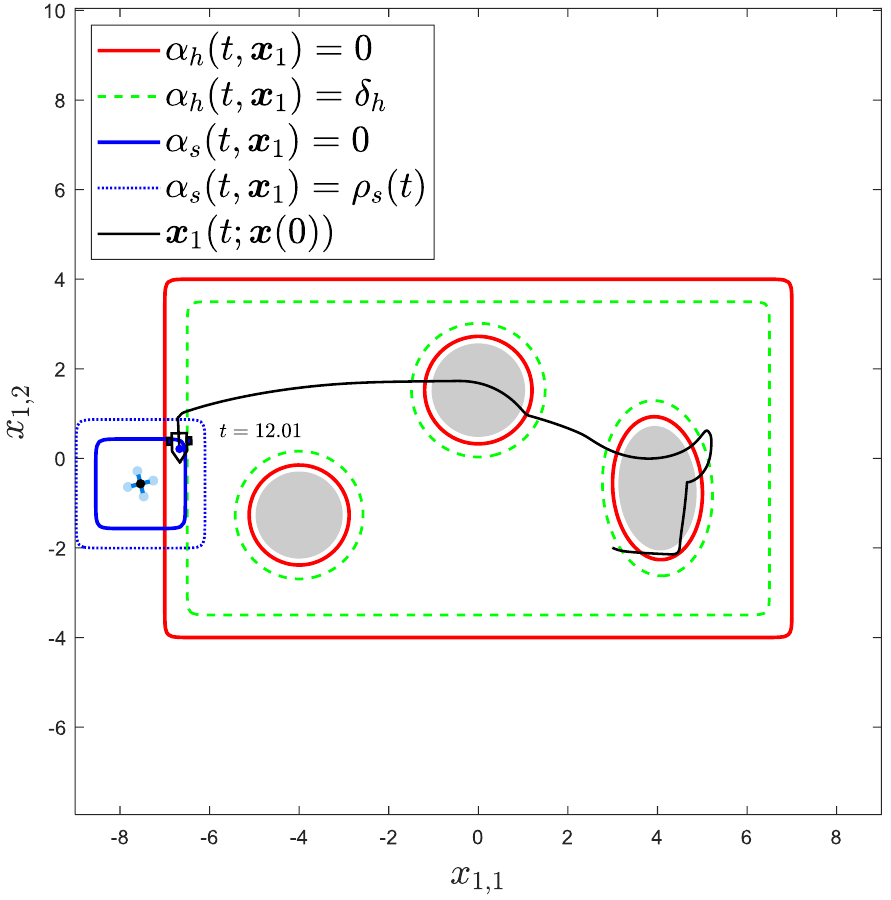}\\[2pt]
		\includegraphics[width=\linewidth]{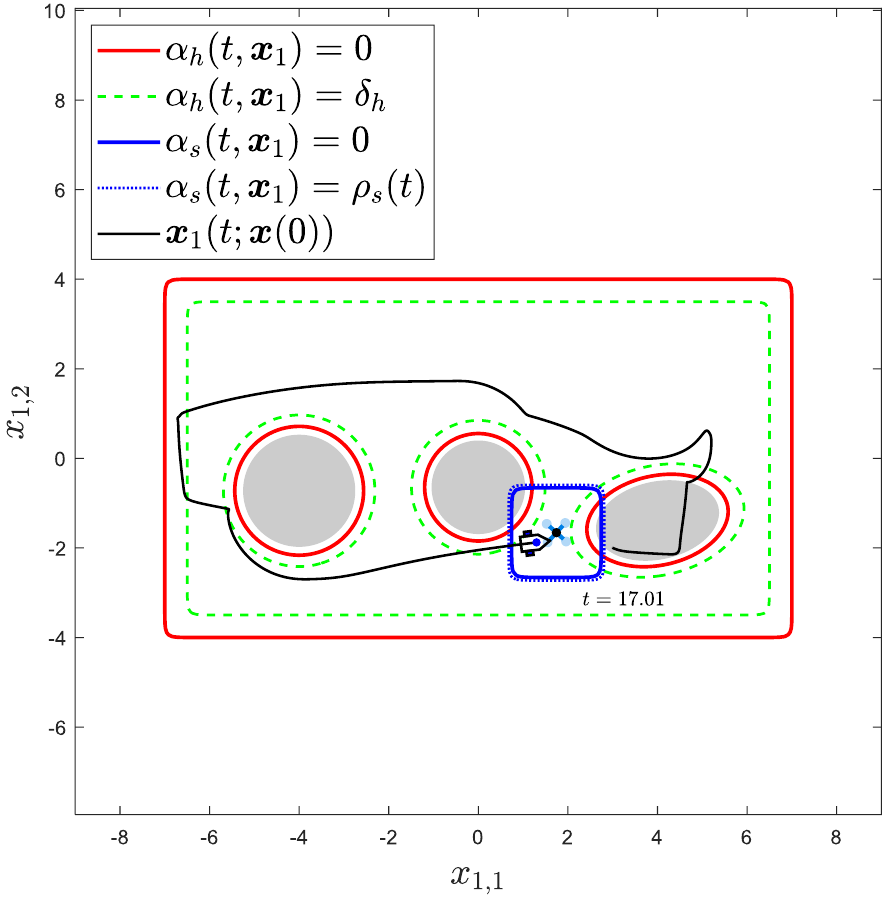}\\[2pt]
		\includegraphics[width=\linewidth]{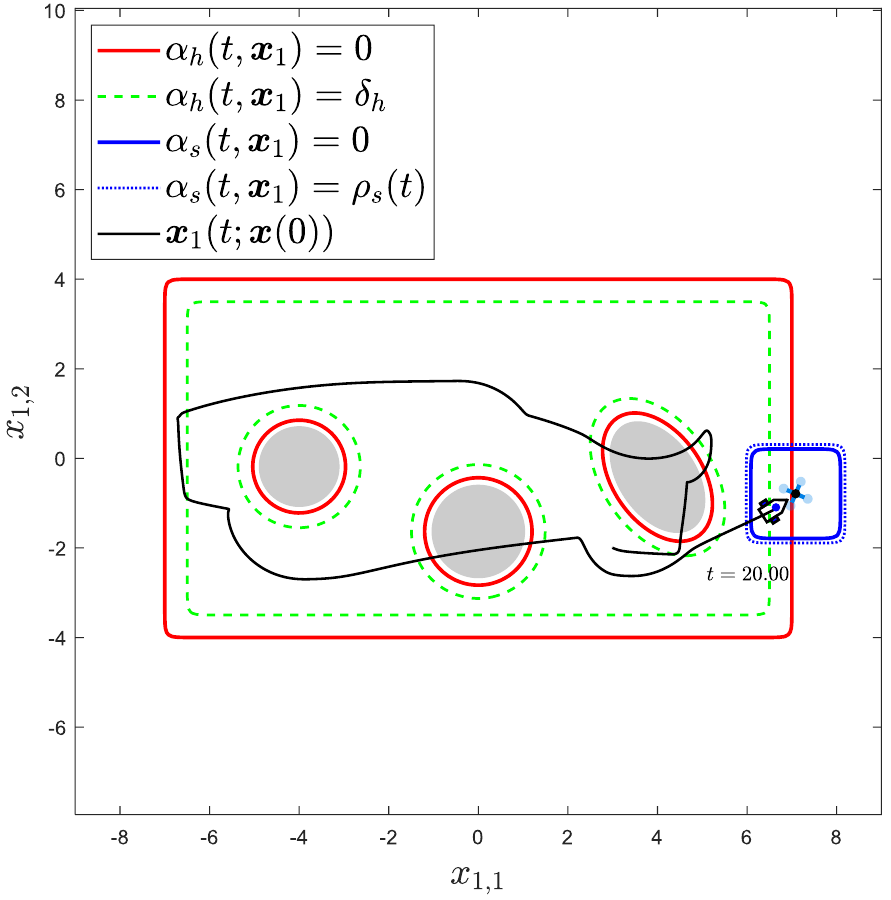}\\[2pt]
		\includegraphics[width=\linewidth]{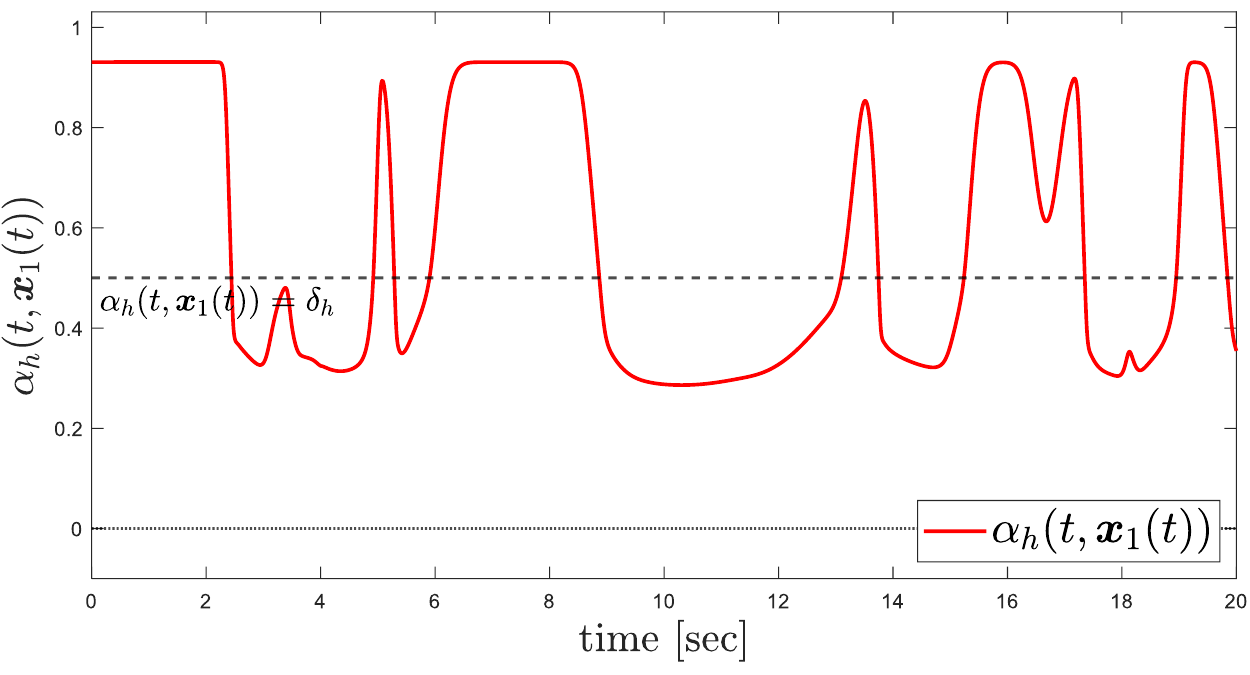}
	\end{subfigure}	
	\caption{Simulation results of Example 2. \vspace{-0.5cm}}
	\label{fig:ex2}
\end{figure}raction of this equilibrium lies to its left, along the line connecting the aerial robot and the obstacle center; no other initial condition drives the robot to it.

\begin{figure}[tbp!]
	\centering
	\begin{subfigure}[t]{0.52\linewidth}
		\vspace{0pt}
		\includegraphics[width=\linewidth]{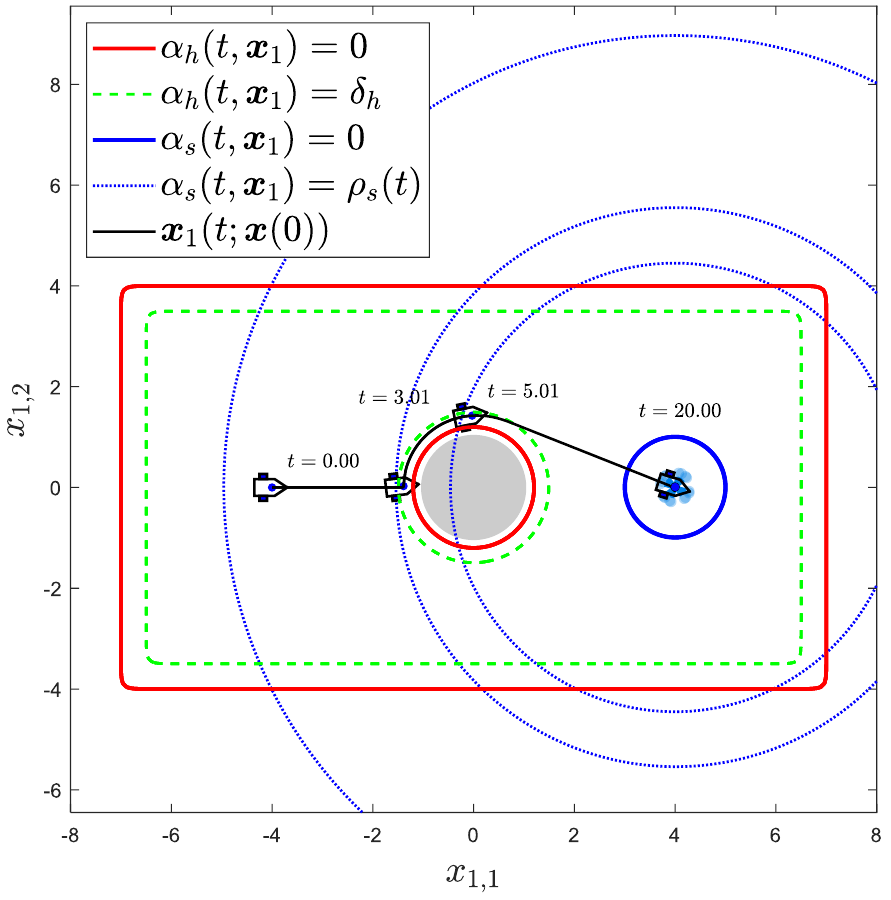}
	\end{subfigure} 
	\begin{subfigure}[t]{0.45\linewidth}
		\vspace{0pt}
		\includegraphics[width=\linewidth]{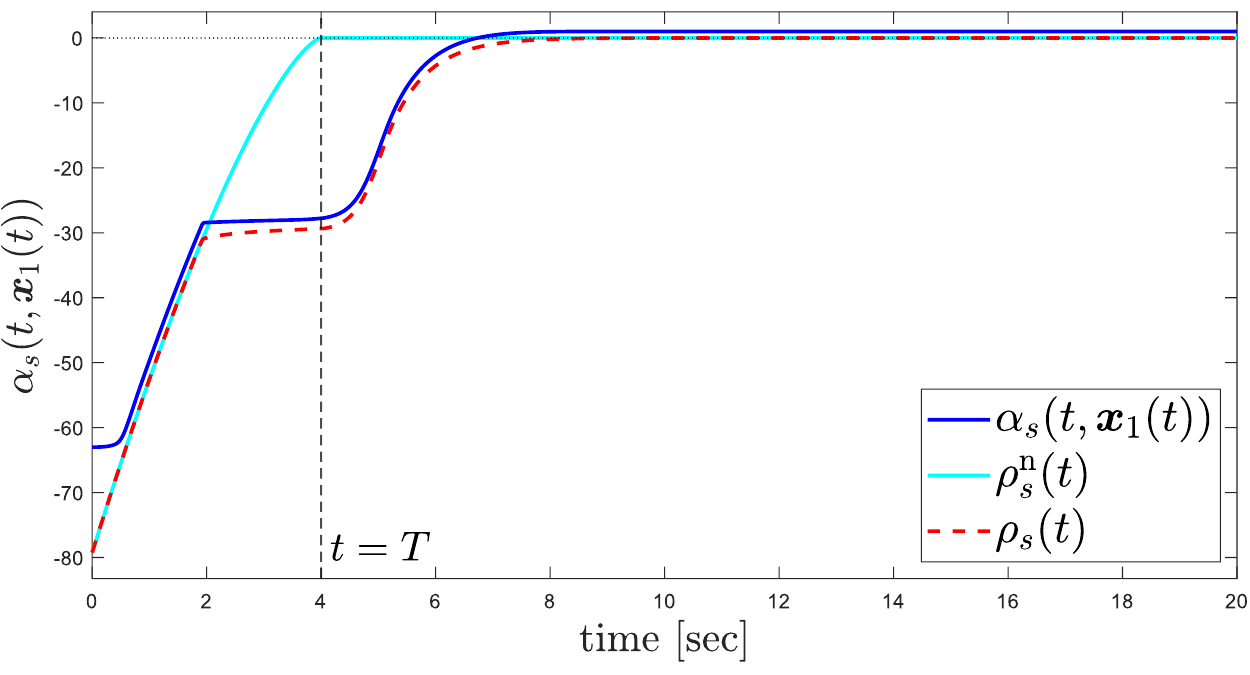}\\[7pt]
		\includegraphics[width=\linewidth]{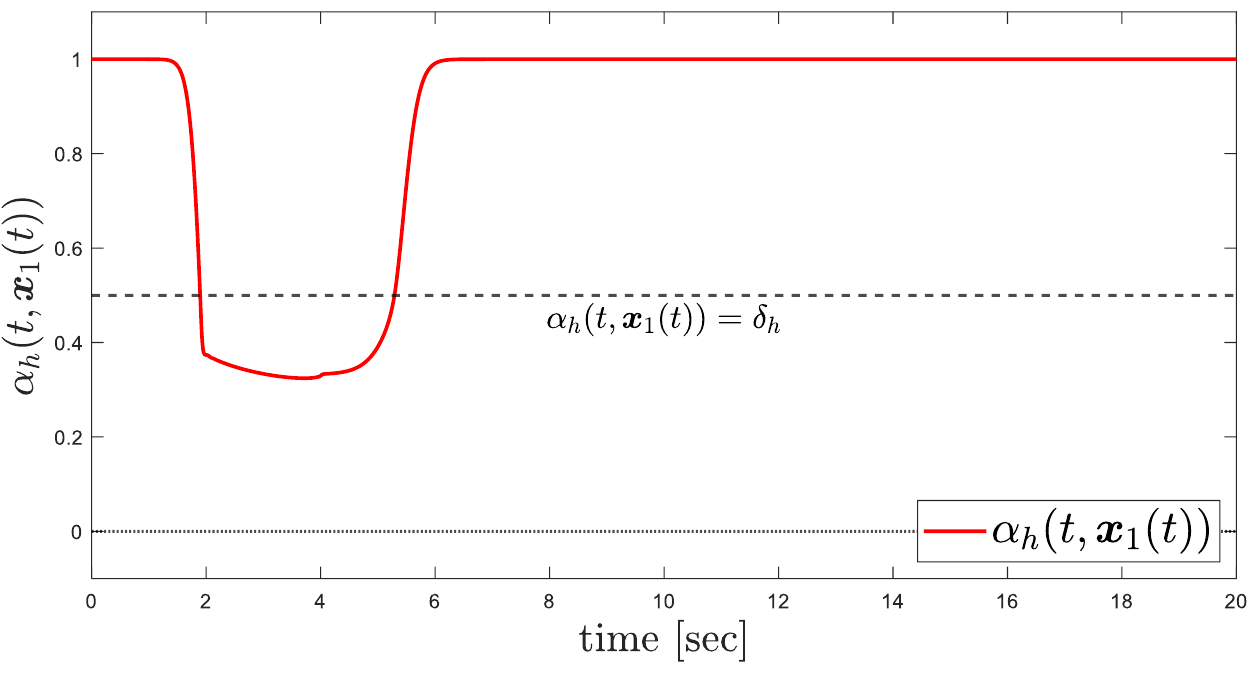}
	\end{subfigure}	
	\caption{Simulation results for Example 3: escaping the undesired equilibrium under a mildly time-varying soft constraint. \vspace{-0.1cm}}
	\label{fig:ex3_2}
\end{figure}

Finally, such undesired equilibria are effectively unstable in practice: even small disturbances in the ground robot's dynamics, or slightly time-varying constraints, make them behave unstably. To demonstrate this, let the aerial robot exhibit small oscillations, $\bx_{\mathrm{a}}^{\mathrm{p}}(t) = [4 + 0.01\sin(t), 0.01\cos(0.3t)]^\top$, which makes the soft constraint time-varying but with negligible magnitude. Repeating the simulation under this condition yields the results in Fig.~\ref{fig:ex3_2}. The ground robot escapes the undesired equilibrium and ultimately completes its task. This example shows that, when undesired equilibria exist, a mild time-varying behavior in the constraints or internal dynamics of the system can mitigate their impact on the closed-loop system behavior.

\begin{figure}[!tbp]
	\centering
	\begin{subfigure}[t]{0.55\linewidth}
		\vspace{0pt}
		\includegraphics[width=\linewidth]{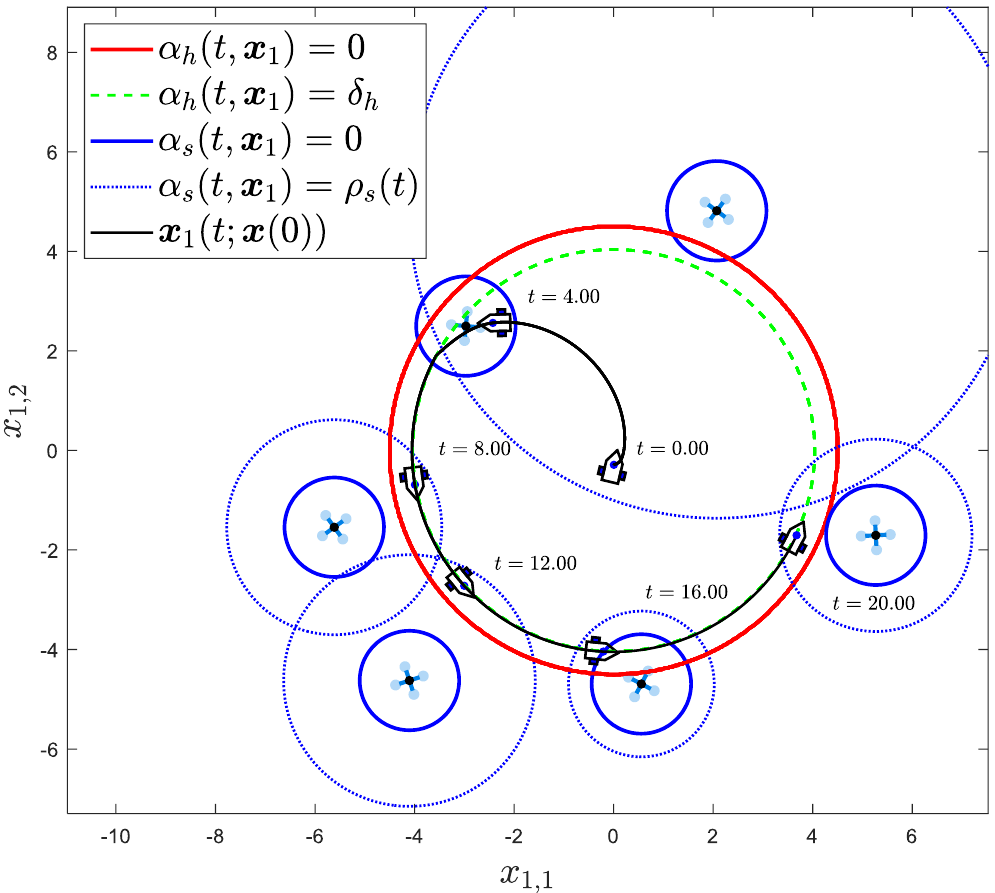}
		\label{fig:confined_robot_ex4}
	\end{subfigure} 
	\begin{subfigure}[t]{0.43\linewidth}
		\vspace{0pt}
		\includegraphics[width=\linewidth]{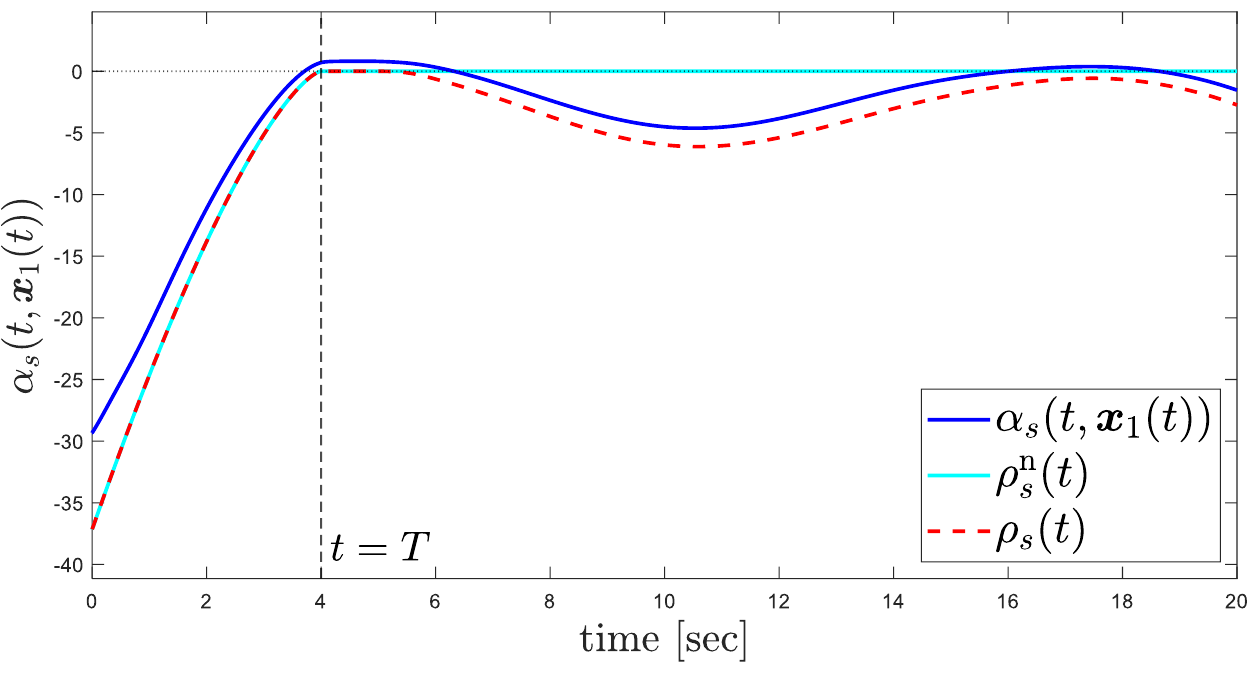}\\[8pt]
		\includegraphics[width=\linewidth]{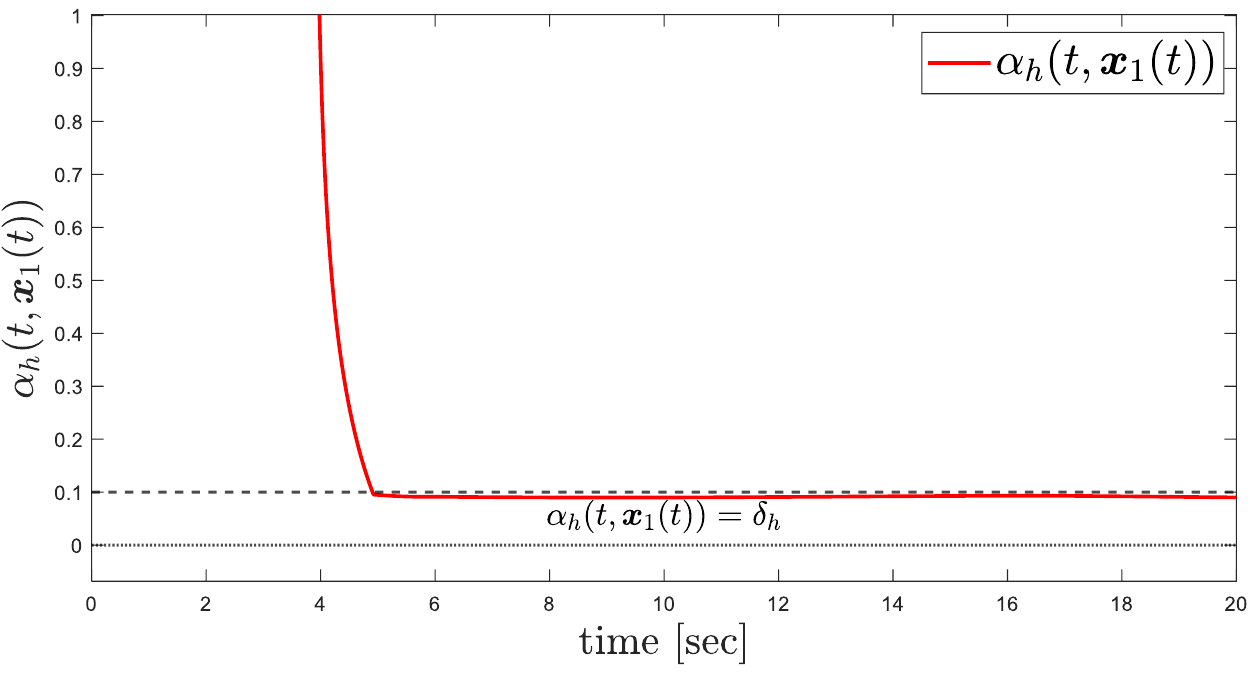}
		\label{fig:confined_alpha_s_h_ex4}
	\end{subfigure}	
	\vspace{-0.5cm}
	\caption{Simulation results of Example 4. \vspace{-0.4cm}}
	\label{fig:ex4}
\end{figure}

\textbf{Example 4}:
Following Remark \ref{rem:sufficient_assum_modified}, to clarify that relaxing Assumption \ref{assu:hard_set_boundary_non_zero_grad} may not influence the proposed control law's effectiveness, we consider a simulation setup akin to Example 1 with the difference that this time the mobile robot is subjected to the following hard constraint $\psi_h(\bx_{1}) = (4.5-\|\bx_{1}\|)^3 > 0$. Notice that $\psi_h(\bx_{1}) > 0$ represents a stationary circular region and we also have $\alphh(\bx_{1}) = \psi_h(\bx_{1})$. It is easy to verify that the gradient of $\alphh(\bx_{1})$ vanishes everywhere on the boundary of the hard constraint where $\alphh(\bx_{1}) = 0$. Indeed, one can verify that the level curve $\alphh(\bx_{1}) = 0$ is constituted by a continuum of (degenerate) saddle points. For this simulation the numerical values of all controller parameters are the same as in Example 1 except for $\delta_h = 0.1$. As Fig.~\ref{fig:ex4} suggests, the proposed control law is still effective even when Assumption \ref{assu:hard_set_boundary_non_zero_grad} does not hold.

\section{Conclusions}
\label{sec:conclusion}
We presented a robust, closed-form control framework for uncertain, high-relative-degree nonlinear MIMO systems subject to time-varying hard (safety) and soft (performance) constraints. All constraints are first condensed into scalar hard- and soft-constraint functions whose time-varying zero super-level sets delineate the regions where the corresponding constraints hold. Reciprocal barrier functions then yield robust explicit velocity-level control laws enforcing both constraint classes. When safety and performance requirements contravene, an innovative dynamic-relaxation scheme temporarily loosens the soft constraints, ensuring that hard constraints are never violated. Finally, tailored nonlinear transformations and a low-complexity backstepping-like design extend the velocity-level law to the actual control input without approximation schemes for the system uncertainties. We also showed that embedding shifting functions decouples controller-parameter tuning from initial conditions, upgrading the guarantees from semi-global to global, deferring only the soft-constraint guarantee to a user-chosen $\Ts \leq T$. In the future, we plan to relax Assumption \ref{assu:hard_set_boundary_non_zero_grad},
extend the results to cooperative control of multi-agent systems, relax the regularity
	conditions on the system dynamics towards weaker continuity classes and
	state-discontinuous (Filippov) dynamics, and characterize and systematically
	resolve the deadlock configurations discussed after Theorem \ref{th:main}.

\appendix
\section{Definition of $\Ccal^1$ switch function}
\label{appen:cont_dif_switch_fun}

The $\Ccal^1$ switching function $\varphi: \mathbb{R}^3 \rightarrow [0,1]$ is defined as
\begin{equation} \label{eq:smooth_switch_def}
	\varphi(\chi,\ub,\lb) \coloneqq
	\begin{cases}
		0 &  \chi > \ub \\
		a_3 \chi^3 + a_2 \chi^2 + a_1 \chi + a_0 &  \lb \leq \chi \leq \ub\\
		1 & \chi < \lb \\
	\end{cases}
\end{equation}
where $\ub > \lb$. The coefficients of the cubic polynomial can be obtained via solving the following linear matrix equation for given $\ub$ and $\lb$; see \cite[Section 7.5.1]{spong2020robot}.
\begin{equation} \label{eq:coef_system_linear_equ}
	\begin{bmatrix}
		1 & \ub & \ub^2 & \ub^3 \\
		0 & 1 & 2\ub & 3\ub^2 \\
		1 & \lb & \lb^2 & \lb^3 \\
		0 & 1 & 2\lb & 3\lb^2
	\end{bmatrix}
	\begin{bmatrix}
		a_0 \\
		a_1 \\
		a_2 \\
		a_3			
	\end{bmatrix}
	=
	\begin{bmatrix}
		0 \\
		0 \\
		1 \\
		0			
	\end{bmatrix}.
\end{equation}  
In general, solving \eqref{eq:coef_system_linear_equ} yields:  
\begin{equation*}
	a_0 = \tfrac{\ub^2 (3\lb - \ub)}{(\lb - \ub)^3}, \; a_1 = \tfrac{-6 \lb \ub}{(\lb - \ub)^3}, \; a_2 = \tfrac{3(\lb + \ub)}{(\lb - \ub)^3}, \; a_3 = \tfrac{-2}{(\lb - \ub)^3}.
\end{equation*}

\section{Gradients and partial time derivatives of $\alphh$ and $\alphs$}
\label{appen:gradients}

The gradients of functions $\alphh$ and $\alphs$ in \eqref{smooth_alph} are
\begin{equation*}
	\begin{aligned}
		\grad{\alpha_{\star}}(t,\bx_1) &= \frac{\sum_{j = 1}^{m_{\star}} \frac{\partial \psi_{\star_j}^\top(t,\bx_1)}{\partial \bx_1} e^{-\nu \psi_{\star_j}(t,\bx_1)}}{\sum_{j = 1}^{m_{\star}} e^{-\nu \psi_{\star_j}(t,\bx_1)}} \\
		& = \left( \sum_{j = 1}^{m_{\star}} \grad{\psi_{\star_j}} \, e^{-\nu \psi_{\star_j}(t,\bx_1)} \right) e^{\nu \alpha_{\star}(t,\bx_1)},
	\end{aligned}
\end{equation*} 
where $\star \in \{h,s\}$. The partial time derivatives of $\alphh$ and $\alphs$ are
\begin{equation*}
	\begin{aligned}
		\gradt{\,\alpha_{\star}(t,\bx_1)} &= \frac{\sum_{j = 1}^{m_{\star}} \frac{\partial \psi_{\star_j}(t,\bx_1)}{\partial t} e^{-\nu \psi_{\star_j}(t,\bx_1)}}{\sum_{j = 1}^{m_{\star}} e^{-\nu \psi_{\star_j}(t,\bx_1)}} \\
		& = \left( \sum_{j = 1}^{m_{\star}} \gradt{\psi_{\star_j}} \, e^{-\nu \psi_{\star_j}(t,\bx_1)} \right) e^{\nu \alpha_{\star}(t,\bx_1)}.
	\end{aligned}
\end{equation*} 

\section{Proof of Theorem \ref{th:main}}
\label{appen:main_theorem_proof}

From \eqref{eq:intermediate_err} we have $\bx_2 = \be_2 + \bs_1(t,\bx_1)$, $\bx_3 = \be_3 + \bs_2(t,\bxb_2)$ and $\bx_i = \be_i + \bs_{i-1}(t,\bxb_{i-1}), i \in \mbI_4^r$. Hence, abusing notation slightly, \eqref{eq:normalized_err_vec} gives
\begin{subequations} \label{eq:x_rewritten}
	\begin{align}
		\bx_2 &= \bvarTheta_2(t) \, \beh_2 + \bs_1(t,\bx_1), \label{eq:x_2_rewritten} \\
		\bx_3 &= \bvarTheta_3(t) \, \beh_3 + \bs_2(t,\bx_1, \beh_2), \label{eq:x_3_rewritten} \\
		\bx_i &= \bvarTheta_i(t) \, \beh_i + \bs_{i-1}(t,\bx_1,\beh_2,\ldots,\beh_{i-1}), \; i \in \mbI_4^r.   
	\end{align}
\end{subequations}

Now from \eqref{eq:sys_dynamics_mimo} and \eqref{eq:x_2_rewritten} we have
\begin{align} \label{eq:x_1_dot}
	\dot{\bx}_1 &\coloneqq \bphi_1(t,\bx_1,\beh_2) \\ 
	&= \bof_1(t,\bx_1) + \bG_1(t,\bx_1) \left(\bvarTheta_2(t) \, \beh_2 + \bs_1(t,\bx_1)\right).  \nonumber 
\end{align}

Taking the time derivative of \eqref{smooth_alpha_hard_def} leads to 
\begin{equation} \label{eq:alpha_h_dot}
	\dalphh \coloneqq \phi_{h}(t,\bx_1,\beh_2) = \frac{\partial \alphh(t,\bx_1)}{\partial \bx_1} \dot{\bx}_1 + \frac{\partial \alphh(t,\bx_1)}{\partial t}. 
\end{equation}

Using \eqref{eq:soft_const_set_lower_bound} and \eqref{eq:rho_relax_dyn}, the time derivative of \eqref{eq:es} is
\begin{align}
	\des &\coloneqq \phi_{s}(t,\bx_1,\beh_2, \rhosr) \label{eq::alpha_s_dot}\\
	& = \frac{\partial \alphs(t,\bx_1)}{\partial \bx_1} \dot{\bx}_1 + \frac{\partial \alphs(t,\bx_1)}{\partial t} - \drhosn(t) + \phi_{\rho}(t, \bx_1, \rhosr). \nonumber
\end{align}

Moreover, differentiating \eqref{eq:normalized_err_vec} with respect to time and substituting equations \eqref{eq:sys_dynamics_mimo}, \eqref{eq:x_rewritten}, and \eqref{eq:control_law} results in: 
\begin{subequations}
	\begin{flalign}
		\dot{\beh}_2  &\coloneqq \bphi_2(t,\bx_1,\beh_2, \beh_3) & \label{eq:dot_ehat_2} \\
		&= \bvarTheta_2^{-1}(t) \Big[ \bof_2(t,\bx_1,\beh_2) + \bG_2(t,\bx_1,\beh_2) \big(\bvarTheta_3(t) \, \beh_3 & \nonumber \\
		& \quad + \bs_2(t,\bx_1, \beh_2)\big) - \dot{\bs}_1(t,\bx_1) - \dot{\bvarTheta}_2(t) \beh_2 \Big], & \nonumber
	\end{flalign} 
	\begin{flalign}
		\dot{\beh}_i  &\coloneqq \bphi_{i}(t,\bx_1,\beh_2,\ldots,\beh_i) & \\ 
		&= \bvarTheta_i^{-1}(t) \Big[ \bof_i(t,\bx_1,\beh_2,\ldots,\beh_i) + \bG_i(t,\bx_1,\beh_2,\ldots,\beh_i) & \nonumber \\ 
		& \quad  \times \big(\bvarTheta_{i+1}(t) \, \beh_{i+1} + \bs_i(t,\bx_1, \beh_2,\ldots,\beh_i) \big) \nonumber \\ 
		& \;- \dot{\bs}_{i-1}(t,\bx_1,\beh_2,\ldots,\beh_{i-1}) - \dot{\bvarTheta}_i(t) \beh_i \Big], \quad i \in \mbI_3^{r-1}, \nonumber
	\end{flalign}
	\begin{flalign}
		\dot{\beh}_r  &\coloneqq \bphi_{r}(t,\bx_1,\beh_2,\ldots,\beh_r) & \\ 
		&= \bvarTheta_r^{-1}(t) \Big[ \bof_r(t,\bx_1,\beh_2,\ldots,\beh_r) + \bG_r(t,\bx_1,\beh_2,\ldots,\beh_r) & \nonumber \\ 
		& \quad \quad \times \bu(t,\bx_1,\beh_2,\ldots,\beh_r) - \dot{\bs}_{r-1}(t,\bx_1,\beh_2,\ldots,\beh_{r-1})  \nonumber \\
		& \quad \quad - \dot{\bvarTheta}_r(t) \beh_r \Big], \nonumber 
	\end{flalign}
\end{subequations}

Now, let us define $\bz \coloneqq [\bx_1^\top, \alphh, \rhosr, \es, \beh_2^\top, \ldots, \beh_r^\top]^\top \in \mbR^{nr+3}$ and consider the dynamical system
\begin{equation} \label{eq:z_dynamics}
	\dot{\bz} = \bphi(t,\bz) \coloneqq
	\begin{bmatrix}
		\bphi_1(t,\bx_1,\beh_2) \\
		\phi_{h}(t,\bx_1,\beh_2) \\
		\phi_{\rho}(t, \bx_1, \rhosr) \\
		\phi_{s}(t,\bx_1,\beh_2, \rhosr) \\
		\bphi_2(t,\bx_1,\beh_2, \beh_3) \\
		\vdots \\
		\bphi_{r}(t,\bx_1,\beh_2,\ldots,\beh_r)
	\end{bmatrix},
\end{equation} 
as well as the (nonempty) open set: $\Oz \coloneqq \mbR^n \times (0,+\infty) \times \mbR \times (0,+\infty) \times \underset{(r-1)\mathrm{\;times}}{\underbrace{(-1,1)^n \times \ldots \times (-1,1)^n}}$.

In what follows, we proceed in three phases. First, we show that a unique, maximal solution \(\bz: [0,\taum) \to \mbR^{nr+3}\) for \eqref{eq:z_dynamics} exists over \(\Oz\) (i.e., \(\bz(t;\bz(0)) \in \Oz\) for all \(t \in [0,\taum)\)). Next, we prove that the control scheme guarantees for all \(t \in [0,\taum)\) that (a) the closed-loop signals in \eqref{eq:z_dynamics} are bounded, and (b) \(\bz(t;\bz(0))\) remains strictly within a compact subset of \(\Oz\), which leads by contradiction to \(\taum = +\infty\) (i.e., forward completeness) in the final phase. Note that the latter implies \(\alphh\), \(\es\), and \(\beh_i,\ i \in \mbI_2^r\) remain in strict subsets of \((0,+\infty)\), \((0,+\infty)\), and \((-1,1)^n\), respectively, thus satisfying \eqref{eq:hard_const_satisfaction_cond}, \eqref{eq:soft_const_satisfaction_cond}, and \eqref{eq:i-th_inter_funnels}.

\textbf{Phase I}:
The set \(\Oz\) below \eqref{eq:z_dynamics} is nonempty, open, and time-invariant. Moreover, for any initial condition \(\bx(0)\) in \eqref{eq:sys_dynamics_mimo}, Assumption \ref{assu:safe_ini} ensures \(\alphh(0,\bx_1(0)) \in (0,+\infty)\), and by the design of \(\rhos(t)\) in \eqref{eq:soft_const_set_lower_bound} we have $\alphs(0,\bx_1(0)) > \rhos(0)$, hence, \(\es(0,\bx_1(0)) \in (0,+\infty)\). As noted in \textit{Step i-a} of Subsection \ref{subsec:control_design}, \(\vartheta_{i,j}^0\) in \eqref{exponential_performance_fun} is chosen so that \(\vartheta_{i,j}^0 > |e_{i,j}(0,\bxb_{i}(0))|\), ensuring \(\eh_{i,j}(0,\bxb_{i}(0)) \in (-1,1)\) for all \(j \in \mbI_1^n\) and \(i \in \mbI_2^r\). Hence, for all \(i \in \mbI_2^r\) we have \(\beh_i(0,\bxb_i(0)) \in (-1,1)^n\). These collectively imply that \(\bz(0) \in \Oz\). Furthermore, for each \(i \in \mbI_1^r\) the system nonlinearities \(\bof_i(t,\bxb_i)\) and \(\bG_i(t,\bxb_i)\) are locally Lipschitz in \(\bxb_i\) and piece-wise continuous in \(t\), while the intermediate control laws \(\bs_i(t,\bxb_i)\) and \(\bu(t,\bx)\) are continuously differentiable; for \(\bs_1\) this follows from Assumption \ref{assu:known_unknown} and the fact that matrix inversion is smooth on the set of nonsingular matrices. Consequently, \(\bphi(t,\bz)\) in \eqref{eq:z_dynamics} is locally Lipschitz in \(\bz\) over \(\Oz\) and piece-wise continuous in \(t\). Thus, by \cite[Theorem 54]{sontag1998mathematical}, a unique maximal solution \(\bz(t;\bz(0)) \in \Oz\) exists for \(t \in [0,\taum)\).

Because \(\bz(t;\bz(0))\in\Oz\) on \([0,\taum)\), we have \(\alphh, \es \in(0,+\infty)\) and \(\beh_i \in(-1,1)^n\), \(i\in\mbI_2^r\), for all \(t\in[0,\taum)\). Consequently, \(\epih\) in \eqref{eq:mapped_alphh}, \(\epis\) in \eqref{eq:mapped_alphs}, and every \(\epi_{i,j}\) in \eqref{eq:mapping_fun} (thus \(\bepi_i\in\mbR^n,\;i\in\mbI_2^r\)) are well-defined on \([0,\taum)\).

By Assumptions \ref{assu:coercive_alpha_funs}, \ref{assu:feasible_constrained_sets}, \(\alphh(t,\bx_1)\) has compact level sets with finite positive maxima at each \(t\) within $\alphh(t,\bx_{1}) > 0$. Let \(\alphhstr(t)>0\) denote its finite global maximum and set \(b_h\coloneqq\max_{ t\ge0}\alphhstr(t)\). Then we can infer that \(\alphh(t,\bx_1(t;\bz(0)))\in(0,b_h]\) for all \(t\in[0,\taum)\), and the compactness of the corresponding level sets ensures that \(\bx_1(t)\coloneqq\bx_1(t;\bz(0))\in\Linf\) for all \(t\in[0,\taum)\). Similarly, using Assumptions \ref{assu:coercive_alpha_funs}, \ref{assu:feasible_constrained_sets}, together with \(\alphsstr(t)>0\), and defining \(b_s\coloneqq\max_{ t\ge0}\alphsstr(t)\), leads us to \(\alphs(t,\bx_1(t;\bz(0))) \le b_s\) for all \(t\in[0,\taum)\). Note that, unlike \(\alphh\), positivity of \(\alphs\) is not claimed here, since the soft constraints may be temporarily relaxed; nevertheless, \(\bx_1(t)\in\Linf\) together with Assumption \ref{assu:bounded_in_t_psi_fun} implies \(\alphs(t,\bx_1(t))\in\Linf\), \(\forall t\in[0,\taum)\).

In what follows, function arguments are omitted when unambiguous. \textit{Phase II} follows the controller design stages of Section \ref{subsec:control_design}.

\textbf{Phase II (Step 1)}:
Differentiating \eqref{eq:mapped_alphh}, using \eqref{eq:alpha_h_dot}, \eqref{eq:x_1_dot}, and substituting \eqref{eq:1st_intermed_ctrl}, \eqref{eq:uh}, \eqref{eq:us} yields
\begin{flalign}\label{eq:epih_dot}
	\depih &=  -\epih^2 \Big[\grad{\alphh}^\top \! \bof_1 + \grad{\alphh}^\top \bG_1 \left(\bvarTheta_2 \, \beh_2 + \bs_1 \right) \!+\! \gradt{\,\alphh}\Big]& \nonumber \\
	&= -\epih^2 \Big[ k_s \epis^2 \grad{\alphh}^\top \grad{\alphs} \!+\! \varphih k_h \epih^2 \|\grad{\alphh}\|^2 + \sigma_h \Big],&
\end{flalign}
where $\sigma_h \coloneqq  \grad{\alphh}^\top \bof_1  + \grad{\alphh}^\top \bG_1 \bvarTheta_2 \, \beh_2  + \gradt{\,\alphh}$. Similarly, differentiating \eqref{eq:mapped_alphs}, using \eqref{eq::alpha_s_dot}, \eqref{eq:x_1_dot}, and substituting \eqref{eq:1st_intermed_ctrl}, \eqref{eq:uh}, \eqref{eq:us}, \eqref{eq:soft_const_set_lower_bound} and \eqref{eq:rho_relax_dyn}  gives  
\begin{flalign}\label{eq:epis_dot}
	\depis &=  -\epis^2 \Big[\grad{\alphs}^\top \! \bof_1 \!+\! \grad{\alphs}^\top \! \bG_1 \! \left(\bvarTheta_2 \, \beh_2 \!+\! \bs_1 \right) \!+\! \gradt{\,\alphs} \!+\! \drhos\Big]& \nonumber \\
	&=-\epis^2 \Big[ k_s\epis^{2}\|\grad{\alphs}\|^{2}
		+(1-\varphigam)\,\varphih k_h\epih^{2}\,\grad{\alphs}^\top\grad{\alphh}  & \nonumber \\ 
	& \quad \quad \qquad + \sigma_s \Big], &
\end{flalign}
where $\sigma_s \!\coloneqq \! \grad{\alphs}^\top \! \bof_1  + \grad{\alphs}^\top \! \bG_1 \bvarTheta_2 \, \beh_2  + \gradt{\alphs} - \drhosn -\krec \rhosr$.

From \textit{Phase I}, \(\bx_1(t) \in \Linf \), \(\forall t \in [0,\taum)\). Therefore, by continuity of \(\bof_1\) and \(\bG_1\) in \(\bx_1\) and Assumption \ref{assu:bounded_in_t_dyn}, both remain bounded \(\forall t \in [0,\taum)\). Moreover, from the expressions of \(\grad{\alphs}(t,\bx_1)\), \(\grad{\alphh}(t,\bx_1)\), \(\gradt{\,\alphs(t,\bx_1)}\), and \(\gradt{\,\alphh(t,\bx_1)}\) in \ref{appen:gradients}, their continuity in \(\bx_1\), and Assumptions \ref{assu:bounded_in_t_psi_fun} and \ref{assu:bounded_in_t_psi_derivatives}, we conclude that $\grad{\alphs}(t,\bx_1), \grad{\alphh}(t,\bx_1)$, $\gradt{\,\alphs}(t,\bx_1)$, $\gradt{\,\alphh(t,\bx_1)}\in\Linf$, \(\forall t \in [0,\taum)\). Additionally, by construction we have \(\bvarTheta_2(t), \drhosn(t) \in \Linf\), \(\forall t \geq 0\), and \(\beh_2 \in \Linf\), \(\forall t \in [0,\taum)\) by \textit{Phase I}. Collectively, these facts show that $\sigma_h \in \Linf$ in \eqref{eq:epih_dot} for all \(t \in [0,\taum)\). Similarly, one can verify that \(\sigma_s \in \Linf\) in \eqref{eq:epis_dot} for all \(t \in [0,\taum)\) if 
$\rhosr(t) \in \Linf$.     

Recall that \eqref{eq:hard_const_satisfaction_cond} and \eqref{eq:soft_const_satisfaction_cond} hold iff \(\epih\) and \(\epis\) remain bounded. We now show that \(\epih, \epis \in \Linf\) for all \(t \in [0,\taum)\), which in turn guarantees that \(\bs_1(t,\bx)\) in \eqref{eq:1st_intermed_ctrl} and its derivative remain bounded on \([0,\taum)\). To do so, we establish boundedness of \(\epih\) and \(\epis\) for each mode of the smooth switch functions \(\varphih\) and \(\varphigam\):
\begin{itemize}
	\item[(A)] When \(\varphih(t) = 0\), i.e., when \(\bx_1(t)\) evolves in the interior region \(\Ohi(t)\) defined in \eqref{eq:interior_region_hard}
	\item[(B)] When \(\varphih(t) > 0\), i.e., when \(\bx_1(t)\) evolves in the boundary region \(\Ohb(t)\) defined in \eqref{eq:boundary_region_hard}, with:
	\begin{itemize}
		\item[(B1)] \(\varphigam(t) = 0\) (i.e., the virtual relaxation of soft constraints is inactive).
		\item[(B2)] \(\varphigam(t) > 0\) (i.e., the virtual relaxation of soft constraints is active).
	\end{itemize}
\end{itemize}

Let $\Ical_{A},\Ical_{B1},\Ical_{B2}\subseteq[0,\taum)$ be the intervals on which modes \textit{A, B1}, and \textit{B2} are active. Note that $\Ical_{A}\cup\Ical_{B1}\cup\Ical_{B2}=[0,\taum)$ and that each interval may consist of several disjoint sub‑intervals.

\underline{\textit{Mode A ($\varphih(t) = 0$)}}:
Based on Assumption \ref{assu:feasible_constrained_sets}, for a sufficiently small \(\delta_h > 0\), \(\Ohi(t)\) in \eqref{eq:interior_region_hard} is non-empty. When \textit{Mode A} is active we have $t \in \Ical_{A}$. In this scenario, since \(\alphh(t,\bx_1(t)) \geq \delta_h\) is strictly above zero $\forall t \in \Ical_A$, from \eqref{eq:mapped_alphh} we have \(\epih \in \Linf\),  $\forall t \in \Ical_A$. Consequently, there exists a \(\epibh^A>0\) such that \(0 < \epih < \epibh^A\), $\forall t \in \Ical_A$. It remains to show that \(\epis \in \Linf\) whenever \(\varphih(t) = 0\). To this end, we first show that \(\sigma_s\) in \eqref{eq:epis_dot} is bounded $\forall t \in \Ical_A$ and then use proof-by-contradiction to establish \(\epis \in \Linf\), $\forall t \in \Ical_A$.

Note that since \(\varphih(t) = 0\), from \eqref{eq:rho_relax_dyn} we have \(\drhosr = -\krec \rhosr\), which implies that \(\rhosr \in \Linf\), $\forall t \in \Ical_A$. Hence, following the  arguments below \eqref{eq:epis_dot}, \(\sigma_s \in \Linf\), $\forall t \in \Ical_A$. 

From \textit{Phase I}, we have \(\es(t) \in (0,+\infty)\) for all \(t \in [0,\taum)\); thus, from \eqref{eq:mapped_alphs}, \(\epis\) can only diverge upward, i.e., \(\epis \rightarrow +\infty\). Suppose \(\epis(t) \rightarrow +\infty\) as \(t \rightarrow t^\prime\) for some \(t^\prime \in \Ical_A\) with \(\varphih(t^\prime) = 0\). Since \(\varphih = 0\), \eqref{eq:epis_dot} becomes \(\depis = -\epis^2 \big( k_s \epis^2 \|\grad{\alphs}\|^2 + \sigma_s \big)\). Given that \(\|\grad{\alphs}\|^2 > 0\) whenever \(\grad{\alphs} \neq 0\) and \(\sigma_s\) is bounded, if \(\grad{\alphs} \neq 0\) then \(\depis \rightarrow -\infty\) as \(t \rightarrow t^\prime\), contradicting \(\epis(t) \rightarrow +\infty\). Moreover, by Assumption \ref{assu:alpha_globalmax}, \(\grad{\alphs} = 0\) occurs only when \(\bx_1(t)\) lies inside \(\Os(t)\), i.e., when \(\alphs(t, \bx_1(t)) > 0\), for which $\es(t) > 0$ holds by construction of $\rhos(t) \leq 0$ in \eqref{eq:es}. Thus, \(\grad{\alphs} = 0\) guarantees that \(\epis\) is bounded, again contradicting the occurrence of \(\epis(t) \rightarrow +\infty\). Overall, \(\epis\) remains bounded for all \(t \in \Ical_A \). Therefore, there exists a \(\epibs^A > 0\) such that \(0< \epis \leq \epibs^A\) for all \(t \in \Ical_A\) whenever \(\varphih(t) = 0\).

\underline{\textit{Mode B1 ($\varphih(t) > 0$ \& $\varphigam(t) = 0$)}}:
Here $t\in\Ical_{B1}$. By definition of the smooth switch function, $\varphigam=0$ iff $\gamma = \epih\,\grad{\alphs}^\top\grad{\alphh}\ge0$. Since $\epih>0$ on $[0,\taum)$ (\textit{Phase I}), $\grad{\alphs}^\top\grad{\alphh} \ge0$ whenever \textit{Mode B1} is active. Unlike \textit{Mode A}, here $\alphh(t,\bx_1(t))<\delta_h$, so $\epih$ in \eqref{eq:mapped_alphh} could grow unbounded, violating the hard constraints. We prove by contradiction that $\epih,\epis\in\Linf$ on $\Ical_{B1}$.

Since $\epih>0$ on $[0,\taum)$, $\epih$ can only grow unbounded from above. Suppose $\epih\to+\infty$ as $t\to t'$ for some $t'\in\Ical_{B1}$. Since $\sigma_h\in\Linf$ and $\grad{\alphs}^\top\grad{\alphh}\ge0$, if $\grad{\alphh}\neq0$ the bracketed term on the RHS of \eqref{eq:epih_dot} stays positive, so $\depih\to-\infty$, contradicting $\epih\to+\infty$. Furthermore, by Assumption \ref{assu:hard_set_boundary_non_zero_grad}, $\grad{\alphh}=0$ only when $\alphh(t,\bx_{1}(t))>0$, so \eqref{eq:mapped_alphh} gives $\epih \in \Linf$ whenever $\grad{\alphh}=0$, again contradicting $\epih\to+\infty$.

Overall, we have $\epih\in\Linf$ on $\Ical_{B1}$, so there exits an $\epibh^{B1}>0$ with $0<\epih<\epibh^{B1}$ for all $t\in\Ical_{B1}$.

Since \(\varphigam(t)=0\), \eqref{eq:rho_relax_dyn} yields \(\drhosr = -\krec \rhosr\) so \(\rhosr\in\Linf\) on \(\Ical_{B1}\). By the arguments below \eqref{eq:epis_dot}, this also gives \(\sigma_s\in\Linf\), \(\forall t\in\Ical_{B1}\). Moreover, recall that \(\es(t) \in (0,+\infty)\), \(\forall t \in [0,\taum)\) and thus \(\epis\) can only diverge upward. Now suppose \(\epis\to+\infty\) as \(t\to t'\) for some \(t'\in\Ical_{B1}\). Since \(\sigma_s,\epih\in\Linf\),  $\varphigam = 0$, and \(\grad{\alphs}^\top\grad{\alphh}\ge0\), if \(\grad{\alphs}\neq0\) the bracketed term on the RHS of \eqref{eq:epis_dot} remains positive, so \(\depis\to-\infty\), contradicting \(\epis\to+\infty\). In addition, under Assumption \ref{assu:alpha_globalmax}, similarly to \textit{Mode A} one can conclude that \(\grad{\alphs} = 0\) ensures the boundedness of \(\epis\), contradicting \(\epis\to+\infty\). Hence, \(\epis\in\Linf\) on \(\Ical_{B1}\), and there exists a \(\epibs^{B1}>0\) such that \(0 < \epis \leq \epibs^{B1}\), $\forall t \in \Ical_{B1}$.

\underline{\textit{Mode B2 ($\varphih(t) > 0$ \& $\varphigam(t) > 0$)}}:
Here $t \in \Ical_{B2}$ and, since $\varphigam > 0$, $\gamma = \epih\,\grad{\alphs}^\top \grad{\alphh} < 0$. Consequently, as $\epih>0$ for all $t\in[0,\taum)$, we have $\grad{\alphs}^\top \grad{\alphh} < 0$ whenever \textit{Mode B2} is active. We use proof-by-contradiction to establish \(\epih, \epis \in \Linf\), $\forall t \in \Ical_{B2}$. 

From \eqref{eq:mapped_alphs}, letting \(\epis\to+\infty\) gives \(\es=\alphs(t,\bx_1(t))-(\rhosn(t)-\rhosr(t))\to0\). Since \(\rhosn(t)\in \Linf\), \(\forall t\geq 0\) and, by \textit{Phase I}, \(\alphs(t,\bx_1(t))\in \Linf\), \(\forall t \in [0,\taum)\), it follows that \(\rhosr(t)\), and thus \(\sigma_s\) in \eqref{eq:epis_dot} remain bounded as \(\epis\to + \infty\).

Moreover, with \(\gamma=\epih\, \grad{\alphs}^{\top} \grad{\alphh}\) and \(\grad{\alphs}^{\top} \grad{\alphh}<0\), \(\gamma\) becomes arbitrarily negative for large \(\epih\). Because the smooth switch is defined by \(\varphigam\coloneqq\varphi(\gamma,0,-\delta_\gamma)\) with fixed \(\delta_\gamma>0\), we have \(\varphigam(t)=1\) whenever \(\epih\) is sufficiently large (i.e., when \(\epih\to+\infty\)).

Suppose both \(\epis, \epih \to +\infty\) as \(t \to t'\) for some \(t' \in \Ical_{B2}\). Because \(\varphigam(t)=1\), the factor \(1-\varphigam\) in
\eqref{eq:epis_dot} vanishes, i.e.\ the relaxation dynamics
\eqref{eq:rho_relax_dyn} exactly offsets the hard-constraint cross term
\(\varphih k_h \epih^2 \grad{\alphs}^\top \grad{\alphh}\), leaving
\(k_s \epis^2 \|\grad{\alphs}\|^2 + \sigma_s\) as in \textit{Mode A}; since
\(\sigma_s \in \Linf\), this implies \(\depis \to -\infty\) whenever
\(\grad{\alphs}\neq \mathbf{0}_n\). This contradicts our assumption, so \(\epis\) and \(\epih\) cannot diverge simultaneously. Next, assume \(\epis \to +\infty\) while \(\epih \in \Linf\).  The RHS of \eqref{eq:epis_dot} again forces \(\depis \to -\infty\) when \(\grad{\alphs}\neq \mathbf{0}_n\), contradicting \(\epis \to +\infty\). Conversely, if \(\epih \to +\infty\) while \(\epis \in \Linf\), the RHS of \eqref{eq:epih_dot} yields \(\depih \to -\infty\) whenever \(\grad{\alphh}\neq 0\), which is likewise contradictory. Similarly to the discussions provided in the analysis of previous modes one can claim that \(\epis\) and \(\epih\) remain bounded when \(\grad{\alphs}=0\) and \(\grad{\alphh}=0\), respectively. Hence we conclude \(\epis, \epih \in \Linf\) for all \(t \in \Ical_{B2}\). Therefore, constants \(\epibh^{B2}>0\) and \(\epibs^{B2}>0\) exist such that $0 < \epih \leq \epibh^{B2}, 0 < \epis \leq \epibs^{B2}, \forall t \in \Ical_{B2}$. 

Define $\epibs \coloneqq \max\{\epibs^{A}, \epibs^{B1}, \epibs^{B2}\}$ and $\epibh \coloneqq \max\{\epibh^{A},$ $\epibh^{B1}, \epibh^{B2}\}$, which are independent of $\taum$. Now combining all the results from \textit{Modes A, B1}, and \textit{B2} gives 
\begin{equation} \label{eq:bounded_epih_epis_step1}
	0 < \epis \leq \epibs, \quad \text{and} \quad 0 < \epih  \leq \epibh, \quad \forall t \in [0,\taum).
\end{equation}

Since \eqref{eq:bounded_epih_epis_step1} gives $\epih>0$ and $\epih\in\Linf$ on $[0,\taum)$,
Lemma \ref{lem:nonnegative_rho_s_relax} in \ref{appen:lemma} yields $\rhosr(t)\ge0$ and
$\rhosr(t)\in\Linf$, $\forall t\in[0,\taum)$. Hence, there exists a constant $\rhosrb>0$, independent of $\taum$ (one may take $\rhosrb = \bar{d}/\krec$, with $\bar{d}$ as in the proof of Lemma \ref{lem:nonnegative_rho_s_relax}), such that
\begin{equation} \label{eq:rhosr_strict_bounds}
	0 \le \rhosr(t) \le \rhosrb, \qquad \forall t \in [0,\taum).
\end{equation}

By \eqref{eq:soft_const_set_lower_bound}, \eqref{eq:es}, \eqref{eq:mapped_alphh},
\eqref{eq:mapped_alphs}, \eqref{eq:bounded_epih_epis_step1}, \eqref{eq:rhosr_strict_bounds},
the \textit{Phase I} results, and $\rho_0 \le \rhosn(t) \le 0$ from \eqref{eq:alpha_lower_bound}, it follows that
\begin{equation} \label{eq:alphas_alphah_strict_bounds}
	\begin{aligned}
		\frac{1}{\epibh} &\leq \alphh(t,\bx_1(t)) \leq b_h, \\
		\frac{1}{\epibs} + \rho_{0} - \rhosrb &\leq \alphs(t,\bx_1(t)) \leq b_s,
	\end{aligned}
\end{equation}
$\forall t \in [0,\taum)$, where $b_h, b_s$ are the constants from \textit{Phase I}. Since $\rho_0 \le 0$ in \eqref{eq:alpha_lower_bound}, \eqref{eq:es}, \eqref{eq:rhosr_strict_bounds}, and \eqref{eq:alphas_alphah_strict_bounds} give
\begin{equation} \label{eq:es_strict_bounds}
	0 < \frac{1}{\epibs} \le \es(t,\bx_1(t)) \le b_s - \rho_{0} + \rhosrb, 
	\; \forall t \in [0,\taum).
\end{equation}

Finally, by \eqref{eq:bounded_epih_epis_step1} and the boundedness of $\grad{\alphh}(t,\bx_1)$, $\grad{\alphs}(t,\bx_1)$ and $\bG_1^{-1}(t,\bx_1)$ (Assumption \ref{assu:known_unknown}), $\bs_1(t,\bx_1)$ in \eqref{eq:1st_intermed_ctrl} is bounded $\forall t \in [0,\taum)$. 

Using the above bounds and Assumptions \ref{assu:known_unknown}, \ref{assu:bounded_in_t_psi_fun} and \ref{assu:bounded_in_t_psi_derivatives}, together with the continuous differentiability of $\bG_1$, one verifies that $\dot{\bs}_1(t,\bx_1)$ is also bounded.

\textbf{Phase II (Step $\mathbf{2 \leq i \leq r}$)}:
Differentiating $\bepi_i = \col{\epi_{i,j}}$ in time and using \eqref{eq:mapping_fun}, \eqref{eq:intermediate_err},  \eqref{eq:sys_dynamics_mimo}, and \eqref{eq:x_rewritten} gives
\begin{flalign} \label{eq:epi_i_dot}
	&\dbepi_i = \bXi_i \Big[\bof_i(t,\bx_1,\beh_2,\ldots,\beh_i) + \bG_i(t,\bx_1,\beh_2,\ldots,\beh_i)& \nonumber \\
	& \quad \quad \quad \quad \times \Big(\bvarTheta_{i+1} \, \beh_{i+1} + \bs_{i}(t,\bx_1, \beh_2,\ldots,\beh_i) \Big)&  \\
	& \quad \quad \quad \;  - \dot{\bs}_{i-1}(t,\bx_1, \beh_2,\ldots,\beh_{i-1}) - \dot{\bvarTheta}_i \beh_i \Big], \quad i \in \mbI_2^{r-1},& \nonumber
\end{flalign}
where for $i=r$, the term $\bvarTheta_{i+1} \, \beh_{i+1} + \bs_{i}$ should be replaced by $\bu$ in \eqref{eq:control_law}. Recall that $\bvarTheta_i \coloneqq \diag{\vartheta_{i,j}}$ and $\bXi_i \coloneqq \diag{\xi_{i,j}}$, in which $\xi_{i,j}$ are given in \eqref{eq:xi_i,j}.

The rest of the proof in \textit{Phase II} continues similarly to steps 2 to $r$ of Theorem 1's proof in \cite{mehdifar2024low}. Therefore for brevity we only provide the proof sketch in the following. For each step $i = 2, \ldots, r$, we consider barrier function $V_i(\bepi_i) = \frac{1}{2} \bepi_i^\top \bepi_i$ as a positive definite and radially unbounded Lyapunov function candidate with respect to $\bepi_i$. Then using the results in \textit{Phase I}, Assumptions \ref{assu:bounded_in_t_dyn}, \ref{assu:symmetric_com_pd}, and the boundedness of $\bs_{i-1}(t,\bx_1,\beh_2, \ldots, \beh_{i-1}), \dot{\bs}_{i-1}(t,\bx_1,\beh_2, \ldots, \beh_{i-1})$, $\forall t \in [0,\taum)$, established in the previous step, and following standard Lyapunov analysis we can show that $\bepi_i$ is Uniformly Ultimately Bounded (UUB) \cite{khalil2002noninear}. Therefore, there exists $\epib_i > 0$ such that $\|\bepi_i\|<\epib_i$, $\forall t \in [0,\taum)$. Hence, from the inverse logarithmic transformation in \eqref{eq:mapping_fun}, we get
\begin{equation} \label{eq:epi_i_strictsubset}
	\!\!\! -1 < \tfrac{e^{-\epib_i} -1}{e^{-\epib_i} + 1} \eqqcolon -\varsigma_{i,j}  \leq \eh_{i,j}(t) \leq \varsigma_{i,j} \coloneqq \tfrac{e^{\epib_i} -1}{e^{\epib_i} + 1} < 1.
\end{equation}
Moreover, at each step one can establish $\bs_i(t,\bx_1, \beh_2,$ $\ldots, \beh_i)$, $\dot{\bs}_i(t,\bx_1,\beh_2, \ldots, \beh_i) \in \Linf$, $\forall t \in [0,\taum)$. 

As a result, by proceeding recursively, we can show that all intermediate control signals $\bs_i$ and the control law $\bu$ remain bounded $\forall t \in [0, \taum)$. Furthermore, from \eqref{eq:x_rewritten} and the results established so far, it follows that all system states $\bx_i$, $i\in\mbI_1^{r-1}$, are also bounded for all $t \in [0, \taum)$.

\textbf{Phase III}: 
To prove that $\taum = \infty$, first collect the bounds in \eqref{eq:rhosr_strict_bounds}, \eqref{eq:alphas_alphah_strict_bounds}, \eqref{eq:es_strict_bounds}, and \eqref{eq:epi_i_strictsubset}, and define  
\begin{subequations}
	\begin{flalign}
		&\Oalph^\prime \coloneqq [\tfrac{1}{\epibh},b_h],& \\
		&\Omega_{\es}^\prime \coloneqq \bigl[\tfrac{1}{\epibs},\, b_s - \rho_0 + \rhosrb\bigr], \\
		&\Orhosr^\prime \coloneqq [0, \rhosrb],& \\
		&\Omega_{\beh_{i}}^\prime \coloneqq [-\varsigma_{i,1},\,\varsigma_{i,1}] \times \dotsb \times [-\varsigma_{i,n},\,\varsigma_{i,n}], \quad i \in \mbI_2^r,& \\ 
		&\Omega_{\beh}^\prime \coloneqq \Omega_{\beh_{2}}^\prime \times \dotsb \times \Omega_{\beh_{r}}^\prime \subset (-1,1)^n \times \dotsb \times (-1,1)^n\!.\!\!\!\!\!\!\!\!&
	\end{flalign}
\end{subequations}

Next, owing to the compactness of the level sets of $\alphh(t,\bx_1)$ and the bounds in \eqref{eq:alphas_alphah_strict_bounds}, we have 
$\bx_1(t) \in \Ox^\prime(t) \subset \mathbb R^n, \forall t \in [0,\taum)$, where
\begin{equation}
	\Ox^\prime(t) \coloneqq \{\bx_1 \in \mathbb R^n \mid \tfrac{1}{\epibh} \le \alphh(t,\bx_1) \le b_h\}.
\end{equation}
Define the time-invariant super-set $\Oxsp \coloneqq \bigcup_{t=0}^{+\infty} \Ox^\prime(t) \subset \mathbb R^n$. Then we have $\bx_1(t) \in \Oxsp \subset \mathbb R^n$ for all $t \in [0,\taum)$.

Finally, define $\Oz^\prime \;\coloneqq\; \Oxsp \times \Oalph^\prime \times \Orhosr^\prime \times \Omega_{\es}^\prime \times \Omega_{\beh}^\prime$, which is a non-empty compact subset of \(\Oz\).  \textit{Phase I} and \textit{Phase II} ensure that $\bz (t;\bz(0)) \in \Oz^\prime, \forall\,t \in [0,\taum)$. Assume, for contradiction, that \(\taum < \infty\). Since \(\Oz^\prime \subset \Oz\), \cite[Prop.~C.3.6, p.~481]{sontag1998mathematical} gives a time \(t^{\prime}\in[0,\taum)\) with $\bz (t^{\prime};\bz(0)) \notin \Oz^\prime$, contradicting the above. Hence \(\taum=\infty\). Consequently, every closed-loop control signal remains bounded \(\forall t\ge 0\).  Moreover, since $\alphh(t,\bx_1(t)) \in [\tfrac{1}{\epibh},b_h] \subset (0,+\infty)$ and $\es(t,\bx_1(t)) \in \bigl[\tfrac{1}{\epibs},\, b_s - \rho_0 + \rhosrb\bigr] \subset (0,+\infty)$, conditions \eqref{eq:hard_const_satisfaction_cond} and \eqref{eq:soft_const_satisfaction_cond} are satisfied for all time. Therefore, the time-varying set \(\Oh(t)\cap\Osv(t)\) is forward invariant, completing the proof.		\hfill\qedsymbol

\section{On Properties of $\rhosr(t)$}
\label{appen:lemma}
The following lemma gives two useful properties of \(\rhosr(t)\).

\begin{lemma} \label{lem:nonnegative_rho_s_relax}
	If $\epih > 0$, the relaxation signal \(\rhosr(t)\) governed by \eqref{eq:rho_relax_dyn} satisfies \(\rhosr(t)\ge 0\) for all \(t\ge 0\) for which the solution exists. If in addition $\epih \in \Linf$, then \(\rhosr \in \Linf\).
\end{lemma}
\begin{proof}
	Note that whenever $\varphih(t)=0$ or $\varphigam(t)=0$, the first term on the RHS of \eqref{eq:rho_relax_dyn} vanishes. When $\varphih(t)>0$ and $\varphigam(t)>0$, \eqref{eq:rho_relax_dyn} gives
	\begin{equation*}
		\drhosr = - \varphigam \varphih k_h \epih^2 \,\grad{\alphs}^\top \,\grad{\alphh} \;-\; \krec\,\rhosr,
	\end{equation*}
	where $\grad{\alphs}^\top \,\grad{\alphh}<0$, $k_h>0$, and $\epih > 0$. As a result, the first term on the RHS of $\drhosr$ is always non‑negative. 
	
	Let $t'\in[0,\infty)$ be arbitrary. If $\rhosr(t')=0$, then $\drhosr(t')\ge0$, so $\rhosr(t)\geq0$ for all $t\ge t'$. Setting $t'=0$ gives $\rhosr(t)\ge 0$ for all $t\ge 0$ whenever the solution of \eqref{eq:rho_relax_dyn} exists.  
	
	Because $\epih > 0$ and $\epih \in \Linf$, \eqref{eq:mapped_alphh} ensures $\alphh(t,\bx_1)>0$. Moreover, $\alphh(t,\bx_1)$ has compact level sets (Assumption \ref{assu:coercive_alpha_funs}) and satisfies $\alphh(t,\bx_1)\le\alphhstr(t)\coloneqq\max_{\bx_1\in\mbR^n}\alphh(t,\bx_1)$, where $\alphhstr(t)$ is finite and positive (Assumption \ref{assu:feasible_constrained_sets}). Thus $\alphh(t,\bx_1)>0$ implies that $\bx_1$ is bounded. By Assumption \ref{assu:bounded_in_t_psi_derivatives}, the expressions in \ref{appen:gradients}, boundedness of $\bx_1$, and continuity of $\grad{\alphs}$ and $\grad{\alphh}$, all these terms are bounded. Hence, if $\epih > 0$ and $\epih \in \Linf$, the first term on the RHS of \eqref{eq:rho_relax_dyn} is bounded and non‑negative. 
	
	Finally, let $\bar{d} > 0$ bound the first term on the RHS of \eqref{eq:rho_relax_dyn}, so $\drhosr \le \bar{d} - \krec\,\rhosr$. Since $\rhosr(0)=0$, the comparison lemma \cite[Lemma 3.4]{khalil2002noninear} yields $\rhosr(t) \le (\bar{d}/\krec)\big(1 - e^{-\krec t}\big) < \bar{d}/\krec$. Hence $\rhosr(t)$ remains bounded.
\end{proof}

\bibliographystyle{elsarticle-num}
\bibliography{Refs} 

\end{document}